\documentclass[a4paper,11pt]{article}
\pdfoutput=1 

\usepackage{jheppub} 
            
\usepackage{bm,amssymb,slashed,graphicx,multirow,soul,mathtools,xspace,array,comment}  
\usepackage{float}                   
\allowdisplaybreaks
\usepackage{ bbold }
\usepackage{color}
\usepackage{subfigure}
\usepackage[printonlyused]{acronym}
\usepackage{hyperref}
\acrodef{PDG}[PDG]{Particle Data Group}
\acrodef{OPE}[OPE]{Operator Product Expansion}
\acrodef{FCNC}[FCNC]{flavour-changing neutral current}
\acrodef{RHC}[RHC]{right-handed currents}
\acrodef{SM}[SM]{Standard Model}
\acrodef{NP}[NP]{New Physics}
\acrodef{MFV}[MFV]{Minimal Flavour Violation}
\acrodef{SD}[SD]{short-distance}
\acrodef{LD}[LD]{long-distance}
\acrodef{DA}[DA]{distribution amplitude}

\newcommand{\matel}[3]{\langle #1|#2|#3\rangle}
\newcommand{\al}{\alpha}
\newcommand{\be}{\beta}
\newcommand{\ga}{\gamma}
\newcommand{\de}{\delta}
\newcommand{\la}{\lambda}
\newcommand{\eps}{\epsilon}
\newcommand{\veps}{\varepsilon}
\newcommand{\gauge}{\xi}

\newcommand{\GeV}{\,\mbox{GeV}}
\newcommand{\MeV}{\,\mbox{MeV}}
\newcommand{\ORD}{{\cal O}}

\newcommand{\BKee}{\Bin \to \Kout e^+ e^-} 
\newcommand{\BKmumu}{\Bin \to \Kout \mu^+ \mu^-}

\newcommand{\Cdot}{\!\cdot \!}
\newcommand{\mi}{\!-\!}
\newcommand{\pl}{\!+\!}

\newcommand{\TAB}{Tab.~}
\newcommand{\FIG}{Fig.~}
\newcommand{\FIGs}{Figs.~}
\newcommand{\SEC}{Sec.~}
\newcommand{\SECs}{Secs.~}
\newcommand{\APP}{App.~}
\newcommand{\APPs}{Apps.~}
\newcommand{\EQ}{Eq.~}
\newcommand{\EQs}{Eqs.~}

\newcommand{\Rea}{\textrm{Re}}

\newcommand{\mlone}{m_{\ell_1}}
\newcommand{\mltwo}{m_{\ell_2}}

\newcommand{\mK}{m_{K}}

\newcommand{\FRone}{(1)}
\newcommand{\FRtwo}{(2)}
\newcommand{\FRthree}{(3)}
\newcommand{\FRfour}{(4)}
\newcommand{\pBbar}{\bar{p}_B}
\newcommand{\pKbar}{\bar{p}_K}

\newcommand{\mga}{m_\gamma}

\newcommand{\KallenB}{{ \lambda_{B}}}

\newcommand{\CV}{C_V}
\newcommand{\CA}{C_A}

\newcommand{\Tg}{\theta_{\ga}}
\newcommand{\Tl}{\theta_{\ell}}

\newcommand{\qz}{q_0}

\newcommand{\Tlz}{\theta_0}

\newcommand{\claa}{c_{\AAA}}

\newcommand{\cl}{c_{\ell}}

\newcommand{\clz}{c_{0}}
\newcommand{\slz}{s_{0}}

\newcommand{\QQ}[1]{\Delta_q^{ #1}}

\newcommand{\Fg}{\phi_{\ga}}

\newcommand{\conetwo}{C_{ij}}
\newcommand{\coneone}{C_{ii}}
\newcommand{\ctwotwo}{C_{jj}}
\newcommand{\Rlog}{R_{ij}}
\newcommand{\Slog}{S_{ij}}
\newcommand{\glogf}{f_{ij}}
\newcommand{\glogg}{g_{ij}}
\newcommand{\glogh}{h_{ij}}

\newcommand{\lonetwo}{\ell_{1,2}}
\newcommand{\ltwoone}{\ell_{2,1}}
\newcommand{\lone}{\ell_1}
\newcommand{\ltwo}{\ell_2}

\newcommand{\bltwo}{\bar{\ell}_2}
\newcommand{\pK}{p_K}
\newcommand{\pB}{p_B}

\newcommand{\loneg}{\ell_{1\ga}}
\newcommand{\ltwog}{\ell_{2\ga}}

\newcommand{\modveclone}[1]{|\vec{\ell}_1^{#1}|}

\newcommand{\veclone}[1]{\vec{\ell}_{1}^{#1}}

\newcommand{\Ampzero}{{\cal A}^{(0)}}
\newcommand{\Ampone}{{\cal A}^{(1)}}
\newcommand{\Amponered}{{a}^{(1)}}

\newcommand{\Amponesred}{{a}^{(1)*}}
\newcommand{\AmponeLow}{{\cal A}^{(1)}_{\textrm{Low}}}
\newcommand{\Amptwo}{{\cal A}^{(2)}}

\newcommand{\rsoft}{r_{\textrm{soft}}}

\newcommand{\reg}{{\cal R}}

\def\Li{\mathop{\mathrm{Li}_2}\nolimits}

\newcommand{\Dep}{\Delta _{\epsilon }}

\newcommand{\Demlh}{\Delta \bar{m}_{\ell}}
\newcommand{\monetwoh}{\bar{m}_{\lone\ltwo}}
\newcommand{\DemBKsq}{\Delta m_{BK}^2}
\newcommand{\DemBKsqh}{\Delta \bar{m}_{BK}^2}
\newcommand{\Kallenl}{\lambda_{\ell}}

\newcommand{\Deps}{\frac{1}{\hat{\eps}}}
\newcommand{\DepsIR}{\frac{1}{\hat{\eps}_{\textrm{IR}} }}
\newcommand{\DepsUV}{\frac{1}{\hat{\eps}_{\textrm{UV}} }}
\newcommand{\epsUV} {\eps_{\textrm{UV}}}

\newcommand{\geff}{g_{\textrm{eff}}}

\newcommand{\EGAmax}{E_{\ga}^\mathrm{max}}

\newcommand{\EGAmaxn}[1]{(E_{\ga }^{(#1)})^{\mathrm{max}}}

\newcommand{\Bga}{B}
\newcommand{\Kga}{K}

\newcommand{\dH}{\Delta{\cal H}}

\newcommand{\dF}{\Delta{\cal F}}

\newcommand{\des}{\de_{\textrm{ex}}} 
\newcommand{\desinc}{\de_{\textrm{ex}}^{\textrm{inc}}}

\newcommand{\AAA}{a }

\newcommand{\Des}{\omega_s}

\newcommand{\Dec}{\omega_c}
\newcommand{\Desc}{\omega_{s,c}}
\newcommand{\MCsoft}{{\Lambda_{s}}}

\newcommand{\ff}{\omega^2}
\newcommand{\TT}{\ell}
\newcommand{\RR}{0}
\newcommand{\LO}{\,  \mathrm{{ LO}}}

\newcommand{\qSaa}{q^2_{\AAA}}
\newcommand{\Tlaa}{\theta_{\AAA}}

\newcommand{\Iijzero}{I_{ij}^{(0)}}
\newcommand{\soft}{{(s)}}

\newcommand{\hc}{{(hc)}} 
\newcommand{\hca}{{(hc)(a)}} 
\newcommand{\tH}{{\tilde{\cal H}}} 
\newcommand{\tF}{{\tilde{\cal F}} }

\newcommand{\Bin}{{\bar{B}}}
\newcommand{\Kout}{{\bar{K}}} 

\renewcommand{\simeq}{\approx} 

\definecolor{violet}{rgb}{0.94, 0.2, 0.8}
\definecolor{lightblue}{rgb}{0.39, 0.58, 1.00} 
\definecolor{lightgreen}{rgb}{0.1, 0.73, 0.33}

\numberwithin{equation}{section}

\newcommand*{\mathcolor}{}
\def\mathcolor#1#{\mathcoloraux{#1}}
\newcommand*{\mathcoloraux}[3]{%
  \protect\leavevmode
  \begingroup
    \color#1{#2}#3%
  \endgroup
}

\title{ \boldmath QED Corrections in $\bar{B} \to \bar{K} \ell^+ \ell^-$ at the Double-Differential Level}
\author[1]{Gino Isidori,}
\author[2]{Saad Nabeebaccus,}
\author[2]{Roman Zwicky}

\affiliation[1]{Department of Physics, Universit\"at Z\"urich, 
Winterthurerstrasse 190,  CH-8057 Z\"urich, Switzerland}
 \affiliation[2]{Higgs Centre for Theoretical Physics, School of Physics and Astronomy, University of Edinburgh, Edinburgh EH9 3JZ, Scotland}

\emailAdd{isidori@physik.uzh.ch}
\emailAdd{saad.nabeebaccus@ed.ac.uk}
\emailAdd{roman.zwicky@ed.ac.uk}

\abstract{
We present a detailed  analysis of QED corrections to $\bar{B} \to \bar{K} \ell^+ \ell^-$  decays at the double-differential level. 
Cancellations of soft and collinear divergences are demonstrated analytically 
using the phase space slicing method. Whereas soft divergences are found to cancel at the 
differential level, the cancellation of the hard-collinear logs $\ln m_\ell$ 
require, besides photon-inclusiveness, 
a specific choice of  kinematic variables. In particular, 
hard-collinear logs in the lepton-pair  invariant mass distribution ($q^2$), are sizeable 
and need to be treated with care when comparing with experiment.
Virtual and real  amplitudes are evaluated using an effective mesonic Lagrangian.
Crucially, we  show that going beyond this approximation does not introduce any further
infrared sensitive terms.
All analytic computations are performed for generic charges and are therefore 
adaptable to semileptonic decays such as $\bar B \to D \ell \bar \nu$. 
}

\begin{document}
\preprint{CP3-Origins-2020-07, DNRF90, ZU-TH 30/20 }

\toccontinuoustrue

\maketitle 
\newpage

\flushbottom

\setcounter{tocdepth}{3}
\setcounter{page}{1}
\pagestyle{plain}


\section{Introduction}
\label{sec:intro}

Rare semileptonic $B$ decays of the type $\Bin \to \Kout^{(*)}\ell^+\ell^-$ have received significant 
interest in the last few years because 
of the hints of Lepton Flavour Universality (LFU) violations reported by the LHCb experiment~\cite{Aaij:2014ora,Aaij:2017vbb,Aaij:2019wad} and  \cite{Bifani:2018zmi} for a review, 
which could be due to physics beyond the Standard Model (SM).  
In view of higher statistics results on these modes, a detailed study of this phenomenon requires 
an accurate estimate of all possible sources of LFU violation present within the SM.

Besides trivial kinematic mass effects, 
the only potential large source of LFU violation present in the SM  are hard-collinear singularities 
in QED.
These can induce non-universal corrections  of order $\ORD(\al) \ln (m_\ell/m_B)$
to the physical decay rates (depending on the definition of the observables), which can be large for light leptons.
These effects are well known and, to a large extent, corrected for in the experimental 
analyses through  Monte Carlo simulations (e.g. PHOTOS~\cite{PHOTOS}). 
In order to cross-check 
the reliability of the approximations which are behind this treatment,  a detailed analytic
analysis of QED corrections is desirable. A first step in this direction was undertaken in~\cite{BIP16}, 
where semi-analytic results for the  LFU ratios $R_K$ and $R_{K^*}$ have been presented.
Here we go one step further: we focus our attention on the process 
$\Bin \to \Kout \ell_1 \bar \ell_2$ (which is a good prototype for a wide class of interesting 
semi-leptonic decays, including charged-current transitions such as  $\Bin \to\pi\ell\nu$),
and analyse  QED corrections at a fully differential level in terms of the 
``visible" kinematics (i.e.~in terms of the two variables that fully specify  the kinematics of 
the non-radiative mode).  Moreover, we present a complete analysis of the problem of evaluating  
both real and virtual corrections within an effective meson approach which is an improvement 
over scalar QED.
 As we demonstrate, 
this approach is sufficient to trace back the origin of all  ``dangerous'' collinear singularities. 

While soft QED singularities cancel out at the differential level in any infrared-safe observable,
the cancellation of the collinear singularities, which are actually physical effects regulated by the lepton mass,
is more subtle. As we show,  the choice of kinematic variables plays a key role in obtaining  
a differential distribution that is not only infrared-safe, but also free from the sizeable LFU violating terms
of order $\ORD(\al) \ln (m_\ell)$.
In particular, as far as the invariant mass 
of the two lepton system is concerned, the following two options can be considered:
$ q_\ell^2  = (\lone+\ltwo)^2  $ and $ q^2_0 = (\pB- \pK)^2 $. 
The first case ($q_\ell^2$), which is the natural choice for experiments where the $B$ momentum is not known
(such as those performed at hadron colliders),
corresponds to defining the invariant mass of the charged lepton system  from the measured 
lepton momenta ($\lonetwo$), i.e.~after radiation has occurred. 
Whereas in the  second case ($q_0^2$),   the hadronic momenta ($p_{B,K}$) are used to define the momentum transfer
to the lepton system before radiation. These two choices coincide in the non-radiative case, but are different in the presence of 
radiation. We show that it is only by using $q_0^2$, as the relevant kinematic variable, that
the  differential distribution is free from $\ORD(\alpha) \ln (m_\ell)$-terms. This does not imply
that one cannot perform clean tests of LFU at hadron colliders, but rather that in such cases, 
the collinear singularities are unavoidable and should be properly corrected for. 

The paper is organised as follows. In \SEC\ref{sec:comp} the computation of real and virtual amplitudes  is presented, 
as well as the phase space measure including the physical cut-off on the photon energy.
Treatment and cancellation of infrared divergences is discussed at length in \SEC\ref{sec:IR}. 
Numerical results, in form of plots, showing
the size of the radiative corrections, are presented in \SEC\ref{sec:plots}.
An outlook on open issues and future directions is presented in \SEC\ref{sec:outlook}.
The paper is concluded in \SEC\ref{sec:conclusion}. The appendices contain 
additional plots, comparison with older work,  comments on $R_K$ (\ref{app:fplots}), 
 amplitudes (\ref{app:AMP}), 
  the parametrisation of kinematic variables (\ref{app:kinematics}), 
  the soft integrals (\ref{app:integral}) 
 and the explicit Passarino-Veltman functions (\ref{app:PV}).

\section{Computation}

\label{sec:comp}

\begin{figure}[t]
	\centering
	\includegraphics[width=0.8\linewidth]{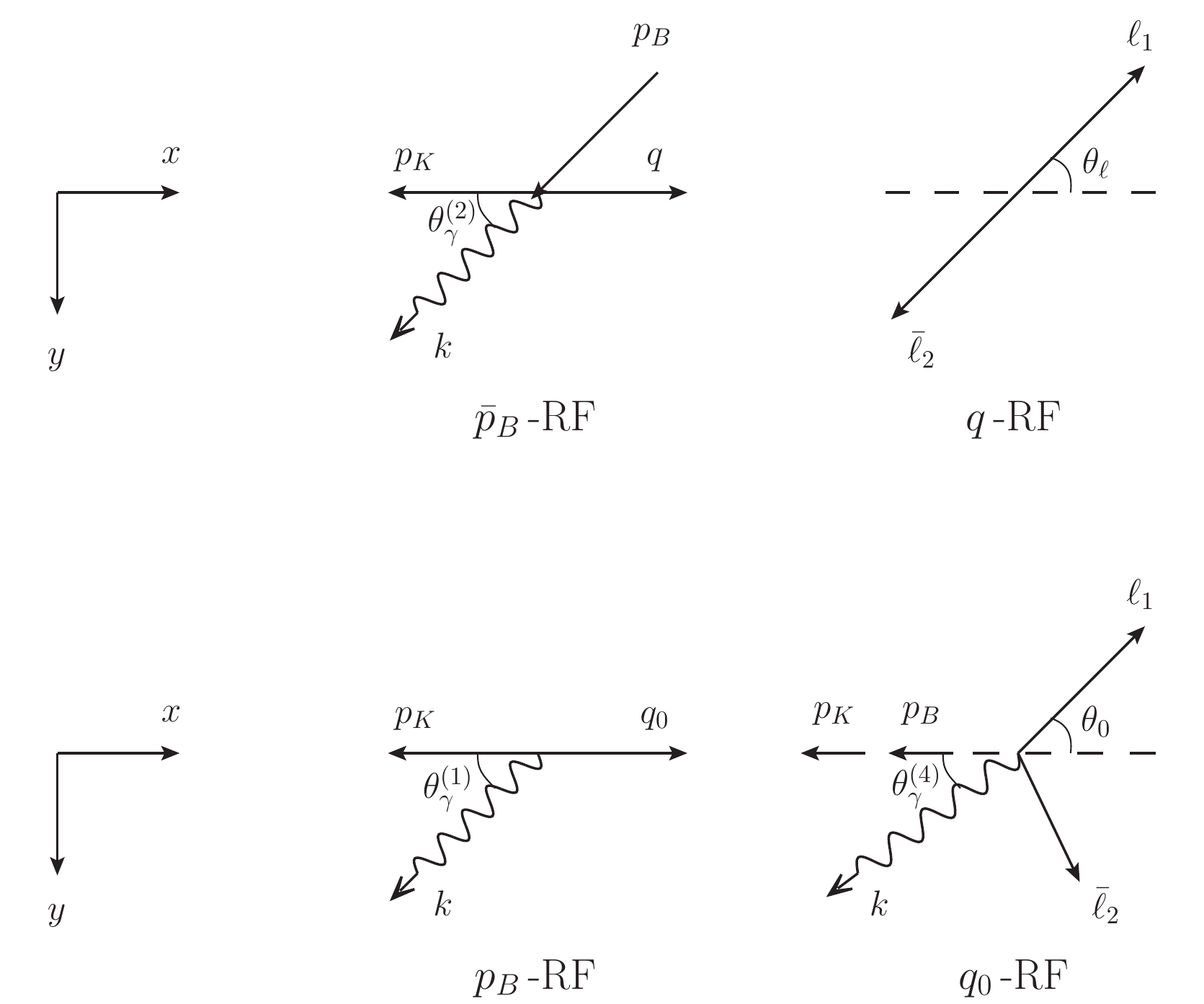}
		\vskip 0.5 true cm
	\caption{\label{fig:angles}
	\small Decay kinematics for the different RFs of interest.	 
	The dashed line corresponds to what is deferred to as the decay axis. 
	For brevity we drop the frame-label on the lepton angles, $\Tl \equiv \Tl^{\FRthree}$ and $\Tlz \equiv \Tlz^{\FRfour}$,  and if no frame-label is indicated on 
	the photon angle,
	$\Tg = \Tg^{\FRtwo}$ is usually understood.}
\end{figure}

The two sets of variables we introduce to describe the differential distribution of the 
$\Bin(p_B) \to  \Kout(\pK)   \lone(\lone)   \bltwo(\ltwo) +  \ga(k)$ process, 
assuming that radiation is not detected, are  
\begin{equation} 
\label{eq:tr}
\{ \qSaa,  \claa \}   = \left\{  \begin{array}{lll}
 q_\ell^2 = (\lone+\ltwo)^2,~ 
 	&  \cl =  - \left(\frac{\vec{\lone}\cdot \vec{p}_K}{ |\vec{\lone}| | \vec{p}_K| } \right)_{q-\textrm{RF}} ~
	&  [\text{``Hadron collider'' variables}]~, \\
q^2_0 = (\pB-\pK)^2~, 
	&  \clz=  - \left(\frac{\vec{\lone}\cdot \vec{p}_K}{ |\vec{\lone}| | \vec{p}_K| } \right)_{q_0 -\textrm{RF}}  ~
	& \textrm{[``B-factory'' variables]} ~,    \end{array} \right. 
\end{equation}
where $q-\textrm{RF}$ and $q_0-\textrm{RF}$ denotes the rest frames of 
 \begin{equation}
q \equiv \lone + \ltwo \;, \quad \qz \equiv  \pB-\pK = q + k \;,
\end{equation}
as illustrated in \FIG\ref{fig:angles} (to conform to standard notations, throughout the paper $q_{\ell} \equiv q$).
As indicated, the set $a=\ell$ is the natural choice for a hadron-collider experiment, while the set $a=0$ 
can be implemented only in an experiment where the $B$ momentum is known. However, as we shall discuss later on,
both sets can be applied to describe appropriate integrated distributions in any kind of experiment. 

A further variable that plays a key role in defining infrared-safe observables is 
\begin{equation}
\label{eq:momC}
\pBbar \equiv p_B - k = \lone + \ltwo + \pK \;,
\end{equation}
which equals the sum of all visible final-state momenta. The kinematic invariant $\pBbar^2$ is the
reconstructed $B$-meson mass in the hadronic set-up, where $\pB$ is not known, and the variable 
\begin{equation} 
\label{eq:des}
\des  \, > \, 1 - \frac{\pBbar^2}{ m_B^2}  \;,
\end{equation} 
provides the most natural choice for the physical cut-off regulating soft divergences.
The complete decomposition of all momenta in the  $p_B$, $\pBbar$, $q$ and $\qz$ RFs
is presented in Appendix~\ref{app:kinematics}, and frames are denoted as $\FRone$, $\FRtwo$, $\FRthree$ and $\FRfour$, 
respectively.

Schematically, we decompose the double differential rate as
\begin{equation}
\label{eq:basicq0}
d^2  \Gamma_{ \Bin \to \Kout   \lone  \bltwo} (\des)  =
\frac{1}{m_B} \left( {\rho}_{\AAA}(m_B^2) |{\cal A}_V|^2 +  
 \int_{\des} d \Phi_\ga  \,{\rho}_{\AAA}(\pBbar^2)  \, |{\cal A}_R|^2  \right) d \qSaa d \claa   \;,
\end{equation}
where  ${\rho}_{\AAA}(m_B^2)$  and ${\rho}_{\AAA}(\pBbar^2)$ denote the 3-body and ``effective-3-body''
phase space factors, and $d \Phi_\ga$ indicates the integration over the undetected photon variables
over a phase space region specified by the physical cut-off $\des$. In the following, we first introduce the 
effective Lagrangians used in our analysis, and then present the calculation of the real emission amplitude (${\cal A}_R$)
and the one-loop virtual corrections to the tree-level 3-body amplitude (${\cal A}_V$),
and finally discuss the corresponding phase space factors. 
Soft divergences and ultraviolet (UV) divergences are regulated in dimensional regularisation (DR).\footnote{~Often in 
QED calculations soft divergences are regulated via an explicit photon mass. For this reason, whenever possible,
we will indicate how results change when using this regulator. However, we found that DR is 
more convenient in performing the soft integrals, this is why we adopt it as default approach.}

\subsection{Mesonic effective Lagrangian}

Generically, we consider non-radiative processes of the type $M_H \to M_L \lone \bltwo$, 
where $M_{H,L}$ are generic scalar mesons (of either parity).
In what follows we take $M_H = \Bin$ and $M_L = \Kout$ and the mediation is described by the following 
  effective partonic Lagrangian 
\begin{equation}
\label{eq:int}
{\cal L}^{\textrm{parton}}_{\textrm{int}} =   \geff   L_\mu V^\mu + \textrm{h.c.}  \,, \quad L_\mu \equiv  \bar \ell_1 \Gamma^\mu \ell_2  \, ,
\quad  V_\mu \equiv \bar q  \ga_\mu  (1-\gamma_5) b \,, \quad \geff \equiv - \frac{G_F}{\sqrt{2}}  \la_{\textrm{CKM}} \,,
\end{equation}
where  $\Gamma^\mu \equiv \ga^\mu( \CV + \CA \ga_5)$. The quark field $q$,  the values of $\CV$ and $\CA$, and $ \la_{\textrm{CKM}}$
can be adapted to describe different processes. Processes mediated by the $b \to u (c) \ell \nu$ charged-current interaction 
correspond to  $q = u(c)$,  with $(C_V,C_A)  =(1,-1)$ and $\la_{\textrm{CKM}} = V_{ub} (V_{cb})$.
Processes mediated by the flavour changing neutral transition
$b  \to (d,s) \mu^+ \mu^-$ are obtained by setting $C_{V(A)}  =  \alpha  C_{9(10)}/(4\pi)$ and  
$\la_{\textrm{CKM}}  = V_{t(d,s)}^*V_{tb}$.\

The corresponding effective mesonic weak Lagrangian
describing the $\Bin \to \Kout \lone \bltwo$ process  
 reads
\begin{equation}
 {\cal L}^{\textrm {EFT}}_{\textrm{int}} = 
  \geff  \,   L^\mu V_\mu^{\textrm {EFT}} + \textrm{h.c.} \;,  \quad  
  V_\mu^{\textrm {EFT}}  = \sum_{n \geq 0} \frac{f_\pm^{(n)}(0)}{n!}  
  (-D^2)^n  [  (D_\mu B^\dagger)  K  \mp       B^\dagger( D_\mu  K)] \;,
  \label{eq:Lint_eff}
  \end{equation}
 where $D_\mu = (\partial  + i e  Q  A)_\mu$ is the covariant derivative
 and  $f_\pm^{(n)}(q^2)$ denotes the $n^{\textrm{th}}$ derivative of the form factor $f_\pm(q^2)$.
 This Lagrangian maps  the $q^2$-dependence of the non-radiative $B\to K$ form factor 
 into a tower of derivative operators, such that the hadronic  matrix element of 
 $V_\mu$ 
 is reproduced to LO in the electromagnetic coupling,
\begin{eqnarray}
H_0^\mu( q_0^2) \equiv  \matel{\Kout }{ V_\mu }{\Bin }  = f_+(q_0^2) (p_B\pl p_K)^\mu + f_-(q_0^2) (p_B\mi p_K)^\mu   = \matel{\Kout }{ V^{\textrm {EFT}}_\mu }{\Bin }   +    {\cal O}(e), \quad
\label{eq:match}
 \end{eqnarray}
where $\matel{0}{B^\dagger }{\Bin (p_B)}  =  e^{- i  p_B \cdot x}$ and
$\matel{\Kout (p)}{K(x)}{0}  = e^{ i  p \cdot x} $, and 
$ f_0 =  f_+ + \frac{q^2}{m_B^2-m_K^2}  f_-$ is the scalar part of the form factor.
The radiative amplitude at ${\cal O}(e)$ is computed by combining the gauge-invariant Lagrangian in  (\ref{eq:Lint_eff})
with the ordinary QED Lagrangian for fermions and mesons,
\begin{equation}
 {\cal L }_{\textrm{QED} }\equiv  {\cal L }_{\xi} (A) + \sum_{\psi = \ell_1, \ell_2} \bar \psi  (i D\!\!\!\!/ -m_\ell) \psi + \sum_{M =B,K}
  (D_\mu M) ^\dagger D^\mu M - m_M^2  M^\dagger M~,
    \label{eq:Lint_QED}
\end{equation}  
where $ {\cal L }_{\xi} (A)$ denotes the Maxwell Lagrangian with the covariant gauge-fixing term, resulting in the photon propagator given in \SEC \ref{sec:virtual}.  
Matters related to going beyond this approximation, at the form factor level,  
are discussed in \SECs \ref{sec:beyondPT} and \ref{sec:SD}.

The non-radiative amplitude is decomposed as
\begin{equation}
\label{eq:amp}
{\cal A}_{\Bin \to \Kout \lone \bltwo} \equiv \matel{\Kout  \lone  \bltwo  }{ (- {\cal L}_{\textrm{int}}) }{\bar{B}} = 
\Ampzero + \Amptwo + {\cal O}(e^4) \;,
\end{equation}
where the superscript indicates the order in the electromagnetic coupling 
and the phase follows the \ac{PDG} convention \cite{PDG}.
The lowest-order (LO) amplitude  reads
\begin{equation}
\label{eq:A0}
\Ampzero_{\Bin \to \Kout \lone \bltwo}   = - \geff\,  L_0 \Cdot H_0    \;,
\end{equation}
with
 \begin{equation}
 \label{eq:L0}
 L_0^\mu \equiv \matel{  \lone  \bltwo}{ L^\mu} {0} =  \bar{u}(\lone) \Gamma^\mu v(\ltwo) \;.
 \end{equation}

For flavour changing neutral currents (FCNCs), such as $\Bin \to \Kout \ell^+ \ell^-$, there are additional contributions
originating from  four-quark operators, dipole and chromomagnetic penguin operators which are apparently not 
described by the mesonic Lagrangian in  \eqref{eq:Lint_eff}. Some of these effects, in particular the long-distance 
contribution associated to the charmonium resonances introduce sizeable distortions of the kinematical distribution
in specific regions of $q^2$. However,  in the case of a  scalar meson final state, 
such effects can partially   be absorbed 
for moderate $q^2 \ll m_{J/\Psi}^2$ into a reparametrisation of the $f_\pm$  form factors.\footnote{~Of course there are 
additional long-distance effect, such as the photon exchange between a charm-loop and the $B$-meson which cannot be captured in this way. We expect the simplified procedure outlined above to absorb 
the main effect at moderate $q^2$.}
Approaches of this type can be found in the literature in the framework of  
e.g.~QCD factorisation \cite{BFS01} and/or light-cone sum rules \cite{DLZ2012,LZ2013,Khodjamirian:2012rm}.

At this point we wish to comment on the QED corrections performed in $K \to \pi \ell^+\ell^-$~\cite{Kubis:2010mp}. 
Formally the main difference is that we perform a form factor expansion \eqref{eq:Lint_eff}
whereas they work with constant form factor which is  a good approximation for 
Kaon physics.  In terms of the kinematics
they directly work with $  q_0^2$-variable (denoted by $s$ in~\cite{Kubis:2010mp}) 
since this variable is accessible in  Kaon experiments.  
Moreover, the photon energy cut-off is implemented  in the $q_0$-RF.

\subsection{Real radiation}
\label{sec:real}

\begin{figure}[t]
\vspace{0cm}
  \includegraphics[width=0.9\linewidth]{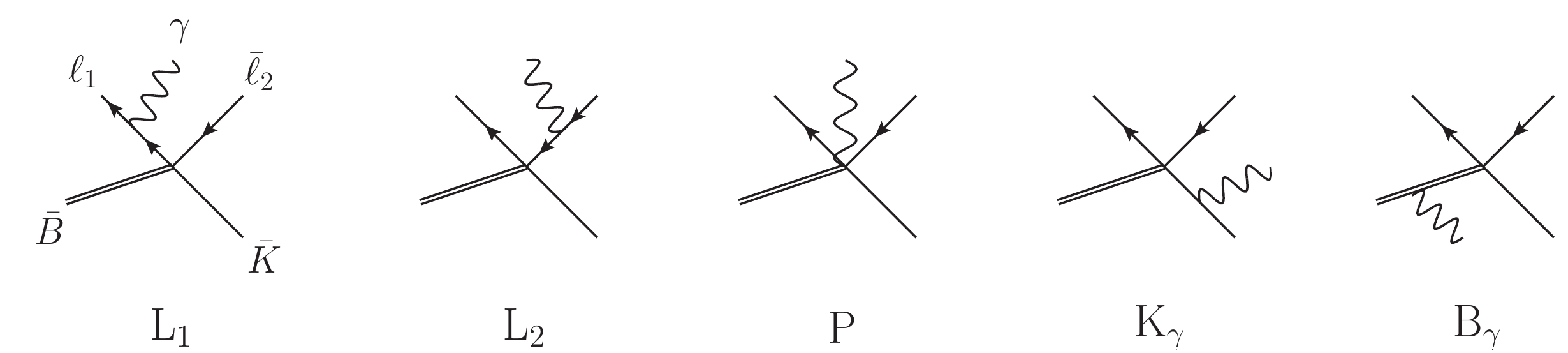}  
  \caption{\small{ ${\cal O}(e)$-graphs with nomenclature referring to photon-emission 
  and the $P$ stands for point-like and can also be interpreted as a contact term.}}
  \label{fig:real}
\end{figure}

The five diagrams  relevant to compute real emission amplitude at  ${\cal O}(e)$ are shown in \FIG\ref{fig:real}.
The result can be expressed as follows\footnote{~Note that, in order to recover the photon mass regularisation, the 
following substitutions in the denominators are sufficient:
$2 k \Cdot p \to 2 k \Cdot p  \pm \mga^2$  with plus sign for outgoing and minus sign 
for incoming momenta.}
 \begin{alignat}{2}
&\Ampone =  - e \geff \Big\{   & &   
    \, \bar u (\lone)  \left[ \hat{Q}_{\lone}\,   \frac{2 \eps^* \Cdot \lone\pl  \slashed{\eps}^* \slashed{k}}{2 k \Cdot \lone }   \Gamma \Cdot H_0(\qz^2)    + 
   \hat{Q}_{\bltwo}  \,   \Gamma \Cdot H_0(\qz^2)  \frac{2 \eps^* \Cdot \ltwo\pl  \slashed{k} \slashed{\eps}^*}{2 k \Cdot \ltwo }  \right] v(\ltwo) \;  + 
   \nonumber \\[0.1cm]
& &&       \hat{Q}_{\Bin} \,   L_0 \Cdot \bar{H}^{(B)}_0(q^2)     \frac{\eps^{*} \Cdot (p_B + \pBbar)}{2k \Cdot p_B}  \; + 
    \hat{Q}_{\Kout} \,  L_0 \Cdot \bar{H}^{(K)}_0(q^2)    \frac{\eps^{*} \Cdot (\pK \pl \pKbar)}{2k \Cdot p_K }     
 \; +
    \nonumber \\[0.1cm]
& &&   ( \hat{Q}_{\Bin} \!  -\!  \hat{Q}_{\Kout} )  \,L_0 \Cdot \eps^* \, f_+(q^2) +  ( \hat{Q}_{\Bin}\!   +\! \hat{Q}_{\Kout})  \,L_0 \Cdot \eps^* \, f_-(q^2) \; + \nonumber \\[0.1cm]   
& && ( \hat{Q}_{\Bin}\!   +\! \hat{Q}_{\Kout} )  L_0 \Cdot (p_B \pm \pK)  (  \eps^* \Cdot (q + \qz)) \sum_{n \geq 1} \frac{f^{(n)}_{\pm}(0)}{n!}  P_{n-1}    \Big\}  \;,
    \label{eq:realAmp}
\end{alignat} 
where $P_n = \sum_{m=0}^n (q^2)^{(n-m)} (q_0^{2})^m$ (with $P_0=1$), 
$\bar{H}^{(X)}_0 = H_0|_{p_X \to \bar{p}_X}$ for $X = B,K$ and $\pKbar \equiv \pK + k$. 

\begin{table}[t]
\centering
\begin{tabular}{ l |  rrrr }
$\Bin\to \Kout \lone \bltwo $ &  $\hat{Q}_{\Bin}$ &  $\hat{Q}_{\Kout}$    &  $\hat{Q}_{\lone}$ & 
   $\hat{Q}_{\bltwo}$    \\   \hline  
 $\Bin^- \to \Kout^- \mu^- \mu^+ $  &  $+1$ & $-1$ & $-1$ & $+ 1$  \\ 
 $ \Bin^{\phantom{-}}_s \to \Kout^- \nu \mu^+  $  &  $0$ & $-1$ & $0$ & $+1$  
\end{tabular}
\caption{\small Example of charge assignment for FCNC and semileptonic decay which obey \eqref{eq:charge}.
Note that generally $Q_P = - Q_{\bar P}$,  rules for the hatted charges are given in the text  and
by convention $\Bin$ and $\Kout$ correspond to mesons with a $b\bar q$ and $s\bar q$ valence quarks.
}
\label{tab:Emax}
\end{table}
 
The rules for the hatted charges are: $\hat{Q}^{\textrm{in} } = - Q^{\textrm{in} } $ 
and $\hat{Q}^{\textrm{out} } =  Q^{\textrm{out} } $. Furthermore we use $ Q_{\bar \ell_2} = - 
 Q_{\ell_2} $ such that $Q_{\bar \ell_2} + Q_{\ell_1} =0$ in the case where the lepton pair is charge neutral, 
 cf. \TAB\ref{tab:Emax} for an illustration.
Charge conservation then implies
\begin{equation}
\label{eq:charge}
  \sum_{i= \Bin,\Kout,\lone,\bltwo } \!\!\!\!\!\! \hat Q_i = 0 \;.
\end{equation}
 Hereafter the $\sum_i$ is defined by the left-hand side (LHS) of the equation above.
 Keeping the leading  terms in the $k \to 0$ limit, i.e.~at $\ORD(1/E_\ga)$, $\Ampone$  assumes the Low or eikonal form,
\begin{equation}
\label{eq:Low}
\AmponeLow = e \Ampzero \sum_i \hat Q_i \frac{\eps^* \Cdot p_i}{ k \Cdot  p_i } \;,
\end{equation}
which is manifestly gauge invariant as a result of \EQ\eqref{eq:charge}.
The subleading terms of $\ORD(E_\ga^0)$ are also universal and are proportional to the angular momentum 
operator (e.g.~$\sigma_{\mu \nu} k^\mu \eps^{*\nu}$ terms in the first line of \eqref{eq:realAmp}).

It is instructive to discuss  gauge invariance of the amplitude  beyond the 
$k \to 0$ limitas it comes in rather disguised form. 
Here we summarise the essence  and defer some detail to \APP\ref{app:GI}.
A gauge transformation ($\eps \to k$) of the first line in \eqref{eq:realAmp}, omitting common prefactors,  
leads to 
\begin{equation}
\left. \Ampone_{\rm{1st\ line}}   \right|_{\eps \to k}  \propto   ( \hat{Q}_{\bltwo}  + \hat{Q}_{\lone}) L_0 \Cdot {H}_0(q_0^2)~,
\end{equation}
 whilst the second and third line 
 combine to  $(\hat{Q}_{\Bin} + \hat{Q}_{\Kout})L_0 \Cdot H_0(q^2)$.  This is would signify gauge invariance if 
 $q^2 = q_0^2$ and that's where the contact term ($P$-graph) comes into play. 
The latter,  fourth line, leads to
 $(\hat{Q}_{\Bin} + \hat{Q}_{\Kout}) L_0 \Cdot [ H_0(q_0^2)-  H_0(q^2)]$, such that $\Ampone|_{\eps \to k}$
is proportional to the sum of charges in \eqref{eq:charge},  
assuring  gauge invariance of the whole amplitude.

\subsection{Virtual corrections}
\label{sec:virtual}

\begin{figure}[t]
  \includegraphics[width=0.9\linewidth]{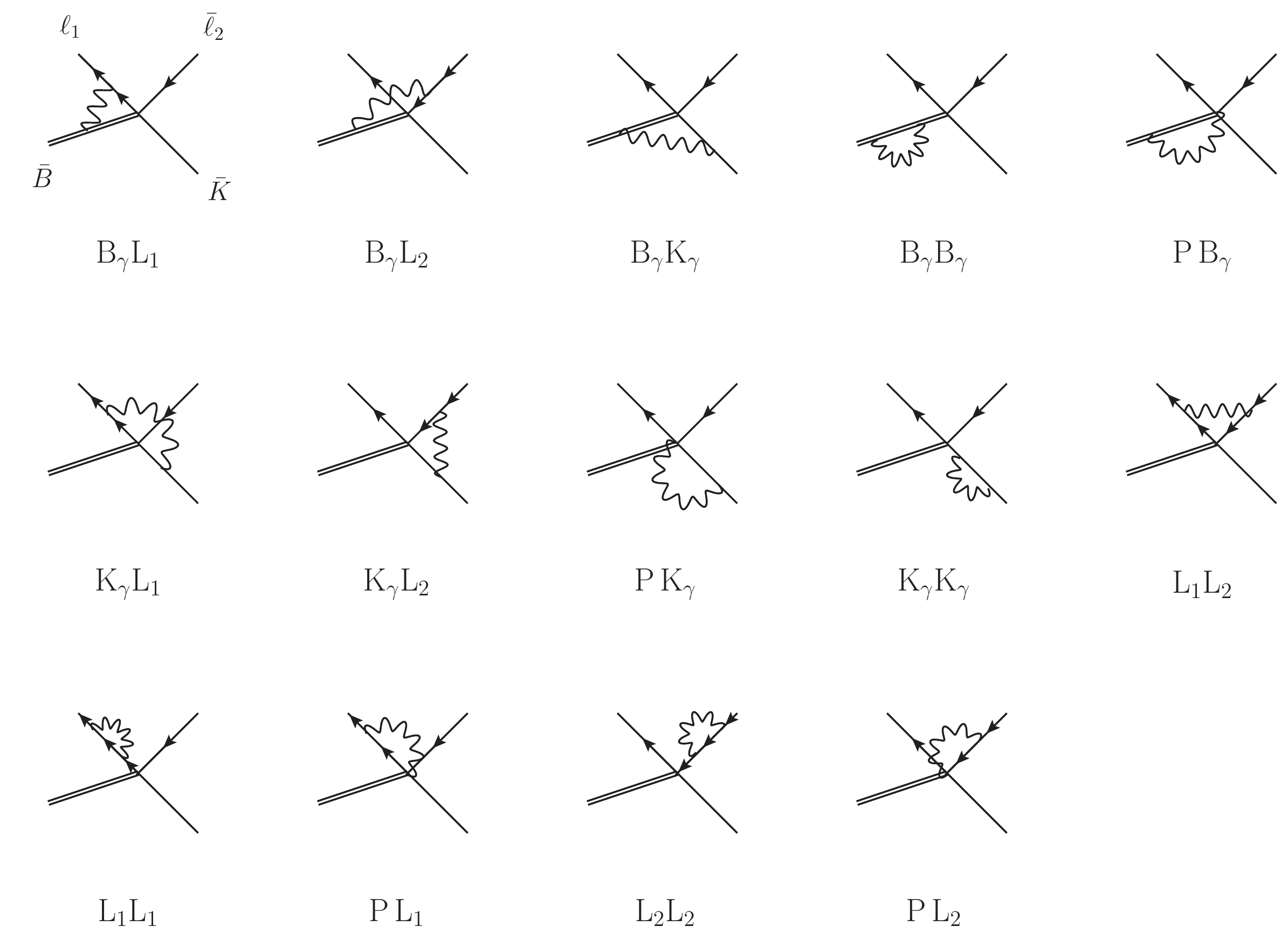}  
  \caption{\small ${\cal O}(e^2)$-graphs with nomenclature adapted for tracking the cancellation of 
  IR-divergences. }
  \label{fig:virtual}
\end{figure}

The diagrams for the virtual corrections are depicted in \FIG\ref{fig:virtual} and decompose into
\begin{eqnarray}
\label{eq:recipe}
\Amptwo  =  \Amptwo_{1PI} + 
  \frac{1}{2} \frac{\al}{\pi} \left[ (Q^2_{\lone} + Q^2_{\bltwo})  \de Z^{(1)}_2   + (Q^2_{\Bin} +Q^2_{\Kout }) \de Z^{(1)}_S \right] \Ampzero \;, 
  \end{eqnarray}
where 1PI stands for one particle irreducible and $\de Z$ correspond to the self-energy corrections.  
The amplitudes for the 1PI graphs are given in  \APP\SEC\ref{app:vir}.
We have explicitly computed corrections up to the second derivative of the form factor but in the actual plots  
 we restrict ourselves to the  first derivative
as they already  are  numerically  time-consuming.

For the $Z$-factors, decomposed as $Z_i = 1 + Q_i^2 \frac{\al}{\pi} \de Z_i^{(1)} + {\cal O}(\al^2) $,  
we find, adapting the on-shell scheme,
\begin{alignat}{2}
\label{eq:Z}
& \de Z^{(1)}_S  &\;=\;&  \frac{1}{4} \left(  (3-\gauge) (  \DepsUV   -  \rsoft) + (1-\gauge)   \right)    \;,    \nonumber \\[0.1cm]
& \de Z^{(1)}_2   &\;=\;& \frac{1}{4}\left(  -    \gauge \DepsUV   - (3-\gauge)    \rsoft    + 3  \ln\left(\frac{m^2}{\mu^2}   \right)  -  
(3 + \gauge)  \right)     \;, 
\end{alignat}
with
\begin{equation}
\label{eq:epshat}
\Deps = \frac{1}{\eps} - \ga_{E} + \ln 4 \pi~.
\end{equation}
The gauge parameter $\gauge$ enters the photon propagator as in $\Delta_{\mu \nu}(k ) =  
 \frac{-1}{k^2-\mga^2}\left(g_{\mu\nu} \mi (1 \mi \gauge) \frac{k_\mu k_\nu}{k^2}\right)$.   
 The factor $\rsoft$  reads 
\begin{equation}
\label{eq:rsoft}
\rsoft   = \left\{  \begin{array}{ll} \ln \left(\frac{\mga^2}{\mu^2}\right)  & \mga \neq 0 \\
\DepsIR  & \mga = 0 
\end{array} \right.~,
\end{equation}
in the case of a photon mass and DR respectively.

As far as $\Amptwo_{1PI}$ is concerned, 
soft singularities can be isolated as follows 
\begin{alignat}{1}
\label{eq:onePI}
\Amptwo_{1PI} = 
\frac{1}{2}\frac{\al}{\pi} \Ampzero
 \sum_{i \neq j}   \hat{Q}_i \hat{Q}_j ( \hat{p}_i \cdot \hat{p}_j)  C_0(m_i^2,m_j^2,(\hat{p}_i +\hat{p}_j)^2, m_i^2 ,\mga^2,m_j^2)  + \textrm{non-soft} \;,
\end{alignat}
where the explicit expression of the $C_0$ function can be found in \APP\ref{app:PV}.
Here  $\hat{p}^{\textrm{in} } = - p^{\textrm{in} } $,  $\hat{p}^{\textrm{out} } =  p^{\textrm{out} } $ in analogy with the hatted charges (and $p_{\lonetwo} \equiv \lonetwo$).  
Note that \eqref{eq:onePI} is  consistent with the crossing rule of reversing momenta and  
charge  when passing from  in(out)- to  out(in)-state.
We explicitly checked that the gauge dependent part of the amplitude vanishes  as a consequence of charge conservation:
\begin{equation}
\Amptwo|_{\gauge}  = \frac{\gauge}{2} \frac{\al}{4 \pi} \Ampzero \left(  \rsoft - \DepsUV - 1 \right) 
 (\sum_i \hat{Q}_i)^2  = 0 \;.
\end{equation}
Let us turn to the UV divergences. 
There are no UV divergences in the neutral meson case   
since the leptonic currents do not renormalise (at our level of approximation).
This does not change when the tower of operators $ {\cal L}^{\textrm {EFT}}_{\textrm{int}} $ \eqref{eq:Lint_eff} is added as the derivatives acts on the mesons only. 
As previously mentioned,  we restrict ourselves to the first form factor derivative approximation or to dimension-eight operators 
(the explicit form factors are given in \SEC\ref{sec:plots}).
In the case of  charged mesons,
there are UV divergences associated
with  operators of dimension six and eight in \eqref{eq:Lint_eff} and there is an additional one 
proportional to $p_B \cdot \ell_1 \, f^{(1)}_{\pm}(0)$ which can be interpreted as a $t$-channel 
operator.\footnote{~The set of operators \eqref{eq:Lint_eff}  does not close under renormalisation and needs to be completed by  the $t$-channel 
operator at dimension eight.} Since $f_{\pm}$ are to be counted separately  this means that there are six counterterms to be fixed 
at our level of approximation.  The appropriate counterterms can be determined by matching to QCD 
which  we hope to address in a forthcoming publication.
In this work, we  treat the divergences with minimal subtraction. We 
comment in \SEC\ref{sec:plots} on the numerical impact of the undetermined finite counterterms.

\subsection{Phase space}

Below we give the $3$- and $4$-particle phase space measures. 
For the photon phase space measure we need a regularised version in order to 
properly account for finite terms. Here, we find it more instructive to discuss explicitly  
results obtained using a non-vanishing photon mass. We refer the 
reader to \APP\ref{app:soft1} for the adaptation to DR.

\subsubsection{Phase space for the radiative and non-radiative decay}
\label{sec:four}

The radiative rate $\Bin\to \Kout \lone \bltwo \ga$, without energy cut-off on the photon,  is 
given by
\begin{alignat}{2}
\label{eq:Rrate}
& d^2 \Gamma_{\Bin\to \Kout \lone \bltwo \ga}   &\;=\;&  
 \frac{1}{   m_B }   
\left( \int \,    {\rho}_{\AAA}   \left[  | \Ampone |^2  + {\cal O}(e^4) \right]d \Phi_\ga \right)  d \qSaa d \claa \;, 
\end{alignat}
where
\begin{alignat}{2}
\label{eq:barred}
&  {\rho}_{\TT} &\;=\;&   \frac{1}{ 2^6  (2 \pi)^3  } \frac{\la^{1/2}(\pBbar^2,q^2,m_K^2) }{ \pBbar^2 q^2 }  \nonumber  \la^{1/2}(q^2,\mlone^2,\mltwo^2) \;,  \\[0.1cm]
&  {\rho}_{\RR}  &\;=\;&   \frac{1}{ 2^6  (2 \pi)^3  } \frac{\la^{1/2}(m_B^2,\qz^2,m_K^2) }{ m_B^2 \qz^2 } 
 \frac{1}{\omega^2}{  \la(\qz^2, \mlone^2, (k+ \ltwo)^2)} \;,
\end{alignat}
with $\la$  the K\"all\'en function  \eqref{eq:Kallen},
and $ \ff$ is given in   \eqref{eq:ff}.  Thus, $ \frac{{\rho}_{\RR}}{ {\rho}_{\TT} } = \det \frac{\partial ( q^2 ,\cl) } {  \partial ( \qz^2,\clz)}$ is the Jacobian which can be computed 
from the defining equation \eqref{eq:tr} and the kinematic parameterisations given in the appendix.
Moreover, the Lorentz-invariant photon phase space integral reads
\begin{alignat}{2}
\label{eq:dphiga}
& \int^{\EGAmax}_{\mga} d \Phi_\ga &\;\equiv\;&  \frac{1}{ (2 \pi)^3} \int_{\mga}^{\EGAmax} \frac{d^3 k}{2 E_\ga}  \nonumber \\[0.1cm]
& &\;=\;&  \frac{1}{2 ( 2\pi)^3}
 \int_{\mga}^{ \EGAmaxn{i} } 
 d E_\ga^{(i)} | \vec{k}^{\; (i)}| \int d\Omega^{(i)}_\ga \Theta\big[ f^{(i)}(E_\ga^{(i)},\theta^{(i)} _\ga,\phi_\ga^{(i)} )\big]  \;,
\end{alignat}
with 
\begin{equation}
\EGAmaxn{1} = \frac{m_B^2+ \mga^2 - ( q+m_K)^2}{2 m_B } \;,\quad
\EGAmaxn{4} = \frac{\qz^2+ \mga^2 - ( m_{\lone} + m_{\ltwo})^2}{2 \qz } \;,
\end{equation}
where the former and the latter correspond to the $\{q^2,\Tl\}_{a = \TT}$  and $\{\qz^2,\Tlz\}_{a = \RR}$ variables respectively, and  $q_a\equiv \sqrt{q_a^2}$ is understood in this context.
The restriction on the angles is
\begin{equation}
\Theta \big[ f^{(i)}(E_\ga^{(i)},\theta^{(i)} _\ga,\phi_\ga^{(i)} )\big] = \left\{ \begin{array}{ll}  1  &\quad  i = 1 \\  \Theta[D(
E_\ga^{\FRfour},\theta^{\FRfour} _\ga,\phi_\ga^{\FRfour},\qz^2,\clz)   ]  &\quad i = 4 \end{array} \right. \;,
\end{equation}
with the function $D$ defined in \eqref{eq:ABCD}. The reason why the restriction in the $\FRfour$-RF, 
appropriate for the $\{ \qz^2, \clz\}$-variables, is non-trivial is that for certain given values of $\{ \qz^2, \clz\}$, the true maximum photon energy is a function of the photon angles and is in general below $\EGAmaxn{4}$. We find it most convenient to implement the kinematic restrictions via the step-function $\Theta(x)$.\footnote{~In the limit of $m_{\ell_1}\to 0$, the step-function $\Theta(x)$ becomes redundant, since the function $D$ is then  positive for all kinematic configurations, as can be seen from \eqref{eq:ABCD}.}

In the $\{q^2,\Tl\}_{a = \TT}$  case, one can conveniently work 
with the Lorentz invariant variable $\pBbar^2$, related to $E_\ga^{\FRone}$ as  $2 m_B E_\ga^{\FRone} = 
m_B^2 + \mga^2 - \pBbar^2$. Moreover, since the passage from $E_\ga^{\FRone}$ to $E_\ga^{\FRtwo}$ is independent 
of the photon angles the replacement  
$d\Omega^{\FRone}_\ga \to d\Omega^{\FRtwo}_\ga$ is allowed.
The photon phase space then assumes the form
\begin{equation}
\label{eq:practice}
 \int^{\EGAmax}_{\mga} d \Phi_\ga \to  \frac{1}{2^3 (2 \pi)^3} \int^{(m_B-\mga)^2}_{ (q+ m_K)^2 } d \pBbar^2 
\frac{ \la^{1/2}(m_B^2,\pBbar^2,\mga^2)}{m_B^2}  \int d\Omega^{(1,2)}_\ga \;.
\end{equation}
The non-radiative $\Bin\to \Kout \lone \bltwo$ rate is
given by
\begin{alignat}{2}
\label{eq:d2Ga0}
& d^2 \Gamma_{\Bin\to \Kout \lone \bltwo}    &\;=\;&  
\frac{\rho_{\TT}|_{\pBbar^2 \to m_B^2}}{m_B}   \left\{  | \Ampzero |^2  +   2 \Rea[ \Ampzero (\Amptwo)^* ]    + {\cal O}(e^4)  \right\} {d q^2 d  \cl} \; .
\end{alignat}
Since there is no photon-emission, in this case there is no difference between  the 
$\{q^2,\cl\}$- and $\{\qz^2,\clz\}$-variables.

\subsubsection{Introduction of a physical photon energy cut-off}
As anticipated, to match experimental observations, we 
introduce a cut-off on the maximal value of $\pBbar^2$ via the 
parameter $\des$,  defined in \eqref{eq:des},  satisfying
\begin{equation}
\label{eq:des2}
\qquad  0 <  \des <  \des^{\textrm{inc}} = 1 - \left( \frac{q+ m_K}{m_B}  \right)^2~.
 \end{equation}
The value $\des^{\textrm{inc}}$ corresponds to the minimal value of $\pBbar^2$ in a  fully photon-inclusive decay. 
This definition translates to the following photon-energy 
cut-off\,\footnote{~When evaluating the photon phase space variable in the \FRfour-RF, appropriate for the $\{ d\qz^2, d \clz \}$-variables, the cut-off can be converted by using
$E_\ga^{\FRone}= \ga_{\qz}  E_\ga^{\FRfour}(1 - \be_{\qz} \cos \theta^{\FRfour}_\ga)$ cf. \eqref{eq:boost4} for the Lorentz boost factors.}
\begin{equation}
\label{eq:Emax}
{ \EGAmax}^{\FRone} = \des \frac{m_B}{2} \;, \quad
\end{equation}
A typical choice for $\des$ in realistic experiments is $\des={\cal O}(0.1)$.
With the inclusion of $\des$, the integral \eqref{eq:dphiga} assumes the form
\begin{equation}
\label{eq:dphigad}
\int_{\des} d \Phi_\ga =    \frac{1}{2^3 (2 \pi)^3} \int^{(m_B-\mga)^2}_{ m_B^2(1- \des) } d \pBbar^2 
\frac{ \la^{1/2}(m_B^2,\pBbar^2,\mga^2)}{m_B^2} \int d\Omega^{\FRtwo}_\ga     \;.
\end{equation}

\section{Cancellation  of Infrared Divergences}
\label{sec:IR}

In order to track the IR divergences it is convenient to split the differential rate as follows
\begin{alignat}{2}
\label{eq:Delta}
&  d^2 \Gamma_{ \Bin\to \Kout \lone \bltwo}(\des )   &\;=\;& 
   {d^2 \Gamma^{\LO}} +     \frac{\al}{\pi} \sum_{i, j }   \hat{Q}_i \hat{Q}_j   
  \left( {\cal H}_{ij} + {\cal F}^{(\AAA)}_{ij}(\des )   \right)    \, d  \qSaa d \claa   + {\cal O}(e^4) \;, \nonumber 
  \\[0.1cm]
 &  &\;=\;&  {d^2 \Gamma^{\LO}} \left[ 1 +  \Delta^{(\AAA)} (\qSaa,  \claa; \des)   \right]    \, d  \qSaa d \claa   + {\cal O}(e^4) \;,
 \end{alignat}
where $ d^2 \Gamma^{\LO} $ corresponds to the zeroth order term in \eqref{eq:d2Ga0}, 
the sums on the charges is understood as in \eqref{eq:charge}, 
and ${\cal H}$ and ${\cal F}$ stand for the virtual and real contributions respectively.  
More precisely,   
${\cal H}_{ij}$ and ${\cal F}_{ij}$ are related to the amplitudes as follows
\begin{eqnarray}
 \frac{\al}{\pi}  \sum_{i, j }   \hat{Q}_i \hat{Q}_j  {\cal H}_{ij}  &=&
 \frac{1}{m_B}  \rho_{\TT}|_{\pBbar^2 \to m_B^2} 2 \textrm{Re} [ {\cal A}^{(2)*} {\cal A}^{(0)}]~, \nonumber\\
 \frac{\al}{\pi}  \sum_{i, j }   \hat{Q}_i \hat{Q}_j  {\cal F}^{(\AAA)}_{ij} &=& 
  \frac{1}{m_B} \int d \Phi_\ga \, {\rho}_{\AAA} \, |{\cal{A}}^{(1)}|^2~,
  \end{eqnarray} 
 where $d \Phi_\ga$ and $ {\rho}_{\AAA}$ are defined in \eqref{eq:barred} and  \eqref{eq:dphiga} respectively.
 
In standard fashion, the integrals are split into  divergent parts 
which can be done analytically and a necessarily regular part which is dealt with numerically. 
We parameterise this decomposition as follows
\begin{alignat}{2}
\label{eq:tactic}
& {\cal H}_{ij}     &\;=\;\;&    
 \frac{d^2 \Gamma^{\LO}}{d q^2 d \cl}  \left(  \tH^\soft_{ij} + \tH^\hc_{ij} \right)  +\Delta {\cal H}_{ij}  
    \;, \nonumber \\[0.1cm]
& {\cal F}^{(a)}_{ij}(\des ) &\;=\;\;& \frac{d^2 \Gamma^{\LO}}{d q^2 d \cl}    \tF^\soft_{ij}  ( \Des ) +
 \tF^\hca_{ij}(\underline{ \de} ) +  \Delta {\cal F}^{(a)}_{ij}(  \underline{\de} )  \;,
\end{alignat}
\noindent
with $\tH^\soft_{ij}$ ($\tH^\hc_{ij}$) and $\tF^\soft_{ij}$ ($\tF^\hca_{ij}$), 
to be defined further below,  containing
all  soft (hard-collinear) singularities, whereas $\dH$  and $\dF$ are  regular
(even in the limit $m_{\lonetwo} \to 0$).  
In order to split the real emission part, besides the previously introduced physical    cut-off $\des$, we adopt the phase space slicing method \cite{Ellis:1980wv}, which requires the introduction of two auxiliary (unphysical) cut-offs $\Desc$, 
\begin{equation}
\underline{\de} \equiv \{ \des,\Des,\Dec\}\;, \quad \Des \ll  1  \;, \quad 
\frac{\Dec}{\Des}  \ll 1  \;.
\end{equation}
We remind the reader that $\des$ has been introduced for meaningful comparison with 
experimental data and mention for clarity that $\tF^\hca_{ij}$ is
 singular in the $m_{\lonetwo} \to 0$  limit but finite for $m_{\lonetwo} \not= 0$.

As already implicit in the decomposition \eqref{eq:tactic}, soft divergences cancel at the 
differential level independently from the choice of variables. This is not the case for 
hard-collinear singularities, given that the hard-collinear integral ($\tF^\hca_{ij}$) 
is not proportional to the non-radiative kinematics. 
Without the physical cut-off $\des$,  the cancellation of both types of 
divergences proceeds as in standard in textbooks discussions 
(see e.g.~\cite{Weinberg:1995mt,Muta:1998vi,Sterman:1994ce}). 
However, the choice of a photon energy cut-off, associated with a preferred frame, makes it 
significantly more involved  compared to the semileptonic case \cite{Ginsberg:1969jh}.
A detailed discussion of the soft singularities and collinear logs 
follows below,
along with the definitions of the $\tilde{\cal F}$ and $\tilde{\cal H}$.
Particular emphasis is given to single out which observables are IR-safe and not.

\subsection{Cancellation of soft divergences at differential level}
\label{sec:soft}

The soft singularities in the virtual corrections are encoded in the triangle functions  $C_0$ in \eqref{eq:onePI}  and the self energy contributions  in \eqref{eq:Z}. Combining them, we define 
\begin{eqnarray}
\tilde{\cal H}^{\soft}_{ij}   &  \! \stackrel{\textrm{def}}{=} \!   & ( 1- \de_{ij}) ( \hat{p}_i \cdot \hat{p}_j) \Rea[ C_0(m_i^2,m_j^2,(\hat{p}_i \pl \hat{p}_j)^2, m_i^2 ,\mga^2,m_j^2) ]  + \de_{ij} \times \de Z_i^{(1)}      \label{eq:Hlead}  \nonumber \\
 & = &  - \rsoft  \left\{ ( 1- \de_{ij})  \frac{ \hat{p}_i \cdot \hat{p}_j } {m_i m_j}
 \frac{x_{ij}} {(1-x_{ij}^2)} 
 \ln|x_{ij}|  + \de_{ij} \frac{1}{2}  \right\}+  {\cal O}(f_\reg)~,\label{eq:Hlead2}
\end{eqnarray}
where $f_\reg$ stands for IR finite terms, including regularisation-dependent ones. 
 The $x_{ij} $-variables are given by 
\begin{equation}
 \label{eq:ditt}
 x_{ij} \equiv  \frac{\sqrt{ y_{ij} }-1}{\sqrt{ y_{ij} }+1}  \;, 
\quad y_{ij} \equiv \frac{  (\hat{p}_i \pl \hat{p}_j)^2 \mi (m_i \pl m_j)^2\pl i0} {(\hat{p}_i \pl \hat{p}_j)^2   \mi (m_i\mi m_j)^2 \pl i0}  \;.
 \end{equation}
Considering the soft part of the real emission amplitude, namely the Low part of the amplitude in \eqref{eq:Low}, 
we define 
 \begin{eqnarray}
  \tilde{\cal F}^{\soft}_{ij}(\Des) & \! \stackrel{\textrm{def}}{=} \!  &   (2 \pi)^2 \int_{\Des}  \frac{-  {p}_i \Cdot {p}_j }{( k \Cdot{p}_i )( k \Cdot {p}_j )  }  d\Phi_\ga = 
  (2 \pi)^2 \int_{\Des}  \frac{-  \hat{p}_i \Cdot \hat{p}_j }{( k \Cdot \hat{p}_i )( k \Cdot \hat{p}_j )  }  d\Phi_\ga    \nonumber  \\[0.1cm]
   &   = &  - K_\reg(\Des) \Iijzero   + {\cal O}( f_\reg)  
  \label{eq:F0} \nonumber  \\
  &   = &  \left[\rsoft  -2 \ln ( \Des) \right]\left\{ ( 1- \de_{ij})  \frac{ \hat{p}_i \Cdot \hat{p}_j   } {m_i m_j}
 \frac{x_{ij}} {(1-x_{ij}^2)} 
 \ln|x_{ij}|  + \de_{ij} \frac{1}{2}  \right\} + {\cal O}( f_\reg)~, \qquad     
 \label{eq:F02}
 \end{eqnarray}
 where the $ {\cal O}( f_\reg)$ terms can be found in \APP\ref{app:SI}. 
As can be checked, the sum  $\tilde{\cal H}^{\soft}_{ij} + \tilde{\cal F}^{\soft}_{ij}(\Des)$ is free from 
soft divergences and this ensures their cancellation  at the differential level.\footnote{~Note that $x_{ij} < 0$ 
as the momenta $p_i$ are assumed to be timelike with positive energy.
Moreover,  the individual ${\cal F}_{ij}$ are gauge dependent (the result is presented  for $\xi =1$),
whereas in the sum over all charges  gauge dependence disappears.}
This includes 
 $\ln^2 m_{\lonetwo}$-terms 
which cancel when the  real and virtual terms 
are summed up: 
these are genuine soft-collinear terms, which cancels as a result of the cancellation of the 
soft divergences.\footnote{~In order to track the $\ln (m_\ell)$ terms, note that  $x_{ij} \to - m_i m_j/(\hat{p}_i \pl \hat{p}_j)^2$ 
 for   $(\hat{p}_i \pl \hat{p}_j)^2 \gg m^2_{i,j} $. Moreover it is worth pointing out that one can write,
 $I_{ij}^{(0)} =  \frac{1}{2 \be_{ij}} \ln \left( \frac{1 + \be_{ij}}{1 - \be_{ij}}  \right  )$, in terms of physically transparent variables with $\be_{ij}  
= \frac{\be_i + \be_j}{1+ \be_i \be_j} $ is the relativistic addition of the velocities of the two particles
$  \be_i \equiv  |\vec{p}_i|/E_i$ in the $ij$-RF.}
We note that as a result of these cancellations  scheme dependent terms due to IR regularisation disappear as well.

The crucial step in evaluating \eqref{eq:F0} is that, neglecting finite terms,  
the integral over the photon energy and the photon angles factorises: 
the angular integral $\Iijzero$ alone becomes separately 
Lorentz invariant (i.e.~frame independent) and can be performed in the 
RF of the radiating pair, where it is particularly simple 
(see \APP\ref{app:integral} for more details). The energy 
and angular integral evaluate to
\begin{eqnarray} 
\label{eq:old} 
 K_\reg(\Des) 
&\;=\;&  - \frac{1}{2} \rsoft + \ln \left( \frac{m_B}{\mu} \right) 
 + \ln ( \Des)  + {\cal O}(\Des) \;, 
 \end{eqnarray}
and 
\begin{equation}
\label{eq:Iijzero}
\Iijzero =  \left\{  \begin{array}{ll} 1 &\; i= j~, \\
 2 \frac{ \hat{p}_i \cdot \hat{p}_j }{m_i m_j}  \frac{x_{ij}}{1-x_{ij}^2} \ln |x_{ij}| &\; i \neq j~.    \end{array} \right.
\end{equation}

 We wish to emphasise that there 
 are single collinear logs, $\ln m_{\lonetwo}$, in $\tilde{\cal H}^{\soft}_{ij} + \tilde{\cal F}^{\soft}_{ij}(\Des)$ which  match up  with corresponding terms in 
 $\tilde{\cal H}^{\hc}_{ij} + \tilde{\cal F}^{\hc}_{ij}(\underline{\de})$.
 The procedure is therefore well set-up for tracking analytically after what phase space integration  
 IR sensitive terms cancel against each other.

 However, since there remain $\ln \Des$-terms in the analytic expression one might wonder whether this leads to a numerically stable integral. We have found that  
 the phase space integral is stable when using a Monte Carlo integration on the photon variables. Alternatively, one might use the  dipole subtraction  method \cite{Catani:1996vz} 
 as applied to QED \cite{Dittmaier:1999mb,Dittmaier_2008,Sch_nherr_2018}.

\subsection{Hard-collinear virtual contribution $\tilde{\cal H}^\hc$}

The hard-collinear virtual contribution, after summing over charges, is given by
\begin{alignat}{2}
\label{eq:hcvirt}
&\tilde{\cal H}^\hc\;&\stackrel{\textrm{def}}{=} &\;
 \sum_{i, j }   \hat{Q}_i \hat{Q}_j   \tilde{\cal H}_{ij}^\hc=   2 \hat{Q}_{\lone}\left ( \hat{Q}_{\bltwo} \pl \hat{Q}_{\Bin}\pl \hat{Q}_{\Kout} \right )  \ln \left(\frac{\mlone}{\mu} \right)  + \{ 1 \leftrightarrow 2 \} \nonumber \\[0.1cm]
& & = & - 2 \hat{Q}_{\lone} ^2  \ln \left(\frac{\mlone}{\mu} \right)  + \{ 1 \leftrightarrow 2 \},
\end{alignat}
where $\left(\frac{\mu^2}{4\pi^2}\right)^{\epsUV}  B_0(m^2,0,m^2)  =  \DepsUV- 2 \ln (m/\mu) +2 + {\cal O}(\eps) $  was used and charge conservation was used in going from the first to the second line.

\subsection{The hard-collinear integral $\tilde{{\cal F}}^{(hc,a)}$}
\label{sec:HC}

We evaluate the  hard-collinear integral using the phase space slicing method \cite{Ellis:1980wv}
following the specific recipe in Ref.~\cite{Harris:2001sx}.
The integral is given by
\begin{eqnarray}
\label{eq:fhc}
\frac{\al}{\pi}  \tilde{{\cal F}}^{(hc,a)}  (\underline{\de}) 
 &\;=\;& 
 \frac{\al}{\pi}  \sum_{i, j }   \hat{Q}_i \hat{Q}_j   \tilde{\cal F}_{ij}^{(hc,a)} (\underline{\de})   \nonumber \\[0.1cm]
   &\;=\;& 
 \frac{1}{m_B} \int^{\des}_{\Des}   \rho_{\AAA}^{\lone || \ga} (\Dec) \,|{\cal A}^{(1)}_{\lone || \ga} |^2 d\Phi_\ga   \; +  \;
 \{ 1 \leftrightarrow 2 \} \;,
\end{eqnarray}
where  $|{\cal A}^{(1)}_{\lone || \ga} |^2$ is  the part of $|{\cal A}^{(1)} |^2$ proportional to $1/(k\cdot \lone)$ when $m_{\lone}\to 0$ which includes contributions beyond the Low term.
Note that the photon-energy integral runs from $\Des$ till $\des$, 
consistent  with \eqref{eq:tactic} where the soft  modes are absorbed into 
$\tF^\soft_{ij}  ( \Des )$.
The phase space factor $ \rho_{\AAA}^{\lone || \ga}  (\Dec) $ is defined as
\begin{equation}
\label{eq:kdotl1dc}
 \rho_{\AAA}^{\lone || \ga}  (\Dec)  =  \rho_{\AAA}  \Theta (  \Dec m_B^2 -k\cdot \lone) \;,
\end{equation}
where $\rho_{\AAA}$ is defined in \eqref{eq:barred}
, the meaning of the integration boundaries can be inferred from \eqref{eq:dphigad}, 
and the step-function encodes the   phase space slicing.
The quantity  $ \Dec \ll 1$ then implies that $k$ and $\lone$ are nearly collinear.

\subsubsection{Phase space slicing of the hard-collinear integral}
\label{app:comp}

In the phase space slicing method, the photon and the light particle it is emitted from, are effectively treated 
as a single particle.
This follows up on the intuitive picture that a particle and its collinear photon
are hard to disentangle. Below, we  give the explicit expressions for $\lone || \ga$, and the $\ltwo || \ga$ case is obtained in a completely analogous fashion. Formally, one  decomposes  the phase space as follows
\begin{align}
\label{eq:gen}
d\Phi_ {\Bin \to \Kout  \lone \bltwo \ga} =d\Phi_ {\Bin \to \Kout  \loneg \bltwo} d \Phi_\ga  \frac{E_{\loneg}}{E_{\lone}} \;.
\end{align}
The collinear region is parameterised by $ \lone  = z \loneg$, where $\loneg \equiv   \lone + k$,
assuming that the transverse part can be neglected in order to extract the collinear logs.
The two parts in \eqref{eq:gen} then assume 
the form 
\begin{alignat}{2}
\label{eq:col-measure}
& d \Phi_\ga  \frac{E_{\loneg}}{E_{\lone}} &\;\to\;& \frac{1}{16\pi ^2}dz\;d \loneg^2 \;,\nonumber  \\[0.1cm]
& d\Phi_ {\Bin \to \Kout  \loneg \bltwo} &\;\to\;& \frac{1}{2^5 ( 2\pi)^3} \frac{\la^{1/2}(m_B^2, \qz^2, m_K^2)}{m_B^2} d \qz^2 d \clz \;,
\end{alignat}
In those variables, the amplitude squared assumes the form (in the $\xi =1$ gauge)
\begin{align}
|{\cal A}^{(1)}_{\lone  || \ga}|^2 &=\frac{e^2}{\left ( k\Cdot \lone \right ) }\hat{Q}_{\lone}\left [ \hat{Q}_{\lone} \left ( 1-z\right)-\frac{2z}{1-z}\left ( \hat{Q}_{\bltwo} \pl \hat{Q}_{\Bin}\pl \hat{Q}_{\Kout} \right ) -\hat{Q}_{\lone}\frac{m_{\lone}^2}{ k \cdot \lone}    \right ] | \Ampzero_{\lone \parallel \ga}|^2 + {\cal O}( m_{\lone}^2)  \nonumber \\[5pt]
\label{eq:A1sqhardcoll}
&=\frac{e^2}{\left ( k\Cdot \lone \right ) }\hat{Q}_{\lone} ^2 
\left( \tilde{P}_{f \to f \ga}(z)  - \frac{m_{\lone}^2}{ k \cdot \lone}  \right) 
 | \Ampzero_{\lone \parallel \ga}(\qz^2, \clz)|^2
\Big{|}_{\Bin \to \Kout \loneg \bltwo} + {\cal O}( m_{\lone}^2) \;,
\end{align}
where $|\Ampzero_{\lone \parallel \ga}|^2 =  |{\cal A}^{(0)}_{\Bin \to \Kout \loneg \bltwo}|^2$ 
and $\tilde{P}_{f \to f \ga}(z)$ is the collinear emission part of the splitting function for a fermion to a photon\footnote{~No prescription is required when $z \to 1$, in our case, as this soft region has been treated in another section and is cut off
by $\Des$. C.f. \APP\ref{app:compare} for a discussion involving the full splitting function.}
\begin{equation}
\label{eq:splitting}
\tilde{P}_{f \to f \ga}(z) \equiv   \left( \frac{1 + z^2}{1- z^{\phantom{2}}}  \right) \;,
\end{equation}
and the $m_{\lone}^2/(k \Cdot \lone)$ term is immaterial for the hard-collinear logs per se but 
of importance for the numerics as it captures $\ln \Des$ terms.
The LO order matrix element squared in \eqref{eq:A1sqhardcoll}, is given in   \APP\ref{rate:LO}.
The first line in \eqref{eq:A1sqhardcoll} is gauge dependent whereas the second is not 
 since  charge conservation has been applied. This is further manifested by the appearance of 
 the splitting function which is a universal object.

\subsubsection{$\tilde{{\cal F}}^{(hc,0)}$, structure of collinear singularities in $d \qz^2 d \clz$}
\label{sec:Fhc0}

Taking \eqref{eq:fhc} and 
using the integration measure $d \Phi_\ga$ in \eqref{eq:col-measure} one arrives at 
\begin{alignat}{2}
\label{eq:FhcMaster}
& \tilde{{\cal F}}^{(hc,\RR)}  (\underline{\de}) = \frac{1}{2^{10}  \pi^3 m_B^3} 
& & \left(  \hat{Q}_{\ell_1}^2 \int^{\max(z(\Des),0)}_{\max(z(\des),0)}      
   \tilde{f}^{(hc)}  (\clz,  m_{\lone} ,\Dec) dz    \; + \right.  \nonumber \\[0.1cm]
   & & & 
 \left.  \hat{Q}_{\bltwo}^2  \int^{\max(z'(\Des),0)}_{\max(z'(\des),0)}      \tilde{f}^{(hc)}  (-\clz,  m_{\ltwo} ,\Dec)  dz   \right)  \;,
\end{alignat}
where  the boundaries on $z$ are determined by the phase space slicing  cut-off $\Des$ 
and the real photon energy cut off $\des$ \eqref{eq:des},
\begin{equation}
\label{eq:z}
z(\de) = z(\de,\qz^2, \clz) = 1-\frac{\de}{1-\hat{s}_{K \ltwo}(\qz^2, \clz) }  \;,
\end{equation}
with $\hat{s}_{K\ltwo} \equiv ( \hat{p}_K+ \hat{\ell}_2)^2 =  \frac{1}{2}\left ( 1-\hat{q}_0^2+ \hat{m}_K^2- \clz  \,\lambda ^{1/2}\left ( 1,\hat{q}_0^2,\hat{m}_K^2 \right )    \right )$  and $z' = z|_{\clz \to - \clz}$.\footnote{~Note that in the $\FRfour$-frame, the collinear limit forces the pair of particles (either $\loneg$ and $\ltwo$, or $\ltwog$ and $\lone$) to move in opposite directions. Since $\clz$ is defined w.r.t. $\lone$, this explains 
the $\clz \to - \clz$ procedure to obtain the corresponding formulae for $\ltwo || \ga$.}
The integrand in \eqref{eq:FhcMaster} reads
\begin{alignat}{2}
&  \tilde{f}^{(hc)}  (\clz,m_{\ell},\Dec)  &\;=\;&    
\la^{1/2}(m_B^2, \qz^2, m_K^2)    |\Ampzero_{m_{\ell}\to 0}(\qz^2, \clz)  |^2  \left( \tilde{P}_{q \to q \ga} (z) \, j^{hc}   - j_{(m_{\lone})} ^{hc}   \right)  \;,
\end{alignat}
with the LO amplitude squared given in \APP\ref{rate:LO} (in terms of $q^2,\cl$ though) and  
the $j^{hc}$'s are functions of $z$, $m_{\lone}$ and the collinear scale $\Dec$,
\begin{alignat}{2}
\label{eq:hc}
& j^{hc}(z,\Dec,m_{\lone})  &\;=\;&  \int^{ \Dec m_B^2}_{ \frac{1-z}{2z} m_{\lone}^2}  
  \frac{d  (k \Cdot \lone) }{ k \Cdot \lone}  = \ln \frac{ 2 \Dec  z}{\hat{m}_{\lone}^2 (1\mi z)} \;, \nonumber 
  \\[0.1cm] 
  & j^{hc}_{(m_{\lone})}(z,\Dec,m_{\lone})  &\;=\;&  
   \int^{ \Dec m_B^2}_{ \frac{1-z}{2z} m_{\lone}^2}  
 \frac{ m_{\lone}^2 \, d  (k \Cdot \lone) }{ (k \Cdot \lone)^2} = \frac{2z}{1-z} - \frac{\hat{m}_{\lone}^2}{\Dec} \;, 
\end{alignat}
and  the integration boundaries on $d \loneg^2$ correspond to \eqref{eq:kdotl1dc}.
Here and below, hatted quantities are normalised w.r.t. the $m_B$ mass, i.e. 
$\hat{m}_K = m_K/m_B$.

In the case of the $\{ \qz^2, \clz\}$-variables, as adapted in this section,   \eqref{eq:FhcMaster} 
can be simplified considerably   
\begin{equation}
\label{eq:FhcR}
\tilde{{\cal F}}^{(hc,0)}  (\underline{\de})  =  \frac{ \la^{1/2}(m_B^2, \qz^2, m_K^2)}{2^{10}  \pi^3 m_B^3}
\left(   |{\cal A}^{(0)}(\qz^2, \clz) |^2 \, \hat{Q}_{\ell_1}^2  J^{(hc,0)}  (\underline{\de})+  \;
 \{ 1,\clz \leftrightarrow 2,-\clz \}  \right) \;, 
\end{equation}
and the remaining  hard-collinear integral $ J^{(hc,0)}$ is easily evaluated\,\footnote{~Note that the $z$-integration strictly speaking involves $\max$ conditions, c.f.~\eqref{eq:FhcMaster}, and this is how we have performed the integral. However for $\Des,\des \ll 1$ the $z$'s are always larger than zero, hence the simplification.}
\begin{alignat}{2}
\label{eq:Ihc}
& J^{(hc,0)} (\underline{\de})  &\;=\;&  
\int^{z(\Des)}_{z(\des)} dz \left( \tilde{P}_{q \to q \ga} (z) \, j^{hc}   - j_{(m_{\lone})} ^{hc}   \right)
  \nonumber  \\[0.1cm]
&  &\;=\;&   A(\des,\Des) \ln \frac{ m_{\lone}^2}{ 2 \Dec m_B^2}+ B(\des,\Des,m_{\lone}) \;,
\end{alignat}
where 
\begin{alignat}{2} 
& A(\des,\Des)  &\;\stackrel{\phantom{z(\Des)\to 1}}{=}\;& \frac{1}{2}  (z(\Des)-z(\des))  (2+ z(\Des) + z(\des) ) + 2 \ln  \frac{ \bar z(\Des)} { \bar z(\des)}\nonumber\\[0.1cm]
&  &\;\stackrel{z(\Des)\to 1}{\to}\;& \frac{1}{2}  \bar z(\des)  ( 3 + z(\des) ) + 2 \ln  \frac{ \bar z(\Des)} { \bar z(\des)}
\; \stackrel{ z(\des) \to 0}{\rightarrow} \;\frac{3}{2} + 2 \ln \bar z(\Des)   \;,\nonumber \\[0.1cm] 
& B(\des,\Des,m_{\ell}) &\;\stackrel{\phantom{z(\Des)\to 1}}{=}\;& \frac{1}{2}\Big[ \left(z(\des)^2+2z(\des)\right) \ln \frac{ {z}(\des)}{ \bar z(\des)} -  \ln  \bar{z}(\des) - 4 \Li  \bar{ z} (\des) \nonumber \\[0.1cm]
&  &\;\phantom{=}\;&-2 \ln^2 \bar z(\des)  -\left(3+2\frac{\hat{m}_{\ell}^2}{\Dec }\right)z(\des)  -\{\des \leftrightarrow \Des\} \Big]  \\[0.1cm]
& &\;\stackrel{z(\Des)\to 1}{\to}\;&\frac{1}{2}\Big[ \left(z(\des)^2+2z(\des)\right) \ln \frac{ {z}(\des)}{ \bar z(\des)}  - \ln  \bar{z}(\des) - 4 \Li  \bar{ z} (\des) 
 \nonumber \\[0.1cm]
&  &\;\phantom{=}\;&   + 
\left(3+2\frac{\hat{m}_{\ell}^2}{\Dec }\right) 	\bar{z}(\des)  -2 \ln^2 \bar{ z}(\des) \;   + 2 \ln^2 \bar{z}(\Des) +4\ln \bar{z}(\Des)  \Big]  \;,
\end{alignat}
with $\bar z \equiv 1- z$ and the $z(\Des) \to 1$ limit has been used since $\Des \ll 1$.
Moreover, $A$ is the coefficient of the collinear log, for which we have also indicated the result for the photon-inclusive limit (i.e.~ $z(\des) \to 0$).
The hard-collinear logs from $\tilde{{\cal F}}^{(hc,0)}$ integrated over the full rate, starting from the soft cut-off  $\Des$, becomes  
\begin{align}
\label{eq:hcl1k}
d^2\Gamma^{(\RR)} \Big{|}^{(hc)}_{\lone || \ga ,\, \ln m_{\lone}}=d^2\Gamma^{\LO} _{\Bin \to \Kout \loneg \bltwo}\left ( \frac{\alpha }{\pi } \right ) \hat{Q}_{\lone}^2 \left [ \frac{3}{2}+2\ln   \bar z(\Des)   \right ] \ln m_{\lone} + \textrm{reg.~terms} \;,
\end{align}
where ``reg. terms'' stands for terms which are finite in the $m_{\lone} \to 0$ limit.

We are now ready to show the cancellations of the $\ln m_{\lone}$-terms by 
assembling all pieces. Defining 
\begin{equation}
\label{eq:hcCancel0}
\left. \frac{d^2\Gamma}{d \qz^2 d \clz} \right|_{\ln m_{\lone}} =
\frac{d^2\Gamma^{\LO}}{d \qz^2 d \clz}\left(  \frac{\al}{\pi} \right) \hat{Q}_{\lone}^2  \ln m_{\lone}  \times C^{(0)}_{\lone} \;,
\end{equation}
we find 
\begin{equation}
C^{(0)}_{\lone} = 
  \left[  \frac{3}{2}  +  2\ln   \bar z(\Des) \right]_{\tilde{{\cal F}}^{(hc)}}  
  + \left[  - 1 - 2\ln   \bar z(\Des) \phantom{ \frac{1}{2} } \!\!\!\! \right]_{\tilde{{\cal F}}^{(s)}}   +
  \left[ \frac{3}{2} - 2   \right]_{\tilde{{\cal H}}}      =0  \;,
\label{eq:hcCancel}
\end{equation}
complete cancellation. 
As explicitly indicated, the first term in square brackets comes from the hard-collinear integral, 
\eqref{eq:hcl1k}, the second term from the  soft integral in \EQ\eqref{eq:Fsoftcoll} of \APP \ref{app:SI}, and the 
last term from the virtual corrections  (here the $\frac{3}{2}$ originates from the $Z$-factors
 and the $-2$ from the $B_0$-functions in \eqref{eq:hcvirt}).  
 Note that the passage from 
$\Bin \to \Kout \loneg \bltwo$ in
\eqref{eq:hcl1k} to $\Bin \to \Kout \lone \bltwo$ in \eqref{eq:hcCancel} is justified since the lepton 
and the photon are collinear and can thus be treated as a single particle. 
The cancellation for the lepton $\bltwo$ is of course completely analogous.
It is worthwhile to point out that the hard-collinear logs, as well as the soft divergences, do cancel 
charge by charge as explicitly shown in \APP\ref{app:charge}.
Note, that in general the cancellation at the differential level
is spoiled by  non photon-inclusiveness  ($\des  < \des^{\textrm{inc}}$) and/or 
going over to the $\{ q^2 ,\cl\}$-variables.

\subsubsection{$\tilde{{\cal F}}^{(hc, \TT )}$, structure of  collinear singularities in  $d q^2 d \cl$}
 
We now proceed to analyse the analogous question for  the $\{q^2,\cl\}$-variables.
Setting $\mK \to 0$, for simplicity, we have (for  lepton $\ltwo$, $\cl \to - \cl$)
\begin{equation}
\label{eq:qreplace}
\qz^2 = \frac{q^2}{z} \;, \quad \clz|_{m_K \to 0} = \frac{\cl(1+z) + 1-z}{\cl(1-z) + 1+z}  \;,
\end{equation}
and using 
\begin{equation}
\label{eq:Jqztoq}
d \qz^2 d \clz \;=\; 4 ( \cl(1-z) + 1+z)^{-2} dq^2 d \cl\;,
\end{equation}
 the analogue of  \eqref{eq:FhcMaster} becomes 
\begin{alignat}{2}
\label{eq:FhcT}
& \tilde{{\cal F}}^{(hc, \TT )}  (\underline{\de})  &\;=\;&  \frac{  \hat{Q}_{\ell_1}^2 }{2^{8}  \pi^3 m_B^3}
\int_{\max(z_\textrm{inc}(\cl),z_{\des}(\cl) )}^{\max(z_\textrm{inc}(\cl),z_{\Des}(\cl))} dz \, \left[ \right.  \frac{ |{\cal A}^{(0)}(q_0^2, \clz) |^2 \la^{1/2}(\qz^2,m_B^2,0)}{ (\cl(1-z) + 1+z)^2  } \times   \; \nonumber \\[0.1cm]
& &\;\phantom{=} \;&  \left( \tilde{P}_{q \to q \ga} (z) \, j^{hc}   - j_{(m_{\lone})} ^{hc}   \right) \left. \right]  \,  + \,  \{ 1,c_{\ell} \leftrightarrow 2,-c_{\ell}  \}  \;, 
\end{alignat} 
where $\clz = \clz(\cl)$ with regard to the symmetrisation over $\cl$, 
$z_{\des}(\cl) $ implements the photon energy cut \eqref{eq:des}
and the arguments have to be substituted by \eqref{eq:qreplace}.
The boundaries for the $z$-integral are given by\,\footnote{~Note that the photon-inclusive case, $\des^\textrm{inc}$, corresponds to the minimum value of $z$, for a given $q^2$. 
In the limit of $m_K\to 0$, one can deduce, from \eqref{eq:qreplace}, that this corresponds to $\qz^2=m_B^2$, which then leads to $z_{\textrm{inc}}(\cl)|_{m_K \to 0}$ in \eqref{eq:zboundaries}.}
\begin{align}
\label{eq:zboundaries}
z_{\textrm{inc}}(\cl)|_{m_K \to 0} = \hat{q}^2 \;, \quad 
z_{\de}(\cl)|_{m_K \to 0} =\frac{1+\hat{q}^2-\de+\cl (1-\hat{q}^2-\de)}{1+\hat{q}^2+\de+\cl (1-\hat{q}^2-\de)}\;,
\end{align}
and obtained  by solving \eqref{eq:z} for $\de = \des^\textrm{inc},\Des,\des$ as appropriate, with \eqref{eq:qreplace} in place.
 The phase space slicing condition 
is implemented via $z_{\Des}(\cl) < 1$.

The new aspect is that the   $|{\cal A}^{(0)}(\qz^2, \clz) |^2$ cannot be factored out since it depends
on $z$ implicitly through $\qz^2$ and $\clz$.  
However, in the limit of $\mK \to 0$ and $m_{\lonetwo} \to 0$,
the  amplitude squared \eqref{eq:A0} is simple enough, 
\begin{equation}
| {\cal A}^{(0)}(q_0^2, \clz) |^2 =  \geff^2  (|\CV|^2 +|\CA|^2)  \, 2( 1- \clz^2) (1- \hat{q}_0^2)^2 \, f_+^2(\qz^2) \;,
 \end{equation}
 and the integral can be done analytically. Note that above $\{ \qz^2,\clz\}$ are to be substituted as  in \eqref{eq:qreplace}.

Adding all the contributions, real and virtual, that contribute to the hard-collinear logs, one finds
\begin{equation}
\label{eq:new}
\frac{d^2\Gamma}{d q^2 d \cl} \Big{|}_{\ln m_{\lonetwo}} = 
 \frac{\al}{\pi} ( \hat{Q}_{\lone}^2 K_{\textrm{hc}}(q^2,\cl)      \ln m_{\lone} + 
\hat{Q}_{\ltwo}^2 K_{\textrm{hc}}(q^2,-\cl)      \ln m_{\ltwo})     \;,
\end{equation}
where $K_{\textrm{hc}} {(q^2,\cl)}$ is a non-vanishing function (cf. \APP\ref{app:compare} for a non-trivial cross-check).
Plots of this quantity are shown in \FIG\ref{fig:sizehc} for $(\lone,\bltwo=\ell^-,\ell^+)$, with $\ell=e,\mu$. 

At last, we would like to mention that for $q^2 \to (m_{\lone} + m_{\ltwo})^2$ and $\cl \to -1$ the assumption that 
$k \cdot \lone$ is small compared to other scalar products breaks down and this leads to artificial enhancements. For example,  the Jacobian factor 
in \eqref{eq:FhcT} becomes too large when $q^2$ is small and $\cl \to -1$. However, 
for a binned rate this effect is negligible and moreover for 
 the $\{\qz^2,\clz\}$-variables there are no such issues  at all.

At times we have made the $m_K \to 0$ approximation for simplicity in presentation. 
The full expressions of  $\clz$ in terms of $\{q^2,\cl\}$ (\EQ\eqref{eq:qreplace}), the Jacobian from $\{\qz^2,\clz\}$ to $\{q^2,\cl\}$ (\EQ\eqref{eq:Jqztoq}), $s_{K \ltwo}$ in terms of $\{q^2,\cl\}$, the integrand for $\tilde{{\cal F}}^{(hc, \TT )}  (\underline{\de}) $ (\EQ\eqref{eq:FhcT}) and the limits of the $z$-integral (\EQ\eqref{eq:zboundaries}) can all be found in a Mathematica notebook appended to the arXiv version.

\subsubsection{Cancellation of hard-collinear logs for the total differential rate}

It is well-known that all IR divergences and IR sensitive terms ought to cancel at the level of the total,
photon-inclusive, rate \cite{Bloch:1937pw}. 
It is the aim of this section  
to verify this for the case at hand. The hard-collinear part of the total rate given by 
\begin{alignat}{2}
& \tilde \Gamma ^{(hc,\TT)}(\Des)\Big{|}_{\ln m_{\lone}} &\;\equiv\;& \frac{\alpha }{\pi }\int_0^{1}d\hat{q}^2 \int_{-1}^{1}d\cl \;\tilde{{\cal F}}^{(hc,\TT)} \;, \nonumber \\[0.1cm] 
& \tilde \Gamma ^{(hc,\RR)}(\Des)\Big{|}_{\ln m_{\lone}} &\;\equiv\;&  \frac{\alpha }{\pi }\int_0^1 d\hat{q}_0^2 \int^1_{-1} d\clz \;\tilde{{\cal F}}^{(hc,\RR)} \;,
\end{alignat}
where we have assumed the $\mK \to 0$ limit.

In accordance with the general expectation, we find
\begin{equation}
\tilde \Gamma ^{(hc)}\Big{|}_{\ln m_{\lone}} \equiv  \tilde \Gamma ^{(hc,\RR)}\Big{|}_{\ln m_{\lone}}  = \tilde \Gamma ^{(hc,\TT)}\Big{|}_{\ln m_{\lone}} \;,
\end{equation} 
equality at the level of the hard-collinear logs originating from the real radiation
\begin{align}
\tilde \Gamma ^{(hc)}(\Des)\Big{|}_{\ln m_{\lone}} &=\frac{m_B \hat{Q}_{\ell_1}^2 }{2^9\,(9  \pi^3)}f_+^2\geff^2(|\CV|^2 +|\CA|^2) \left [8+6\ln \Des +\ORD(\Des) \right] \ln m_{\lone} \;.
\end{align}
Since we have explicitly shown the cancellation for $\frac{d^2\Gamma}{d\qz^2 d\clz}$, this implies that 
the hard-collinear logs cancel for the integrated $\int \frac{d^2\Gamma}{dq^2d \cl} d q^2 d \cl $.
The $\ORD(\Des)$-terms can be safely neglected, since $\Des\ll 1$, and in any case 
the same approximation has been used when evaluating the soft integrals, c.f.~\APP\ref{app:SI}.

\subsection{On hard-collinear logs and structure-dependent terms}
\label{sec:beyondPT}

We turn to the important question as to whether further hard-collinear logs could be  missing 
due to omitted structure-dependent corrections. 
Using gauge invariance, we are able to show that this is not the case.
In doing so, we will further establish why the hard-collinear logs can be written as a sum of terms proportional 
to $\hat{Q}_{ \lone,\bltwo}^2$. 
At the end of the section, we give a physical argument of the previously established result that hard-collinear logs cancel at differential rate $\frac{d^2}{d\qz^2d \clz}\Gamma$, that is when expressed in 
$\{\qz^2,\clz\}$-variables.

The starting point is to realise that 
hard-collinear logs $\ln m_{\lonetwo}$ are generated by interference of 
\begin{equation}
\frac{1} {k \Cdot \lonetwo} 
\end{equation}
denominators  ($k$ approaching $\lonetwo$) with other terms.
 Without loss of generality, we may focus our attention to lepton $\lone$. 
The real amplitude 
can be  decomposed, 
\begin{equation}
\label{eq:decQ1}
\Ampone = \hat{Q}_{\lone}  \Amponered_{\lone}  +   \de \Ampone \;,
\end{equation}
into a term 
$\hat{Q}_{\lone} \Amponered_{\lone}$ with all terms proportional to   $\hat{Q}_{\lone}$, and the remainder 
$\de \Ampone$.  Note, that at this point we have not yet made use of charge conservation. 
From \eqref{eq:realAmp},  
\begin{equation}
\label{eq:Ampone}
\Amponered_{\lone}  =  - e \geff  \bar u (\lone)  \left[    \frac{2 \eps^* \Cdot \lone\pl  \slashed{\eps}^* \slashed{k}}{2 k \Cdot \lone }   \Gamma \Cdot H_0(\qz^2)   \right] v(\ltwo) \;,
\end{equation}
which contains all $1/(k \Cdot \lone)$-terms.
It is seen that the structure-dependence  of this term is encoded in the form factor $H_0$ 
(defined in  \eqref{eq:match}) only. For our purposes it is convenient 
to write the amplitude square, using \eqref{eq:decQ1},  in terms of three terms 
\begin{equation}
\label{eq:realsquared}
\sum_{\textrm{pol}}  |\Ampone|^2
   =   \sum_{\textrm{pol}}  |\de \Ampone|^2
 -  \hat{Q}_{\lone}^2 \sum_{\textrm{pol}}   |\Amponered_{\lone}|^2  
 + 2 \hat{Q}_{\lone}  \textrm{Re} [ \sum_{\textrm{pol}}  {  \Ampone}  \Amponesred_{\lone} ]    \;,
\end{equation}
where it will be important that $ \Ampone$ is gauge invariant.
By construction, the first term
is manifestly free from hard-collinear logs $\ln m_{\lone}$.  
To simplify the discussion, we may  use 
gauge invariance and set $\xi =1$ in this section under which  the polarisation sum,
$\sum_{\textrm{pol} }\eps^*_\mu \eps_\nu = ( - g_{\mu\nu} + (1-\xi) k_\mu k_\nu/k^2) \to - g_{\mu\nu}$,
collapses to the metric term only.
In this case, the second term evaluates to 
 \begin{alignat}{2} 
 \label{eq:above}
\int d \Phi_\ga\,  \hat{Q}_{\lone}^2 \sum_{\textrm{pol}}   |\Amponered_{\lone}|^2  &=\;& 
   \int d \Phi_\ga  \, \hat{Q}^2_{\lone} \frac{ {\cal O}(m^2_{\lone}) + \ORD( k \Cdot \lone)} { (k \cdot \lone)^2 }   = 
   {\cal O}( 1) \,  \hat{Q}^2_{\lone}   \ln m_{\lone} \;,
  \end{alignat}
where we used $k - \ell_1 = {\cal O}(m_{\lone}^2)$, valid in the collinear region. 
 Note
 that the form factor part  $ H_0(\qz^2)$ does not participate in the photon phase space integration, and factorises  when   working with $d\qz^2$. 
 We now turn to the third term.   Noting that $\Ampone\equiv \eps^{*\mu}\Ampone_{\mu}$, the crucial step in use is that  gauge invariance $k^{\mu} \Ampone_{\mu} = 0$ implies 
 $\lone^{\mu} \Ampone_{\mu} = {\cal O}(m_{\lone}^2)$  in the collinear region  and thus 
 the third term assumes the form
 \begin{alignat}{2} 
\label{eq:third}
 \hat{Q}_{\lone}  \sum_{\textrm{pol}}  { \Ampone} \Amponesred_{\lone}  &=\;& 
c_1 \, \hat{Q}^2_{\lone} \frac{ {\cal O}(m^2_{\lone})+\ORD( k \Cdot \lone)} { (k \cdot \lone)^2 } + c_2 \, \hat{Q}_{\lone} \hat{Q}_{X}  \frac{ {\cal O}(m_{\lone})} { (k \cdot \lone)} +  \dots \;, 
  \end{alignat}
 where $X \in \{\Bin,\Kout,\bltwo\} $ and the ellipses stand for less singular contributions.
 The $c_1$-term   has the same origin as the one in \eqref{eq:above}. 
 The $c_2$-term comes from 
interfering the spin dependent term in \eqref{eq:Ampone} with the $\hat{Q}_{\lone}$-independent 
part of  ${\Ampone}$ and it is  by the use of the 
equation of motion, that one arrives at the ${\cal O}(m_{\lone})$-suppression\footnote{~In fact, this result is true more generally since the 
spin dependent part is proportional to the Lorentz-generator which, by contraction, is a boost into the 
direction of the photon. Let us assume that $m_{\lone} = 0$.
Since in the collinear limit, the photon and the lepton are parallel, the massless lepton is boosted in direction of movement. Since the helicity of a massless particle cannot be changed, the generator has to vanish. 
If the lepton mass is reinstalled, then there 
are terms of the form $m_{\lone} \ln m_{\lone}$ which are however safe.} 
\begin{equation}
\int d \Phi_\ga \frac{ {\cal O}(m_{\lone})} { (k \cdot \lone)}  = {\cal O}(m_{\lone}) \ln m_{\lone} \;,
\end{equation}
as compared to \eqref{eq:above}.
Hence we have established that all  hard-collinear terms $\ln m_{\lone}$ 
can be written as a sum of terms proportional to $\hat{Q}_{\lone}^2$.
It should be added that in making this statement,  charge conservation was used 
since  gauge invariance was assumed. All statements hold irrespective of any photon phase space restrictions 
such as an energy cut-off $\des$ or a photon angle cut (cf.  \SEC\ref{sec:distortion}).
Thus, any gauge invariant addition to the amplitude, 
due to structure-dependent terms, will not give rise to any additional $\ln m_{\lone}$-terms.  

So far, our analysis has been concerned with the real amplitude only.
Assuming that
 hard-collinear logs cancel charge by charge combination 
at the differential level in the $\{\qz^2,\clz\}$-variables, irrespective of the microscopic approach,
the same conclusion applies to each virtual 
diagram.\footnote{~A physical argument of the correctness of this assumption is given in the last of paragraph of this section.
In particular, we have verified this explicitly 
up to the second derivative of the form factor in our approach  and 
produced a formal derivation that holds to all orders.}  
 For  virtual diagrams, there is no distinction between 
$\{\qz^2,\clz\}$- and $\{q^2,\cl\}$-variables and thus the conclusion holds irrespective of the 
differential variables. 
As the reader might suspect, 
 the same conclusions holds for lepton $\ell_2$ by symmetry.
Let us summarise these  findings: 
 \begin{itemize}
\item Additional structure-dependent corrections, which are of course gauge invariant, will not give rise to
any additional hard-collinear logs $\ln m_{\lonetwo}$.\footnote{This applies to either, approaches resolving the mesons by partons or an evaluation of the $B(K)_\ga L_{1,2}$-diagrams, cf. \FIG\ref{fig:virtual}, 
 including higher terms in the expansion \eqref{eq:Lint_eff}.}
 \item  At the double-differential level, hard-collinear logs $\ln m_{\lonetwo}$, real and virtual, 
can be written as a sum of terms proportional to $\hat{Q}_{\lonetwo}^2$  consistent with our explicit evaluation using the phase space slicing method  in Eq.~\eqref{eq:hcl1k}.
\end{itemize}
To this end, let us give a physical explanation as to why
 hard-collinear logs $\ln m_{\lonetwo}$ are to cancel at the differential level in $\{ \qz^2, \Tlz \}$.
In those variables, the  decay corresponds to the disintegration of a   scalar particle of mass 
$\qz^2$ which is an infrared-safe observable.
 Now, the angle $ \Tlz$ has no meaning when the decay axis, cf \FIG\ref{fig:angles},  
 is decoupled and the $\Bin$ and the $\Kout$ are interpreted  as a single particle of mass $\qz^2$. 
 This observation is backed up  by our explicit formal verification in Eq.~\eqref{eq:hcCancel0}. 
 In essence $\qz^2$ is an IR-safe kinematic variable and the entirety of 
 the particles in $\qz^2$ can be viewed as the moral cousin of a jet.

\section{Results for $\Bin \to \Kout e^+ e^-$ and $\Bin \to \Kout \mu^+ \mu^-$}
\label{sec:plots}

The total radiative corrections are  presented in \SEC\ref{sec:total}, 
followed by  a discussion of the distortion of the spectrum due to $\ga$-radiation in \SEC\ref{sec:distortion}. 
The size of the hard collinear logs and some comparison with older work is deferred 
to \APPs\ref{app:sizehc} and \ref{app:BIP} respectively.
Before proceeding thereto, we summarise the input to the numerics below.

For the particles participating in the decay,  the following masses   are assumed:
$m_e = 0.511  \MeV$, $m_\mu = 0.10565 \GeV$,  $m_B = 5.28 \GeV$ and $m_K = 0.495\GeV$. 
Other parameters are  the Wilson coefficients,  
$C_{9} = 4.035$ and  $C_{10} = -4.25$ at $\mu_{\textrm UV} = 4.7\GeV$  (the $b$-quark pole mass) 
and  the fine 
structure constant,   $1/\al = 137.036$.
For the $B \to K$ form factors \eqref{eq:match},  the light-cone sum rules computation \cite{Ball:2004ye}, including radiative correction 
up to twist-$3$,  was used with updated Kaon distribution amplitude parameters\footnote{~For 
the Kaon distribution amplitude, the values 
$a^K_1(1\GeV) =  0.115(34)$ and 
$a^K_2(1\GeV) = 0.090(20) $ taken from the 
$N_f = 2 +1$ lattice computation \cite{Bali_2019} (uncertainties were added in quadrature) were used.
These values are consistent with earlier QCD sum 
rule computations \cite{Braun:2004vf,Ball:2005vx,Ball:2006fz,Chetyrkin:2007vm}.}

\begin{equation} 
\label{eq:ffs}
\{ f_{+ }, f_{- } \} ^{B \to K}(0) = \{ 0.271,-0.206\} \;,\quad \frac{d}{dq^2} \{ f_{+ }, f_{- } \} ^{B \to K}(0)=\{ 0.0151,-0.0109\}\GeV^{-2} \;, 
\end{equation}
where the uncertainty is roughly $15\%$ if one additionally takes into account  the error on the Kaon distribution amplitude. 
For the auxiliary cut-offs of the phase space slicing method,
$\Des(e) = 2.5 \Cdot 10^{-3}$, $\Des(\mu) = 4 \Cdot 10^{-3}$, $\Dec(e) =1 \Cdot 10^{-2}  \Des(e) $ 
and $\Dec(\mu) =2 \Cdot 10^{-2}  \Des(\mu) $ lead to stable results.  The hierarchy $ \Dec/\Des \ll 1$ is 
important since  terms of this order are neglected.\footnote{~We refer the reader to \cite{Harris:2001sx} for an uncertainty analysis involving the auxiliary cut-offs.}  
Here, we refrain  from a complete uncertainty analysis.
 Let us nevertheless mention the sources. 
There are the form factor uncertainties which can be largely reduced by considering correlations amongst 
 the four numbers  \eqref{eq:ffs} entering the computations. Besides a more complete 
structure-dependent approach, cf. \SEC\ref{sec:SD}, there are  missing finite counterterms 
in the charged meson case, which we set to zero and  refer the reader to the  discussion in \SEC\ref{sec:virtual}.
Concerning the latter, one might get a naive dimensional analysis estimate by varying the constant $c$, 
associated with  $1/\epsUV + c$, by a factor of $2$ (or alternatively by varying the renormalisation scale $\mu_{\textrm{UV}}$).

 Adding these effects  in quadrature results in an  $\ORD(1\%)$-variation.

\begin{figure}[h!]
\includegraphics[width=0.5\linewidth]{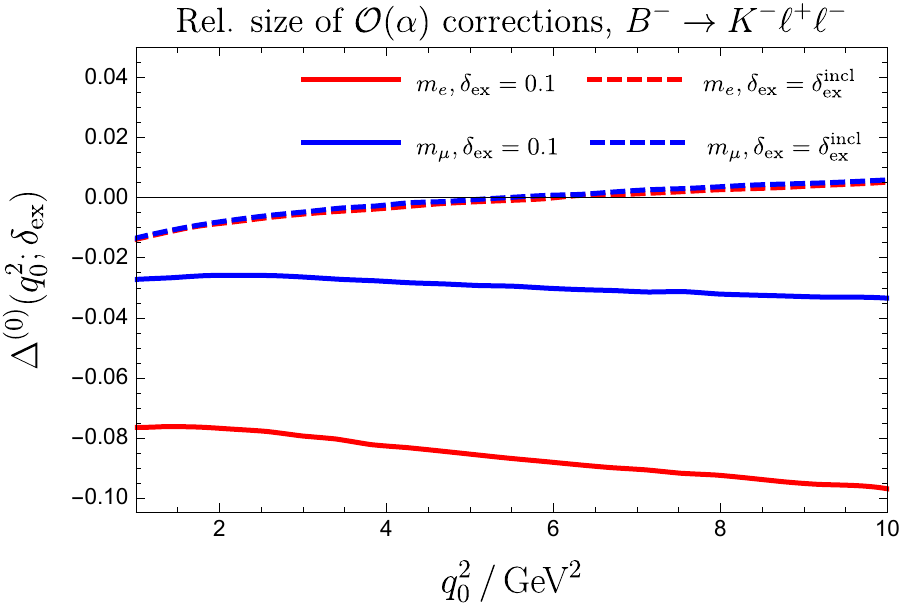}  
        \includegraphics[width=0.5\linewidth]{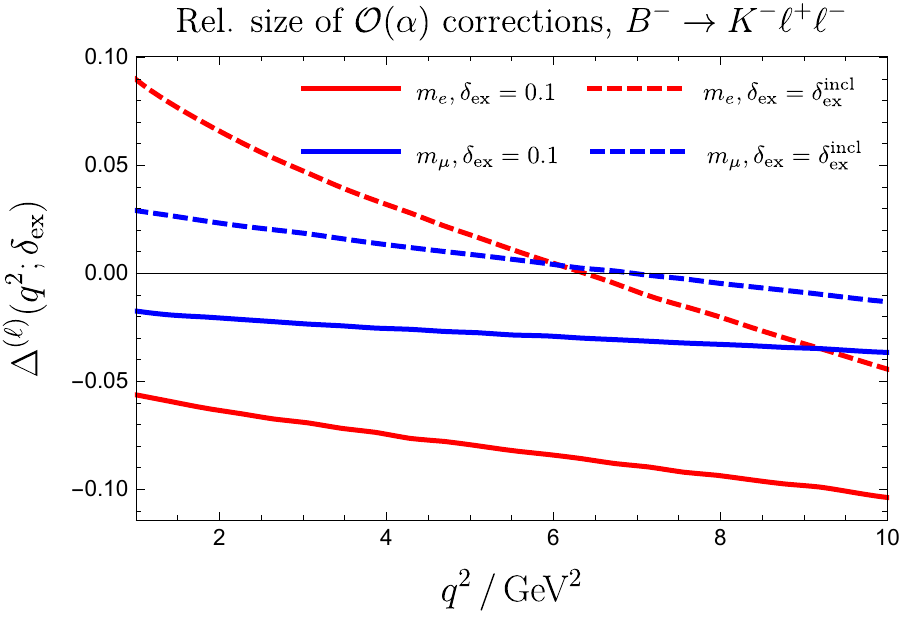}  
        \hfill \\[-0.2cm]
	\includegraphics[width=0.5\linewidth]{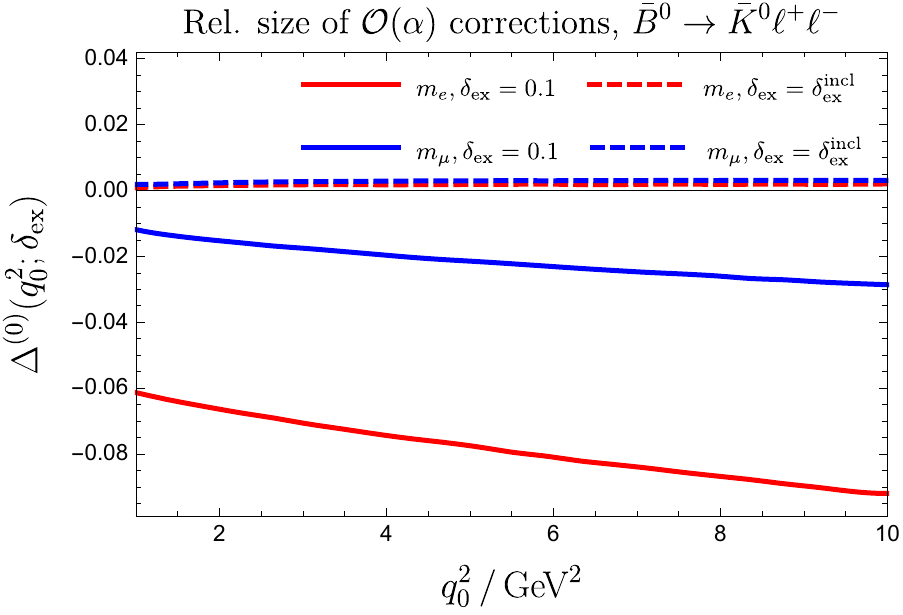}  
        \includegraphics[width=0.5\linewidth]{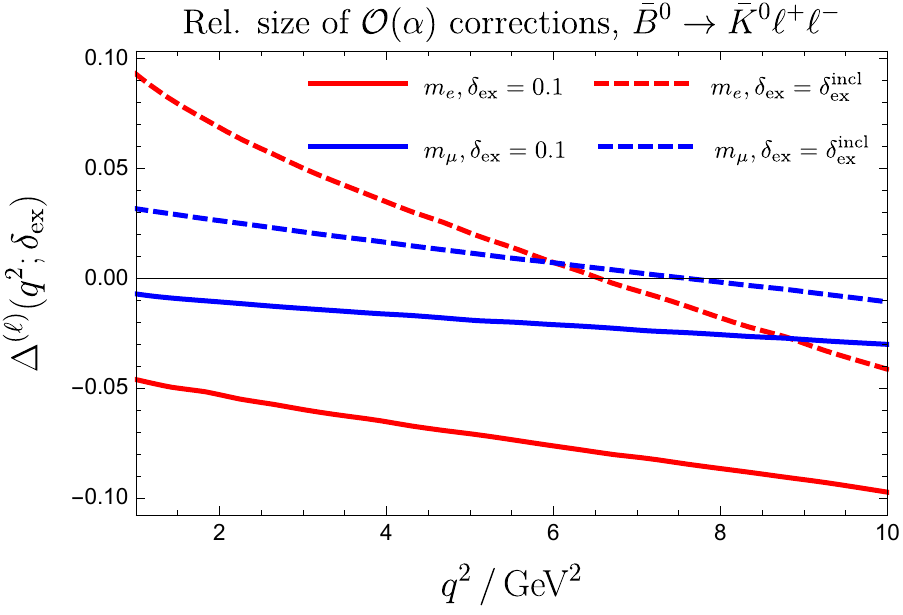}  	
	\caption{\small Total relative QED-corrections, cf.~\eqref{eq:dq2Del} 
	for the definition, including finite terms.
	The upper and lower figures correspond to the charged and neutral modes 
	in the  $\qz^2$- and $q^2$-variables on the left and right  respectively.  
	In the photon-inclusive case ($\des= \desinc$, dashed lines), all IR sensitive terms 
	cancel in the $\qz^2$ variable locally and in the $q^2$-variable when integrated 
	which is nicely visible in both cases. 
	 In the charged case, however, we see finite effects of the $\ORD(2 \%)$  due 
	to $\ln \hat{m}_K$ ``collinear logs" which do not cancel. 
	An important aspect is the (approximate) lepton universality on the plots on the left.       
	As is well-known,  effects due to the photon energy cuts are sizeable
	since  hard-collinear logs do not cancel in that case.  This is in particular for electrons.}
	\label{fig:sizeDel}
\end{figure}

\begin{figure}[h!]
	\includegraphics[width=0.5\linewidth]{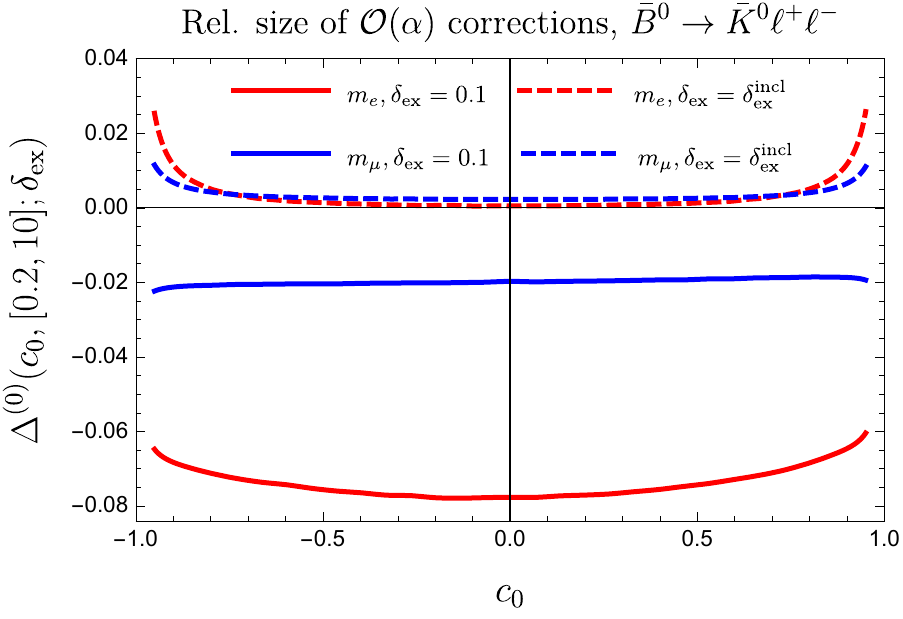}  
        \includegraphics[width=0.5\linewidth]{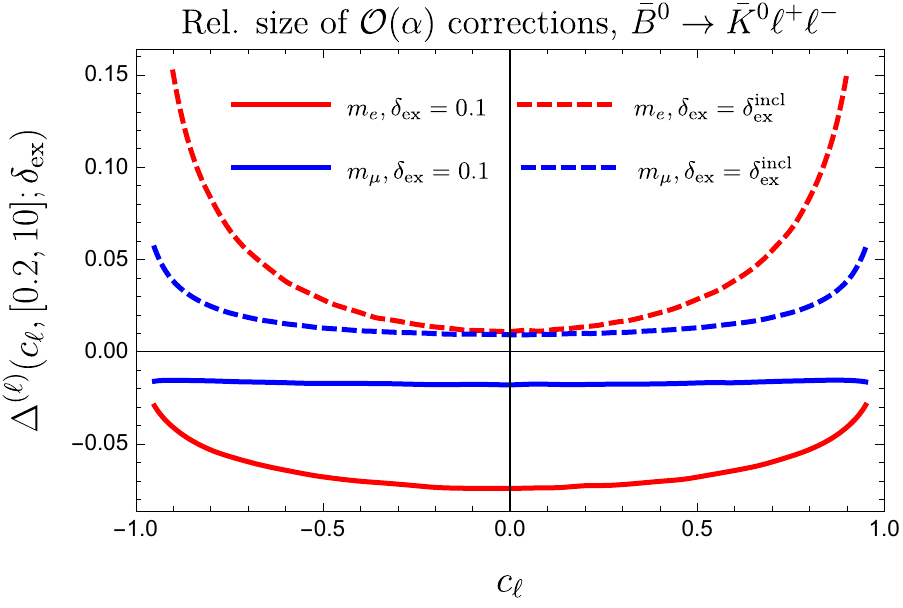}  	%
	\caption{\small Total relative QED-corrections \eqref{eq:dclDel}  in terms of 
	$\clz = \cos (\theta_0)$   $\cl = \cos (\theta_\ell)$ respectively for the electrically neutral hadron case.
	 In the $\clz$-variable effects are small for $\des = \desinc$ cf. comments in text and previous figures. 
	 The enhanced effect towards the endpoints $\{-1,1\}$ in the electron case is, partly, due to the special   behaviour of the LO expression \eqref{eq:dG0} which behaves like $\propto (1- \cl^2) + \ORD( m_{\ell}^2)$ and 
	 explains why the effect is less pronounced for muons.}
	\label{fig:sizeDelANG}
\end{figure}

\subsection{Radiative corrections as a function of $\qz^2,\clz$ and $q^2,\cl$}
\label{sec:total}

 We consider it most instructive to discuss 
 the  relative QED corrections, implicitly defined
in \eqref{eq:Delta},  
\begin{equation}
\label{eq:DeltaAAA}
\Delta^{(\AAA)}(\qSaa,  \claa;\des) =   \left( \frac{d^2 \Gamma^{\LO}}{d\qSaa d \claa}\right)^{-1}  \frac{d^2\Gamma(\des)}{d\qSaa d \claa} \Big{|}_{\al}   \;.
\end{equation}
Above $|_{\al}$ stands for the inclusion of the  ${\cal O}(\al)$-corrections only.  
The LO rate is given  \EQ\eqref{eq:ratell}.
We further consider  the relative single differential in $\frac{d}{d\qSaa}$
\begin{equation}
\label{eq:dq2Del}
 \Delta^{(\AAA)}(\qSaa;\des)  =   \left( \frac{d \Gamma^{\LO}}{ d\qSaa}\right)^{-1}  \frac{d\Gamma(\des)}{d \qSaa } \Big{|}_{\al}   \;,  
\end{equation}
where the numerator and denominator are integrated separately  over
  $\int_{-1}^1 d \claa$ respectively. In addition, we  define the single differential in
$\frac{d}{d\claa}$ 
\begin{equation}
\label{eq:dclDel}
 \Delta^{(\AAA)}( \claa,[q_1^2,q_2^2];\des) =  
  \left( \int_{q_1^2}^{q_2^2}  \frac{d^2 \Gamma^{\LO}}{d\qSaa d \claa}  d\qSaa  \right)^{-1} \int_{q_1^2}^{q_2^2}   \frac{d^2 \Gamma(\des)}{d \qSaa d  \claa } d \qSaa  \Big{|}_{ \al}  \;,
\end{equation} 
where the non-angular variable is binned. 
We would like to stress that it is important to integrate  the QED correction and the LO separately 
as this corresponds to the  experimental situation.

Results for  $\Delta^{(\AAA)}(\qSaa;\des)$ and $ \Delta^{(\AAA)}( \claa,[q_1^2,q_2^2];\des)$ 
are shown in \FIGs\ref{fig:sizeDel} and \ref{fig:sizeDelANG} respectively.
Let us first focus on \FIG\ref{fig:sizeDel} where in the photon-inclusive case ($\des = \desinc$, dashed line),
one observes two important features: Approximate lepton-universality and the cancellation of 
the hard-collinear logs. 
In the $\qz^2$-variable, this happens at the differential level whereas for 
the $q^2$-variable, integration over the entire range is needed (the tendency thereto is  visible in the plot of the RHS). 
To be clear, the cancellation in the later case only occurs upon integration over the full $q^2$-range.
We further remind  the reader 
that in all cases the soft divergences cancel locally as explicitly shown in \SEC\ref{sec:soft}.
It is noticeable that for the charged case,   there are  $\ORD(2\%)$-effects in the  $\qz^2$-variable 
due to  ``collinear logs", $\ln\hat{m}_K\simeq -2.36 $. These  logs, of course,  cancel
upon integration over all differential variables.
The impact of the photon energy cuts are large, cf.~\APP\ref{app:sizehc}, 
and care needs to be taken when considering quantities like $R_K$ for example.
An important physical effect,  visible in the plots on the right in \FIGs\ref{fig:sizeDel}, is the distortion 
of the $q^2$ distribution w.r.t.~the non-radiative case. 
This is particularly prominent in the photon-inclusive limit 
as  discussed in the next section.

The angular differential $ \Delta^{(\AAA)}( \claa,[q_1^2,q_2^2];\des)$ in \FIG\ref{fig:sizeDelANG} 
 shows similar patterns in the photon-inclusive case  ($\des = \desinc$, dashed lines), e.g. 
 lepton universality and small effects in the $\clz$-variable due to the cancellation of hard-collinear logs. 
 In the electron case, there is a significant enhancement towards 
the endpoints $\{-1,1\}$ which is due to the peculiar behaviour of the LO rate  
$d \Gamma^{\LO} \propto (1- \cl^2) + \ORD( m_{\ell}^2 )$    \eqref{eq:dG0}. This is the same effect 
as the helicity suppression in a $\pi^- \to \ell^- \bar\nu$  decay and  further explains why the effect 
is less prominent in the muon case.  A more detailed analysis of the angles will follow in a forthcoming 
paper cf. comments in \SEC\ref{sec:moments}. Cuts on the photon energy are again sizeable and the same 
remarks as before apply.

Plots of the hard-collinear logs $\ln m_\ell$ 
are deferred to \APP\ref{app:sizehc}. Moreover in \APP\ref{app:BIP} our results are compared to 
the  earlier work~\cite{BIP16} where virtual corrections were indirectly inferred
and radiative corrections have been evaluated in terms of a radiator function 
depending  on $q^2$ and $q_0^2$ only, and not on the 
photon-emission angle. 

\subsection{Distortion of the  $\Bin \to \Kout \ell^+ \ell^-$ spectrum due to $\ga$-radiation}
\label{sec:distortion}

\begin{figure}[t]
	\centering
\includegraphics[width=0.49\linewidth]{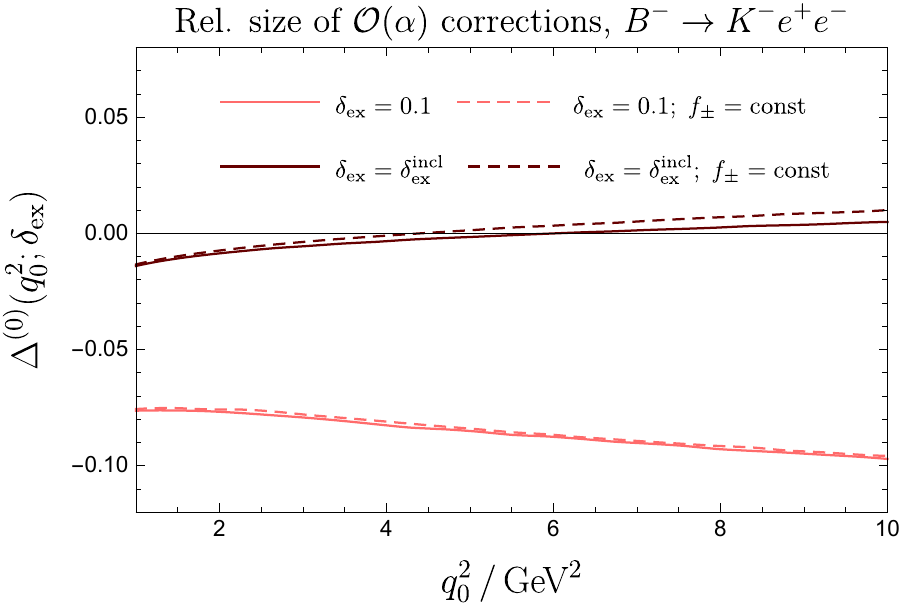}  
\hfill 
        \includegraphics[width=0.49\linewidth]{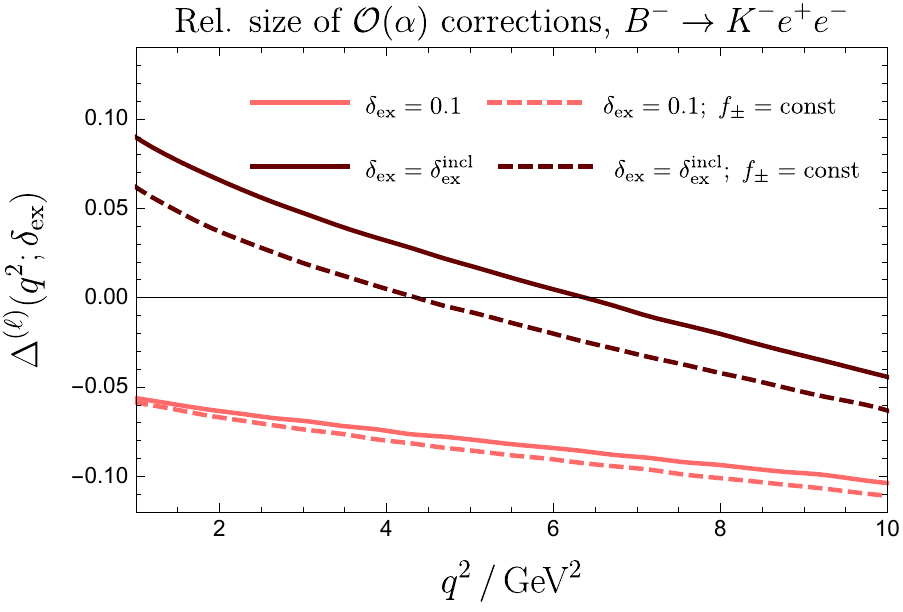}  
		\vskip 0.5 true cm
	\caption{\label{fig:distortion}
	\small Plots of total relative QED corrections \eqref{eq:DeltaAAA} for $B^- \to K^-  \ell^+ \ell^-$
	comparing the constant form factor case versus taking one derivative correction into account 
	with values given in \eqref{eq:ffs} (cf. below \eqref{eq:L0} for further comments). 
	Effects are more prominent in the photon-inclusive case ($\des = \desinc$) since there is more phase space for the $q^2$- and $\qz^2$-variables to differ.
	In the neutral case, we found that the effects are similar albeit slightly smaller.}
\end{figure}

As discussed in \SEC\ref{sec:IR} the  $\{\qz^2 ,\Tlz\}$-variables are safer than the  $\{q^2 ,\Tl\}$-variables 
because of the cancellation of the hard-collinear divergences. 
In this section, we wish to emphasise yet another reason why it is  preferable to use the  
$\{\qz^2 ,\Tlz\}$-variables. This is sometimes called 
the migration of radiation or the distortion of the spectrum: 
at fixed $q^2$, effectively the radiative process is probed at a different $\qz^2 = (q+k)^2$ 
as a result of the photon carrying away momentum. If the spectrum has significant variations in 
$q_0^2$, this implies a significant distortion in the kinematical distribution.  
This effect is indeed well-known from the determination of the $J/\Psi$-pole in 
$e^+ e^- \to \textrm{hadrons}$~\cite{Greco:1975rm}.  
Generically,  the more inclusive one gets in the photon 
energy and angle,
the more pronounced it is,  as in this case  the radiative topologies ($4$-body) 
can be very different from the virtual ones ($3$-body).

Let us illustrate the effect by considering the hard-collinear radiation, $\tilde{{\cal F}}^{(hc, \TT )}  (\underline{\de})  $ given in \EQ\eqref{eq:FhcT}.
Assuming the $m_K=0$ limit, for simplicity, the $dz$-intgegrand contains $ |{\cal A}^{(0)}(\qz^2, \clz) |^2 \propto f_+(\qz^2)^2 = f_+(q^2/z)^2$ (c.f.\,\EQ\eqref{eq:dG0} with $m_\ell=0$)
and $\qz^2 =q^2/z$ from \EQ\eqref{eq:qreplace}.
Since $z<1$ in general, it is clear that momentum transfers of a  higher range are probed.
For  $\cl = -1$, maximising the effect,  one gets 
\begin{equation}
\label{eq:shift}
z_{\des}(q^2)\Big|_{\cl = -1} = \frac{q^2}{q^2 + \des m_B^2} \;, \quad (\qz^2)_{\max} =q^2 + \des m_B^2 \;,
\end{equation}
upon using \eqref{eq:zboundaries}. 
Thus for $\des = 0.15$ and $q^2 = 6 \GeV^2$ one finds $ (\qz^2)_{\max}  = 10.18\GeV^2$ 
which is of course problematic when one wants
 to probe $R_K$ in the $q^2 \in [1,6]\GeV^2$ range, given that the charmonia start to impact more severely 
 well below $10\GeV^2$.
 In the photon-inclusive case, the lower boundary becomes  $z_{\textrm{inc}}(\cl)|_{m_K \to 0} = \hat{q}^2$
 by  \EQ\eqref{eq:zboundaries} and $(\qz^2)_{\max} =m_B^2$.
  Hence, in that case the entire spectrum is probed  for any fixed value of $q^2$ 
  which confirms the earlier statement. As it can easily be understood, this would be rather 
  problematic in $\Bin \to \Kout \ell^+ \ell^-$ decays due to the large charmonia contributions
  (cf.~comments below \eqref{eq:L0}), that would 
  ``contaminate'' all the $q^2$ region below their masses. This is why in experimental analyses, 
  stringent cuts on the photon energy (or the reconstructed $B$-meson mass) 
  and its emission angle are implemented.

The effects described above are visible  in both plots in \FIG\ref{fig:distortion}.  
We stress that they are  underestimated since a) we kept only one power 
in the derivative expansion and b)~one would need to incorporate  long-distance effects in addition.  
Note that for the virtual contributions, 
it is only when both hadrons are neutral that 
the derivative expansion can be avoided. If this is not the case, it is important to take 
into account higher derivative corrections and perform the matching of the finite counterterms from QCD.

As alluded above, besides the cut on the reconstructed $B$-meson mass, in order to reduce 
the migration of radiation (or better the distortion of the $q^2$ spectrum)
one can further restrict the photon's phase space in the 
photon's emission angle.
From $q = q_0 - k$, taking into account  \eqref{eq:(1)a}, one gets $q^2 = \qz^2  - 2E_\ga^{\FRone} 
(E_{\qz} ^{\FRone} + | \vec{q}_0^{\;\FRone}|\cos \Tg^{\FRone})$.
Then using  the expression of the maximum photon energy  in \eqref{eq:Emax}, one arrives at
\begin{equation}
(\qz^2)_{\max} = q^2 + \des m_B (E_{\qz} ^{\FRone} + | \vec{q}_0^{\;\FRone}|\cos \Tg^{\FRone}) \;.
\end{equation}
Assuming again for simplicity the $m_K =0$ limit where $ E_{\qz} ^{\FRone}  = (m_B^2 +\qz^2)/(2 m_B)$ 
and $ | \vec{q}_0^{\;\FRone}| = (m_B^2 -\qz^2)/(2 m_B)$, one finds 
\begin{equation}
(\qz^2)_{\max}   =\left\{  \begin{array}{lll}  q^2 +\des \qz^2    &    \quad   \cos \Tg^{\FRone} = -1 &  \quad \textrm{tight-angle cut}    \\
q^2 +\des m_B^2 &    \quad   \cos \Tg^{\FRone} =+ 1 &  \quad \textrm{max-angle}
\end{array} \right. \;.
\label{eq:A9}
\end{equation}
This means that for fixed $q^2$, and a cut of $\des = 0.15$,
 the radiative process probes values of  
 $(\qz^2)_{\max} = q^2/(1-\des) \simeq 1.18\, q^2$ (tight-angle cut)  
 and $(\qz^2)_{\max} \simeq  q^2 + 4.18\GeV^2$ (max-angle) respectively.
 Note that the maximum angle cut in the photon-emission gives the same result as
 the maximum lepton angle. 
 This is because in the collinear limit ($ \vec{\ell}_1 \propto  \vec{k}$), the maximum lepton angle aligns $\vec{\ell}_1$ and $\vec{k}$
  with the decay axis (x-axis, see \FIG\ref{fig:angles}), and this coincides with the maximum angle cut.
  
 \subsection{Remarks on the Lepton Flavour Universality ratio $R_K$}

LFU ratios, such as $R_K$, are good observables to search for  specific types of physics beyond the 
SM, namely new interactions that are not universal among the different lepton species.  
Owing to the cancellation of many hadronic uncertainties, these ratios can be predicted well 
up to LFU violating interactions. In the SM, LFU is broken by the fermion masses and, as such,
sizeable effect could result from the logarithms of QED
\begin{equation}
\label{eq:RK}
R_K|_{q_0^2 \in [q_1^2,q_2^2]{\small \GeV}^2} = 
\frac{\Gamma[\Bin \to \Kout \mu^+ \mu^-]}{\Gamma[\Bin \to \Kout e^+ e^-]}\big|_{q_0^2 \in [q_1^2,q_2^2]{\small \GeV}^2} \approx 1 + \Delta_{\rm QED} R_K  \;,
\end{equation}
 as first quantified in~\cite{BIP16}.
Whereas for a meaningful comparison to experiment  a purpose build 
Monte Carlo with complete differential treatment is desirable (cf. \SEC\ref{sec:MC}),
 one may already raise the point  that the precise treatment has a relevant 
impact.  

For example,  considering only the cuts on reconstructed $B$-meson mass in \cite{Aaij:2014ora},
the net QED correction that should be applied to $R_K$ according to our analysis amounts to 
\begin{equation}
	\Delta_{\rm QED} R_K \approx   \left. \frac{ \Delta \Gamma_{K\mu\mu}}{ \Gamma_{K\mu\mu} } 
	\right|^{m_B^{\textrm{rec}}= 5.175{\small \GeV}}_{q_0^2 \in [1,6]{\small \GeV}^2}
	-  \left. \frac{ \Delta \Gamma_{K ee }}{ \Gamma_{Kee} } 
	\right|^{m_B^{\textrm{rec}}= 4.88{\small \GeV}}_{q_0^2 \in [1,6]{\small \GeV}^2}
	\approx +1.7\%~ \;,
\end{equation}
whereas the correction has to be compared with the $\Delta_{\rm QED} R_K \approx +3\%$ quoted in~\cite{BIP16} that, 
as explained in~\APP\ref{app:BIP}, takes into account an additional implicit tight cut on the photon-emission angle.  
Note that the different photon energy cuts for muons ($m_B^{\textrm{rec}}= 5.175\GeV \leftrightarrow 
\des= 0.0394 $) and electrons ($m_B^{\textrm{rec}}= 4.88\GeV \leftrightarrow 
\des= 0.1458 $) reduce the effect of QED corrections to $R_K$.  In addition, 
$|\Delta_{\rm QED} R_K^{\textrm{BIP}}| >  |\Delta_{\rm QED} R_K^{\textrm{INZ}}|$  has  to be expected
since the BIP computation~\cite{BIP16} is more exclusive, in view of the tight photon-angle cut, than the explicit computation 
presented here.
However, in both cases  the overall impact of QED corrections in the LFU ratios (currently estimated 
by the experiment using  PHOTOS)  is not exceedingly large
and below the current experimental error 
$R_K = 0.846^{+ 0.060 + 0.016}_{- 0.054 - 0.014}$~\cite{Aaij:2019wad}.

\section{Outlook}
\label{sec:outlook}

 In this section, we briefly address various topics which go beyond the scope of this paper and are worthwhile
 to be pursued in future investigations. 
 
\subsection{Structure-dependent terms}
\label{sec:SD}

In this work, we have treated the mesons as fundamental fields.  The effective Lagrangian 
employed is able to perfectly describe their internal structure up to ${\cal O}(e^0)$. However, 
the electromagnetic probe sees the mesons as a structureless particle.  
Hence our effective Lagrangian corresponds to approximating a multipole expansion by 
the monopole term.

In the language of meson fields, one would need to build a systematic effective field theory 
with  gauge invariant operators out of  covariant derivatives and meson fields. 
This would include, amongst others, terms beyond minimal coupling of the form
 $(D^\mu B)^\dagger  F_{\mu\nu} \, D^\nu K$.
  It is beyond doubt that in full QCD, the meson's partons  give rise to such higher multipole emissions,
which we referred to as structure-dependent terms.\footnote{~The full theory, including QCD and QED, 
is needed to compute the corresponding Wilson coefficients.
and counterterms when involving loops.}
The question is whether they are sizeable.
For light-light systems, such as $K \to \pi$ decays, these terms are known to be small 
e.g. \cite{DAmbrosio:1994bks,DAmbrosio:1996jmq}
(unless the leading amplitude is accidentally suppressed).
For heavy-light systems,  this might change since the masses of the valence quarks introduce 
a sizeable asymmetry that will eventually be resolved.

A result established in this paper provides some protection.  
It was shown in \SEC \ref{sec:beyondPT} that structure-dependent corrections 
do not lead to any additional hard-collinear logs. 
Since  soft divergences cancel at the  differential level, this means that the 
employed approximation   captures all  IR sensitive terms.
However, it cannot be precluded that new and interesting hadronic effects, not 
directly related to infrared effects,  could come into play.  
An example of which is  provided by $B_s \to \mu^+ \mu^-$, where it was found that the 
chirality suppression of the non radiative decay $m_\mu/m_b$ is lifted to $m_\mu/\Lambda_{\textrm{QCD}}$
(``enhanced power corrections") 
when QED corrections are taken into account  \cite{BBS17}.
These authors  develop  QED corrections to $B$ decays within the  soft collinear effective theory (SCET) framework, recently extended to $B \to K \pi$ \cite{Beneke:2020vnb}.
It allows for the   resummation  of different types of logarithms \cite{Beneke:2019slt}
but  does not capture all $1/m_b$ effects.  
To what extent $1/m_b$-corrections are important in QED corrections to  $B$-mesons decays is an interesting and open question.
 Another approach is lattice QCD, where the precision in Kaon physics per se demands the inclusion 
of QED corrections \cite{Carrasco:2015xwa,Sachrajda:2019uhh} with first results in leptonic decays \cite{Giusti:2017dwk,DiCarlo:2019thl,Portelli:2019}.
For $B$ decays, in the region of fast recoiling particles, 
more work is needed  because of too many exponentially growing modes that have to be captured by a fit or dealt with in some other way.
     
\subsection{Moments of the differential distribution}
\label{sec:moments}

A special feature of QED corrections is that they have genuine infrared effects when compared 
to the weak interaction with natural scale $m_W \gg m_B$.  
As pointed out in \cite{Gratrex:2015hna}, this changes the angular distribution in that there is 
not a specific hierarchy of moments in the angles  (cf. section 5  in this reference). 
Without QED corrections, it is the dimension of the effective Hamiltonian that limits
the partial waves to its lowest numbers. Higher moments (in the partial wave expansion)
 are therefore absent or suppressed by further powers 
of $m_b/m_W$. Hence, measuring higher moments allows to measure  QED effects. 
It is therefore interesting to scrutinise the size of these corrections from the theory side in order 
to identify the most sensitive moments and give further motivation to an experimental investigation. 
We will turn to this task in a forthcoming publication.

\subsection{The $\Bin \to \Kout \ell^+ \ell^-$ differential distribution through Monte Carlo}
\label{sec:MC}       

Our results  can be used to estimate  the radiative corrections to the $\Bin \to \Kout \ell^+ \ell^- (\gamma)$ 
differential distribution  semi-analytically.\footnote{~The integration over the photon variables is done numerically and this is why we refer to  them as semi-analytic.}
As demonstrated, the choice of  differential variables (which might
be dictated by their accessibility in a given experiment) that we have introduced \eqref{eq:tr} 
directly impacts in what way hard-collinear logs cancel.   An alternative approach,  more in line with  current
analysis techniques
, is to build a Monte Carlo program 
for the numerical simulation of the radiative and the non-radiative processes, and  evaluate 
the impact of the radiative corrections entirely numerically. 
This happens at an  
even more differential level by taking into account   the photon kinematics  on an event-by-event basis. 
Given the sizeable contributions from hard-collinear logs, it will be an important task to crosscheck the 
purpose-built Monte Carlo against standard tools used in experimental analysis.
In this case, our virtual corrections are  essential in that they   provide   the  normalisation of the Monte Carlo code.\footnote{~The Monte Carlo code requires the introduction 
of an (unphysical) soft cut-off $\MCsoft$, 
below which the mode is treated as a three-body  decay.
The rate ~\eqref{eq:Delta} is then split into, 
 $d^2 \Gamma(\des ) = [d^2 \Gamma(\MCsoft)]_{\textrm{MC}_3} + 
 \big[  \frac{\al}{\pi} \sum_{i, j }   \hat{Q}_i \hat{Q}_j  ({\cal F}^{(\AAA)}_{ij}(\des )  - {\cal F}^{(\AAA)}_{ij}(\MCsoft ) ) \big]_{\textrm{MC}_4} \, d  \qSaa d \claa $, a first term which is done semi-analytically with our computation and simulated with three-body kinematics,
and a second term which is obtained through the simulation of the full four-body kinematics.
Note that both terms are free from soft divergences and $\MCsoft$ is analogous to 
phase space slicing cut-off $\Des$ introduced  in Sect.~\ref{sec:IR}.}
Within this approach, we are free to adopt the $\{ \qz^2, \clz\}$ or the 
$\{q^2, \cl\}$-variables, 
since these are used to describe the simulated events and
the experiment can  produce a distribution in either of the variables 
by using local corrections factors.  
The final  comparison with experiment is performed 
in a subsequent step after taking 
into account experimental efficiencies, resolutions, and cuts to reduce the background.
Given our results in  Sect.~\ref{sec:IR}, it is clear that the choice of $\{ \qz^2, \clz\}$ 
is more convenient, since for each value of $\qz^2$ and $\clz$ the corresponding 
photon-inclusive rate is free from hard-collinear singularities. 
 A detailed Monte Carlo code for 
$\Bin \to \Kout^{(*)} \ell^+ \ell^-(\gamma)$, taking into account  the finite $\ORD(\alpha)$ terms
evaluated in this work, will be presented in a forthcoming publication.

It is worth stressing that most of the considerations presented in this work, 
and particularly the strategy outlined above to build a Monte Carlo code, 
apply if the final-state hadron is a narrow vector resonance 
(such as the $\Kout^{*}$), rather than a stable scalar meson.
In the narrow-width approximation, 
we can neglect the interference of the radiation emitted by the final-state mesons,
produced by the vector-meson decay, with the radiation from the $B$ decay products
(i.e.~the radiation described in this work). In this limit 
(which is a rather good approximation in the  $\Kout^{*}$ case, given that 
$\Gamma_{\Kout^{*}}/m_{\Kout^{*}} \approx 5\%$)
the formalism is essentially identical, up to a richer form factor structure.

\subsection{Remarks on charged-current semileptonic decays}
\label{sec:semileptonic}

In the main section, charges and masses were kept completely general, so that any semileptonic 
decay can be covered, including charged-current processes such as $\bar B \to D \ell \nu$. 
A significant difference to $\Bin \to \Kout \ell^+ \ell^-$ is that the variable 
$\pBbar^2$, defined as in \eqref{eq:momC}, is not observable (because of the unidentified neutrino).  Whereas this does not pose a problem for the Monte Carlo simulation discussed above, this is an issue for the semi-analytical determination  of 
 an $\ORD(\al)$ infrared-safe distribution of $\bar B \to D \ell \nu$.  

One possibility to overcome this problem is to consider $\hat{p}_B^2 \equiv (p_B-p_\nu)^2$ as the effective photon energy variable.  
A photon energy cut-off, similar to \eqref{eq:des}, can be introduced as follows $\de^{\textrm{ex}}_{\textrm{SL}} = 
(1 - p_{D\ell}^2/ \hat{p}_B^2 )$ which translates to $E_\ga^* < \, \de^{\textrm{ex}}_{\textrm{SL}}( p_{D\ell}/2)$  ($E_\ga^*$ is the photon energy in the $D$-lepton RF).  
The new aspect with regards to the FCNC case is that  the lower cut-off on the energy variable, 
$(\hat{p}_B^2)_{\textrm{min}} = p_{D\ell}^2 /( 1- \de^{\textrm{ex}}_{\textrm{SL}})$, 
is dependent on a differential  variable.\footnote{~Alternatively, one could 
trade $\pBbar$ with 
$\bar p_{\rm vis} \equiv p_B - k - p_{\nu}$.  
The upper cut-off on $\pBbar^2$  is then to be replaced by 
a lower cut-off on $\bar p_{\rm vis}^2$ and the adaption of our formalism requires 
to work  with a finite neutrino mass. It is understood that this approach might be challenging on the numerical side.}

Another strategy is 
to impose the minimal kinematic limits on  $\bar p_{\rm vis}^2 \equiv (p_B - k - p_{\nu})^2$ and 
accept all events with $E_D$ and $E_{\nu}$ which lie within the non-radiative Dalitz-plot. 
This is the  ``traditional''  approach adopted in Refs.~\cite{Ginsberg:1969jh,Cetal01,CGH08}. This can work in a clean environment, in $K$-factories, but would 
not be a feasible approach for the LHC collider environment.
Incidentally, we note that the variables ($E_D$, $E_{\nu}$) are an alternative choice to our $\{q^2,\cl\}$-variables.  
We finally stress that the approach followed in~\cite{deBoer:2018ipi}, where an effective cut on the photon
energy is implemented irrespective of the photon-emission angle, might lead to a miss-estimate of the 
hard collinear logs (see the discussion in Appendix~\ref{app:BIP}).

\section{Conclusions}
\label{sec:conclusion}

In this paper we have analysed the $\ORD(\al)$ corrections to a generic $M_H \to M_L \lone \bltwo$ decay, 
where $M_{H,L}$ are  scalar mesons (of either parity).  We have performed a complete calculation of these 
corrections within improved scalar QED, employing a mesonic effective Lagrangian 
(with a tower of effective operators) which provides an accurate description of the non-radiative hadronic form factors.
We have shown by means of explicit computation that all soft divergences cancel at the double differential level (\SEC\ref{sec:soft}), irrespective of the choice of the variables used to describe the ``visible" kinematics. 
On the other hand, we have demonstrated that the hard-collinear logs can survive, 
even in the photon-inclusive limit, depending on the 
variables employed to describe the photon-inclusive distribution. 
More precisely, they cancel in the case of the $\{ \qz^2, \clz\}$- but not the 
$\{q^2, \cl\}$-variables defined in Eq.~\eqref{eq:tr}.

Our analysis goes well beyond, in terms of accuracy and generality,  w.r.t. 
previous analytical treatments of radiative effects in $M_H \to M_L \lone \bltwo$ decays. 
Still, some open issues remain, as discussed in \SEC\ref{sec:SD}. 
In particular the matching of the residual UV ambiguities with QCD (which would allow
the inclusion of  QED effects to the Wilson coefficients  \cite{Bobeth2004,HLMW05}) 
and resolving the 
photon interaction with the quarks themselves. 
As we have shown, gauge invariance ensures that such ambiguities  
cannot induce $\ln m_\ell$-enhanced corrections (\SEC\ref{sec:beyondPT}). 
This implies, in particular, that these corrections
have a negligible impact on the experimental determination of the LFU ratios.

Our analysis indicates that great care must be taken when comparing theoretical with experimental 
data, given that radiative corrections for the electron modes can easily exceed the $10\%$-level (as already
indicated by previous analyses). As discussed in  \APP\ref{app:BIP}, the overall impact of QED corrections on integrated LFU ratios, such as $R_K$, is not too large, especially given the current cuts applied on the reconstructed invariant mass for electron and muon modes~\cite{Aaij:2019wad}. On the other hand,  differential observables are subject 
to  potentially larger effects.\footnote{~Even for total decay rates (non-LFU type) 
there can be relevant effects such as the $\ln m_K$-logs discussed in \SEC\ref{sec:total}.}
In particular, as we have shown in \SEC\ref{sec:distortion}, a sizeable lepton-non-universal 
distortion of the dilepton invariant mass spectrum occurs if the latter is expressed in term of the $\{q^2, \cl\}$-variables.
To overcome this problem the best way to report data is in terms of the of the $\{ \qz^2, \clz\}$ distribution
(as currently done by most experiments), where the ``dangerous" hard-collinear logs ($\ln m_{\ell}$) 
cancel at the differential level. In the case of the LHCb experiment, where $\qz^2$ is not directly measurable,
this is done after comparing the results with a Monte Carlo code and correcting for the effect of the QED radiation.
In this context,  we note that our analysis provides the theoretical groundwork to build a Monte Carlo program with 
a complete differential treatment of radiative corrections and an accurate parameterisation of the hadronic form-factors
(possibly including also long-distance contributions),
which represents a key ingredient for a  precise comparison between data and theoretical predictions 
in the future.

\acknowledgments
We would like to thank Andrea Pattori for collaboration at the very early stages of this work.
Useful discussion with Melissa van Beekveld, Stefan Dittmaier, Giulio Falcioni,  Antonin Portelli, 
Marek Sch\"onherr, Jennifer Smillie  and Gabor Somogyi
are acknowledged. 
RZ and SN are grateful to the UZH and the Pauli Center for the hospitality in Z\"urich
during various stays while working on this project.
RZ is supported by an STFC Consolidated Grant, ST/P0000630/1. SN is supported by a Higgs Scholarship and an Edinburgh Global Research Scholarship.
GI has received funding from the European Research Council (ERC) under the European Union's Horizon 2020 research and innovation programme under grant agreement 833280 (FLAY), and by the Swiss National Science Foundation (SNF) under contract 200021-175940.

\appendix

\section{Additional  Plots and further Numerical Results}
\label{app:fplots}

\subsection{The size of hard-collinear logarithms as a function of $\des$ and $q^2$}
\label{app:sizehc}

\begin{figure}[t]
\includegraphics[width=0.5\linewidth]{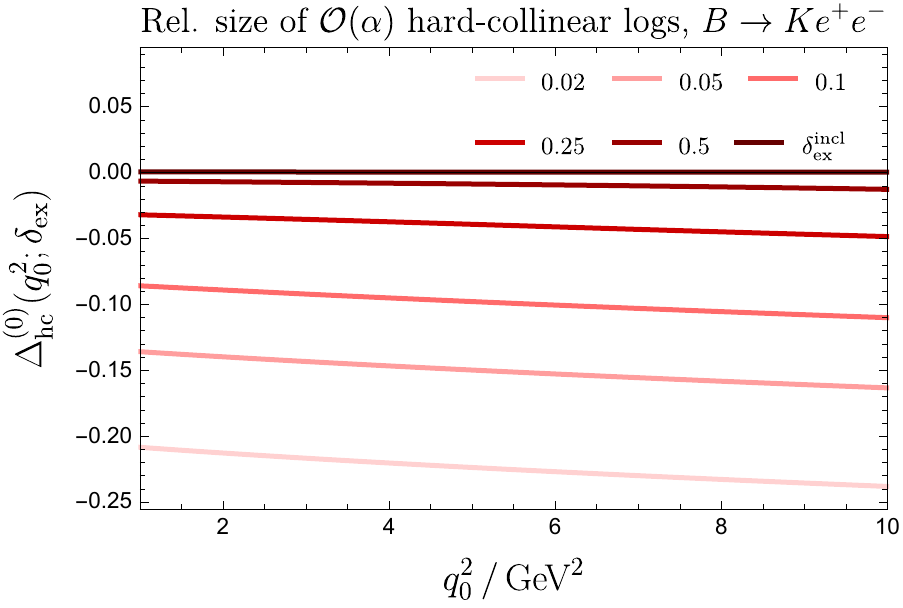}  
\includegraphics[width=0.5\linewidth]{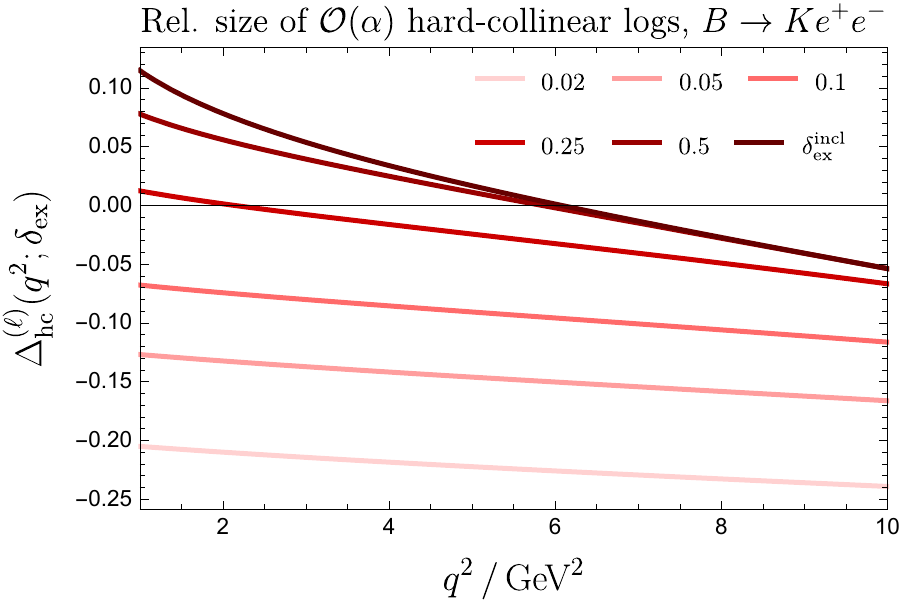}  
		        \hfill \\[-0.2cm]
		        \includegraphics[width=0.5\linewidth]{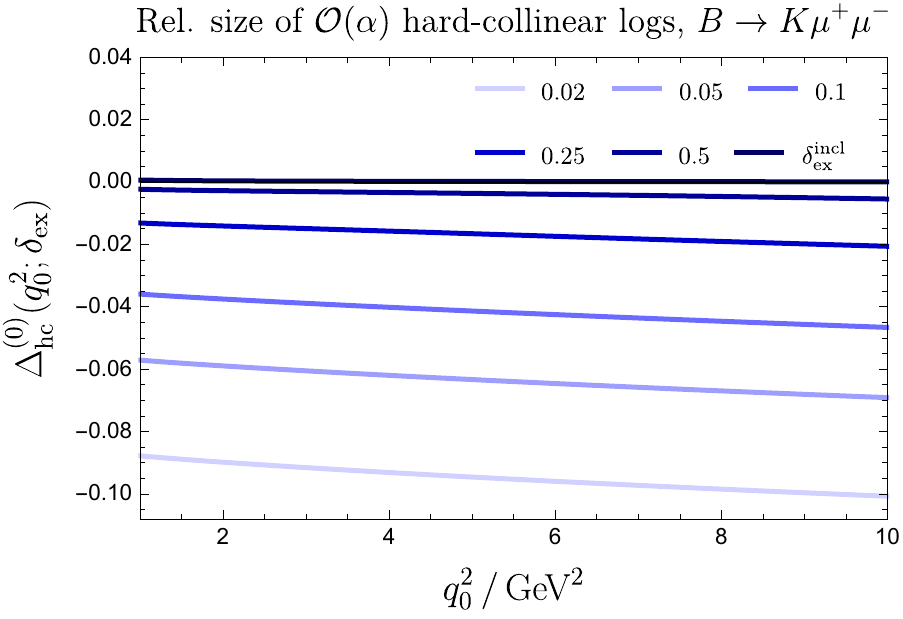}  
          \includegraphics[width=0.5\linewidth]{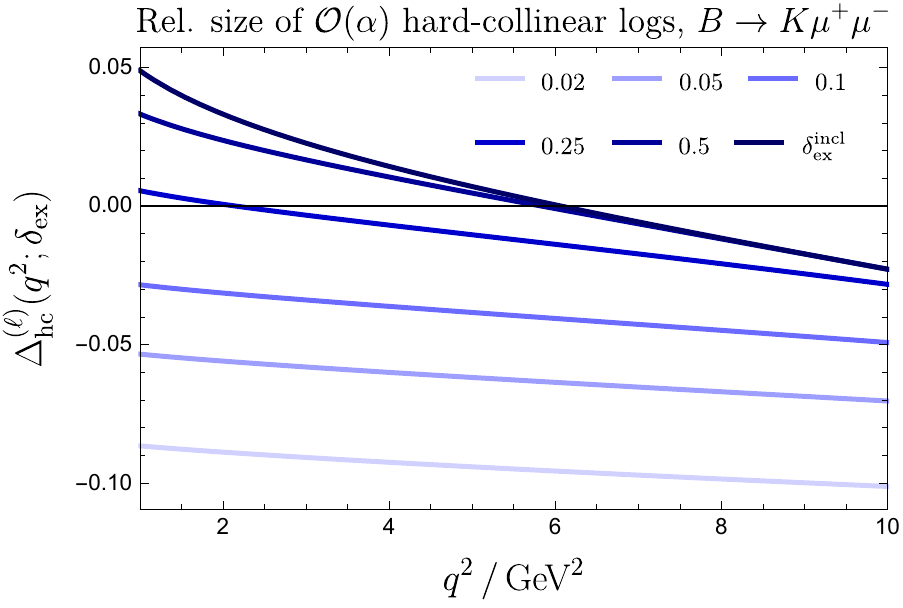}  
        	\caption{\small 
	Hard-collinear logs $\Delta^{(\AAA)}_{\textrm{hc}}(\qSaa;\des)$  as a function of $\qSaa$ 
	for the electron and muon (top) and (bottom) respectively.  
	The quantity is shown for various photon energy cut-offs $\des$ \eqref{eq:des}. It is noted that for 
	$\des = \desinc$, the cancellation of the logs can be seen, though not completely, as we only show
	a restricted interval of $q^2$.
	Bottom and top figures are similar by a scaling factor cf.~\eqref{eq:Rhc} and the explanation above it.}
	\label{fig:sizehc}
\end{figure}

It is of interest to investigate the size of the collinear logs.  
We do this by normalising against the non-radiative differential rate, as done previously in \SEC\ref{sec:total}, 
\begin{equation}
\label{eq:d2Delhc}
 \Delta^{(\AAA)}_{\textrm{hc}}(\qSaa,  \claa;\des) =
 \Delta^{(\AAA)}(\qSaa,  \claa;\des)\Big{|}_{\ln \hat{m}_{\lonetwo}}  =   \left( \frac{d^2 \Gamma^{\LO}}{d\qSaa d \claa}\right)^{-1} \frac{d^2\Gamma(\des)}{d\qSaa d \claa} \Big{|}_{\ln \hat{m}_{\lonetwo}}    \;,
\end{equation}
where  the terms on the RHS
can be found in Eqs.~\eqref{eq:ratell} and \eqref{eq:new} respectively.  
Charged and neutral meson modes are not distinguished as they 
contain the same collinear divergences as the latter are strictly 
proportional to the lepton charges, i.e.~independent of the hadron charges.  Thus, there is only one  
basic mode  of interest for the hard-collinear logs per lepton pair final state.
The integrated quantities $\Delta_{\textrm{hc}}^{(\AAA)}(q^2;\des)$ and 
$ \Delta_{\textrm{hc}}^{(\AAA)}(\cl,[q_1^2,q_2^2];\des)$ are defined in complete analogy to \EQs \eqref{eq:dq2Del} 
and \eqref{eq:dclDel} respectively.

Plots of $ \Delta^{(\AAA)}_{\textrm{hc}}(\qSaa;\des)$ are shown in \FIG\ref{fig:sizehc} for different 
photon energy cuts $\des$ \eqref{eq:des} for electrons and muons with larger effects for the former 
because of the size of $\ln \hat{m}_e$  versus $\ln \hat{m}_{\mu}$ logs. 
In the photon-inclusive case, the cancellation of the hard-collinear logs is visible at the differential level
in the $\qz^2$-variable.
For the $q^2$-variable, the hard-collinear logs cancel when integrated over the entire 
$q^2$-interval, the tendency of which can be inferred from the plots on the reduced 
interval  $q^2_{\textrm{max}} < 10  \GeV^2$.  
 The reader is reminded that hatted quantities are normalised 
w.r.t. to the $B$-mass, $\hat{m}_{\lonetwo}  = m_{\lonetwo}/m_B$.  
Hence one expects 
\begin{equation}
\label{eq:Rhc}
R_{\textrm{hc}} = \frac{  \Delta^{(\AAA)}_{\textrm{hc}}(\qSaa;\des)|_{\BKee}  }{ \Delta^{(\AAA)}_{\textrm{hc}}(\qSaa;\des)|_{\BKmumu}} \simeq \frac{\ln ( \hat{m}_e ) }{ \ln ( \hat{m}_\mu )}  \simeq 2.363 \;,
\end{equation}
with corrections of the order of $\ORD( m_e^2 \ln ( \hat{m}_e ) 
  -m_\mu^2 \ln ( \hat{m}_\mu )   )$. Inspection of the plots shows that this is indeed the case.
  We would like to stress that extracting the hard-collinear logs on their own is slightly ambiguous 
  as one needs to normalise them (hatted notation). The unambiguous way to show them is through the full
  plots in the main text.
  Nevertheless, they illustrate nicely the effect of the photon energy cut.


\subsubsection{Comparison of $\Bin \to \Kout \ell^+ \ell^-$ to the inclusive case $b \to s \ell^+\ell^-$}
\label{app:compare}

\begin{figure}[t]
\begin{center}
\includegraphics[width=0.5\linewidth]{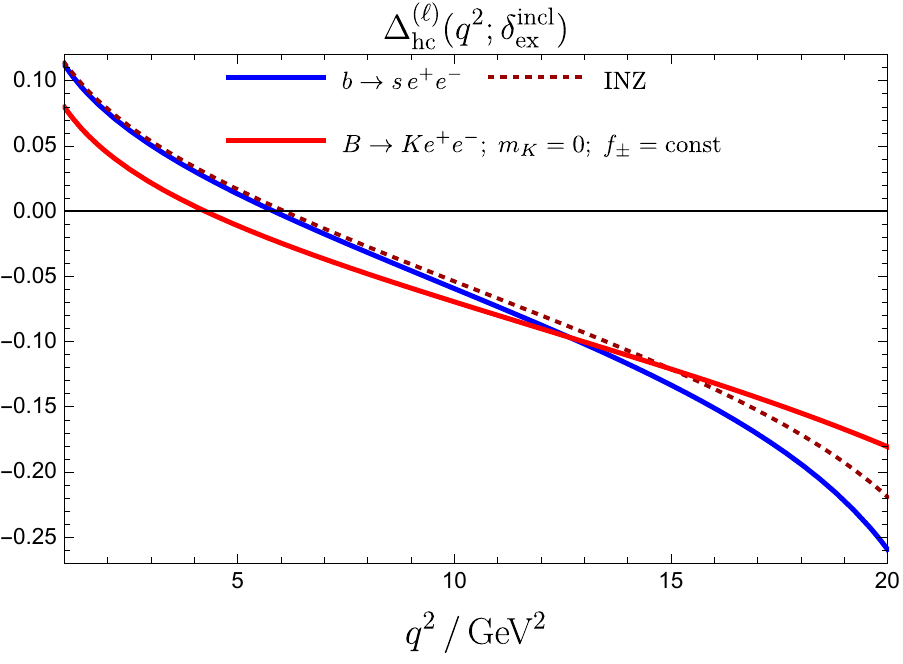}  
\end{center}
        	\caption{\small 
	Comparison of hard-collinear logs in $b \to s e^+ e^-$ (solid blue line) versus 
	$\Bin \to \Kout e^+ e^-(\ga)$ with no photon energy cut, constant form factors and $m_K\to 0$ (solid red line)  corresponding to \EQs\eqref{eq:Delinc}.  This illustrates the spin-dependance of the hard-collinear 
	which can be traced back to the LO differential rates in the case at hand cf. \eqref{eq:LOrates}.
	For further comparison, we have added the full result in the dotted red line 
	for this paper with no photon energy cut either. The agreement at low $q^2$ 
	of the latter with the $b \to s e^+ e^-$  is somewhat accidental.}
	\label{fig:inc_comp}
\end{figure}

It is interesting to compare our results to the inclusive rate in \cite{HLMW05} with regard to the hard-collinear logs.  
Let us define
\begin{equation}
\Delta^{(\TT)}_{\textrm{hc}}(\hat{q}^2)  = 
 \frac{2 \al \,\hat{Q}_{\lone}^2}{\pi \,}
\left( \frac{1}{ \Gamma^{\LO }} \frac{d \Gamma^{\LO}(\hat{q}^2) }{d\hat{q}^2}  \right)^{-1} 
\tilde{\Delta}^{(\TT)}_{\textrm{hc}}(\hat{q}^2) \,.
\end{equation}
where $\hat{Q}_{\lone}^2 = \hat{Q}_{\bltwo}^2$ and $\mlone=\mltwo\equiv m_{\ell}$
have been assumed. 
Then, it is known that the collinear logs of the electron can be extracted from (e.g. chapter 17 \cite{Peskin:1995ev})\footnote{From section 5 in  \cite{HLMW05},  one can extract a similar formula for the collinear logs
\begin{equation}
\label{eq:inc}
\tilde{\Delta}^{(\TT)}_{\textrm{hc}}(\hat{q}^2)  =  \frac{1}{  \Gamma^{\LO}} \left( 
\int_{\hat{q}^2}^{1} \frac{d z}{z}  \tilde{P}_{f \to f \ga}(z)   \frac{d \Gamma^{\LO}(\hat{q}^2/z) }{d\hat{q}^2/z} - 
\int_0^1 d z  \tilde{P}_{f \to f \ga}(z)  \frac{d \Gamma^{\LO}(\hat{q}^2) }{d\hat{q}^2}     \right)    \ln \frac{\Lambda_b}{m_\ell} \;,
\end{equation}
where $\tilde{P}_{f \to f \ga}(z)$  \eqref{eq:splitting} is the collinear emission part of the splitting function. 
Soft divergences at $z \to 1$ cancel between the two integrals. Translating into our notation from  \cite{HLMW05} demands
 $x = 1-z$, $\hat{s}  = \hat{q}^2$ and $\tilde{P}_{f \to f \ga}(z)$  is  the part collinear in $f_\ga^{(m)}$ up to 
 factors of proportionality properly accounted for. Our formula \eqref{eq:magic} can be recovered upon using that 
$ \int d z   P_{f \to f \ga}(z) =0$.}   
\begin{equation}
\label{eq:magic}
\tilde{\Delta}^{(\TT)}_{\textrm{hc}}(\hat{q}^2)  =  \frac{1}{  \Gamma^{\LO}} \left(
\int_{\hat{q}^2}^{1} \frac{d z}{z}  P_{f \to f \ga}(z)   \frac{d \Gamma^{\LO}(\hat{q}^2/z) }{d\hat{q}^2/z} \right)    \ln \frac{\Lambda_b}{m_\ell} \;,
\end{equation}
where $\Lambda_b = \ORD(m_b)$ is some reference  scale, $P_{f \to f \ga}(z) $ is the full leading order splitting function
\begin{equation}
\label{eq:splitting2}
P_{f \to f \ga}(z)  =  \frac{1 + z^2}{(1- z)_+}  + \frac{3}{2} \de(1-z) \;,
\end{equation}
and $1/(1-z)_+$ is the plus distribution $\int_0^1 dz  f(z) /(1-z)_+= \int_0^1 dz ( f(z)- f(1))/(1-z) $. 
Note that by construction, 
 the hard-collinear logs cancel in the total rate. 
This can be seen by  reversing the order of integration and adopting  the change of variable $\hat{q}^2/z = \hat{q}_0^2$
  to arrive at 
 $\int_0^1 d \hat q^2 \tilde{\Delta}^{(\TT)}_{\textrm{hc}}(\hat{q}^2) \propto \int_0^1  dz P_{f \to f \ga}(z) = 0$. 
  Now, the zeroth moment of the splitting function vanishes since it
 corresponds to the anomalous dimension of the (conserved) electromagnetic current. 
 Conversely, \eqref{eq:magic} can be deduced from  \EQ\eqref{eq:A1sqhardcoll} by integrating 
 over $d\clz$, substituting $\qz^2 = q^2/z$ and then integrating over $z$.
 From \eqref{eq:splitting}, the full splitting function is then easily deduced by adding 
 a delta function ansatz $A \de(1-z)$ and regularising the $1/(1-z)$ such that the soft divergences cancel 
 (which leads to the plus distribution).

 The leading order differential rates are given by
 \begin{equation}
 \label{eq:LOrates}
 \frac{1}{  \Gamma^{\LO}}\frac{d \Gamma^{\LO}(\hat{q}^2) }{d\hat{q}^2}  =  \left\{  \begin{array}{ll}  
 2 (1-\hat{q}^2)^2 (2 \hat{q}^2+1)    &  \quad b \to s  \ell^+ \ell^- \\
 4 (1-\hat{q}^2)^3      &  \quad \Bin \to \Kout \ell^+ \ell^-
  \end{array} \right. \;,
 \end{equation}
 where  the $m_s \to 0$ limit is implied in \cite{HLMW05} and for  
 simplicity we have assumed the $m_K \to 0$ limit and a constant form factor.
 Note that the factor $\hat{\la}^{1/2}_B  = \la^{1/2}(1,\hat{m}^2_K,\hat{q}^2)|_{m_K \to 0} = 1- \hat{q}^2$ is the square root of the K\"all\'en-function and as such related 
 to the three velocity of the strange particle in the $B$-meson's RF.  Its power in the rate is determined by the interaction
 and the spin of the particle (e.g. if it were $\Bin \to \Kout^* \ell^+ \ell^-$ then $d \Gamma^{\LO} \propto 
 (1-\hat{q}^2)$ \cite{Hiller:2013cza}). The factor  $2 \hat{q}^2+1$ originates from the  $s$-quark's 
 spin summation.
One finds
\begin{alignat}{2}
\label{eq:Delinc}
& \tilde{\Delta}^{b \to s \ell^+ \ell^-}_{\textrm{hc}}(s) &\,=\,&  
2 \left( (6 s^2 \mi4
   s^3 \mi1 ) \ln 
   s  \pl2 (1 \mi s)^2 (2 s\pl1) \ln
   (1\mi s)  \pl  \frac{12 s^2   \mi 8 s^3 \mi 3 s \mi 1}{3}  \right)   
  \;, \nonumber\\[5pt]
& \tilde{\Delta}^{\Bin \to \Kout \ell^+ \ell^-}_{\textrm{hc}}(s)   &\,=\,&
4 \left( \left(2 s^3\mi6
   s^2\pl3 s\mi1\right) \ln s \pl2
   (1\mi s)^3 \ln (1\mi s) \pl
  \frac{4 s^3\mi6 s^2\pl6
   s\mi6 }{3} \right)   \;,
\end{alignat}
where $s=\hat{q}^2$.
The basic form is similar in both cases and we observe the $\ln  q^2 $-term leading  to enhanced 
 collinear emission at low $q^2$ which can be interpreted as a migration of the photon radiation 
 cf. \SEC\ref{sec:distortion}. 
We wish to stress again that $ \tilde{\Delta}^{\Bin \to \Kout \ell^+ \ell^-}_{\textrm{hc}}$ receives 
corrections due to finite $m_K$ and non-constant form factor and that $\des= \desinc$ was assumed. Both of these features are included 
in the comparison plot \FIG\ref{fig:inc_comp}. We have checked that integrating \eqref{eq:new} over $\int_{-1}^1 d \cl$ reproduces the $\tilde{\Delta}^{\Bin \to \Kout \ell^+ \ell^-}_{\textrm{hc}}$-expression 
  in \eqref{eq:Delinc}.  This comparison 
provides another non-trivial 
cross-check of our analysis.

\subsection{Comparison with earlier work on $\Bin \to \Kout \ell^+ \ell^-$}
\label{app:BIP}

We compare our results to those presented in~\cite{BIP16}.  The analysis of~\cite{BIP16},
which first investigated the impact of LFU breaking  in $\Bin \to \Kout \ell^+ \ell^-$ induced by 
QED corrections, is a simplified analysis based on the following three principles/assumptions:
\begin{itemize} 
\item[i.]  indirect determination of  virtual corrections by imposing the absence of log-enhanced terms 
in the photon-inclusive $d\Gamma/dq_0^2$ spectrum (for any value of $q^2_0$);
\item[ii.]  constructing  a radiator function depending  on $q^2$ and $q_0^2$ only, which describs the 
probability of a dilepton pair (of invariant mass $q^2$) to originate from  
momentum transfer $q_0^2$, after photon-emission;
\item[iii.]  neglecting lepton-flavour universal radiative corrections, such as those induced by the emissions 
from meson legs only.
\end{itemize}
As proved in general terms in this paper, assumption i.~is correct and provides an efficient way to determine the radiator function.  
Our  analysis shows that the non-log enhanced terms 
are  small in the neutral-meson case (as shown in \FIG\ref{fig:sizeDel}). They do exceed the $1\%$ level in the charged-meson case, but this is a lepton-flavour universal effect.

\begin{figure}[t]
\includegraphics[width=0.5\linewidth]{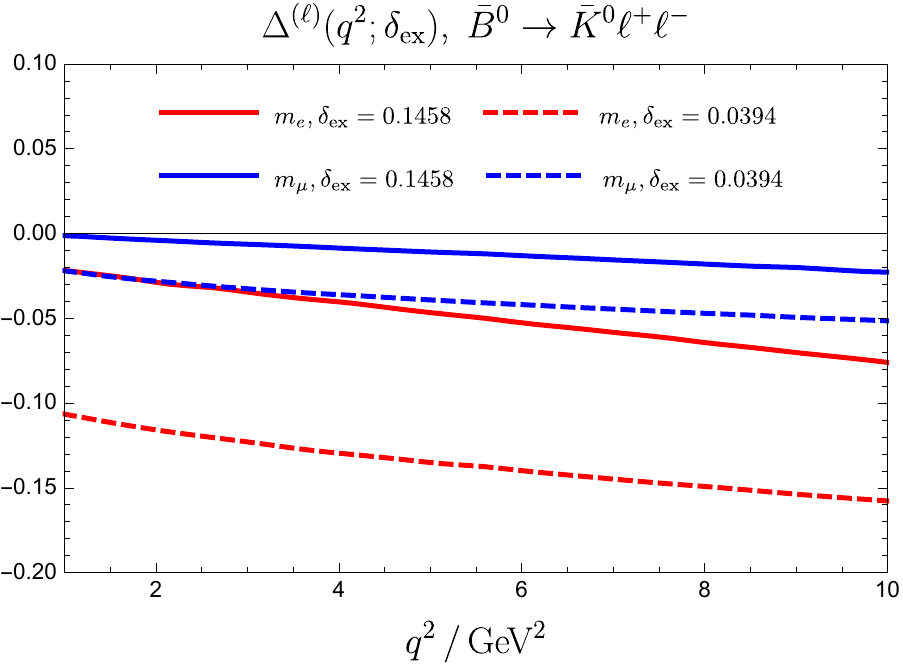}  
\includegraphics[width=0.5\linewidth]{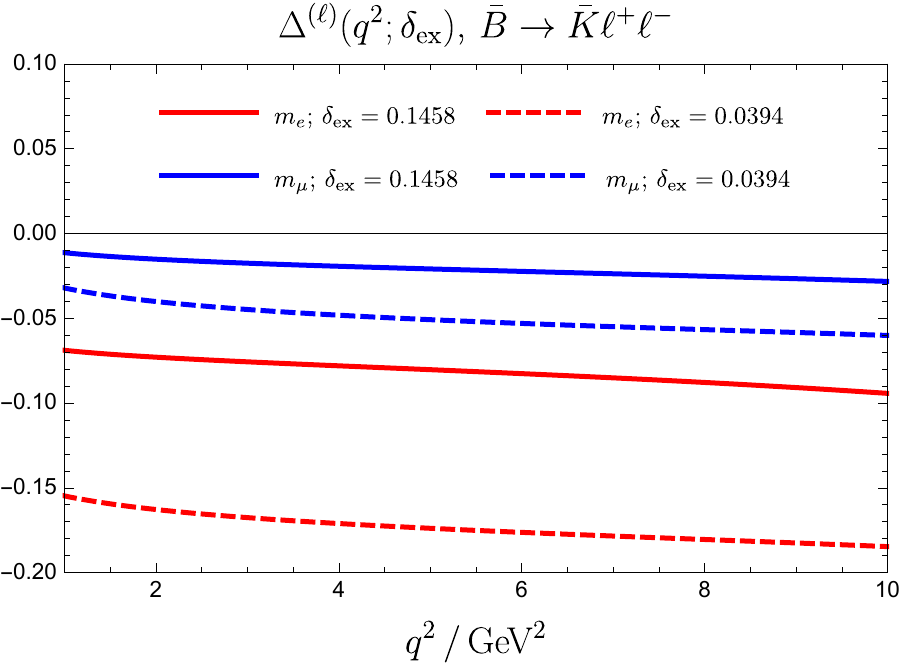} 
 \hfill \\[-0.2cm]
\includegraphics[width=0.5\linewidth]{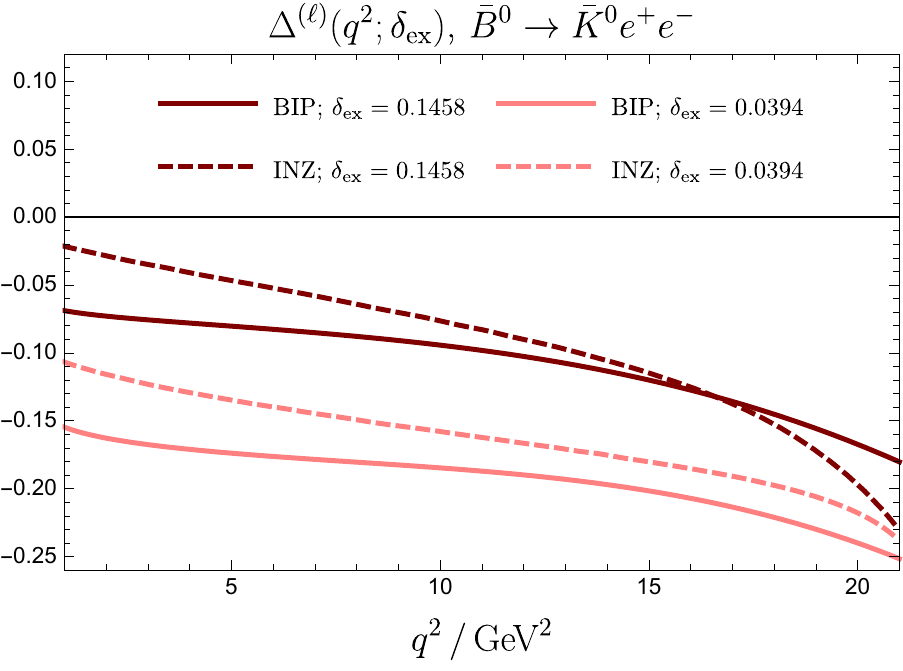}  
\includegraphics[width=0.5\linewidth]{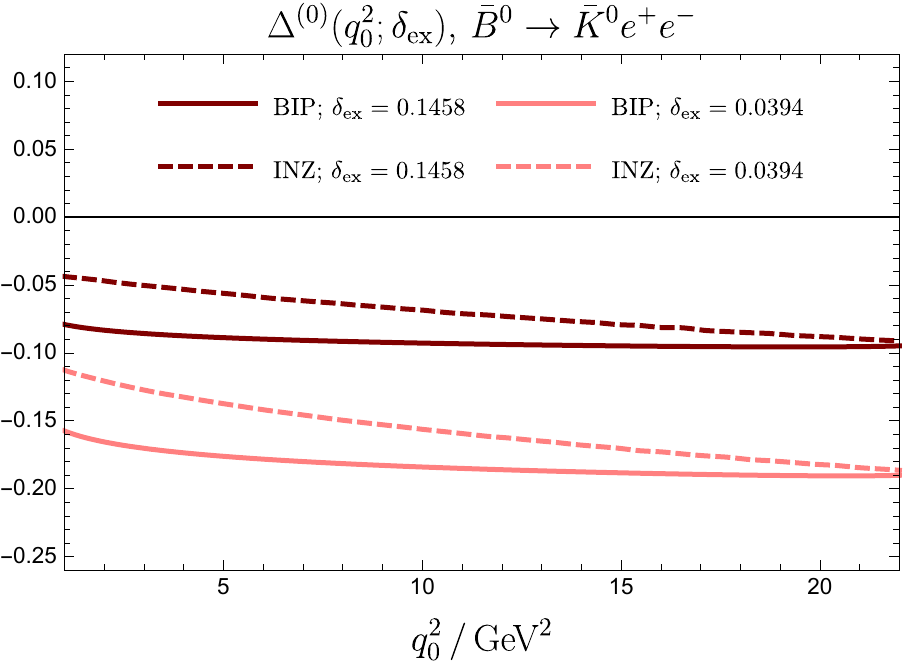}  
        	\caption{\small 
	Relative effects of relative corrections as a function of $q^2$, in the neutral case, with the 
	cuts on $m_B^{\rm rec}$ used in \cite{BIP16} computed in this work (top left) vs.~those presented 
	in~\cite{BIP16} (top right). The bottom left and bottom right plots compare our results with those in \cite{BIP16} for the $q^2$ and $\qz^2$-spectrum respectively.
	The translation between the notation of this reference and ours is 
	$\de_{\textrm{ex}} = 1 -  (m_B^{\textrm{rec}}/m_B)^2$ with 
	($\pBbar = m_B^{\textrm{rec}}$) and $\{ 0.1458, 0.1, 0.0394  \}   \leftrightarrow 
	m_B^{\rm rec}= \{4.88,5.009, 5.175\}\GeV$.}
	\label{fig:BIP-INZ}
\end{figure}

On the other hand, while assumption ii.~is a legitimate choice, 
 it is incompatible with the goal of 
estimating radiative corrections by implementing only a cut on the reconstructed $B$-meson mass:\footnote{~We note that 
a radiator function depending  on $q^2$ and $q_0^2$ only is sufficient to estimate the 
distortion of the $q^2$ spectrum in the absence of a photon-energy cut, as is for instance done in 
Higgs-collider physics~\cite{Bordone:2015nqa}.}
 the radiator 
in~\cite{BIP16} is obtained by integrating over all photon angles; however, as already discussed in 
\ref{sec:distortion}, in the $B$-RF
the relation connecting $q_0^2$ and  $q^2$ does  not only depend on $m_B^{\textrm{rec}}$ but also 
on the photon's emission angle. To overcome this problem, in~\cite{BIP16} the maximal 
$q_0^2$ value affecting the spectrum at a given $q^2$ value has been determined imposing  the tight 
cut defined in \eqref{eq:A9}. This choice corresponds to the {\em minimal} value of  $(q_0^2)_{\rm max}$ obtainable with an experimental cut on photons not emitted forward with respect to  $\vec{q}$ (in the $B$-RF).  Incidentally, we note that a cut of this type is implemented in the experimental analysis to avoid 
a large migration effect (e.g. charmonium resonances at low $q^2$,~cf.~\SEC\ref{sec:distortion}).
This is the most important difference among the two approaches. As illustrated in \FIG\ref{fig:BIP-INZ}, 
the net effect is quite sizeable, especially for the electrons at low values of $q^2$. 

In practice, the implicit cut applied in~\cite{BIP16} on the photon-emission angle 
removes  some hard-collinear logs.
We may track the difference on the collinear logs analytically.
 We demonstrate this for the $q_0^2$-spectrum since the expression \eqref{eq:FhcR} is much simpler 
 than the corresponding one for $q^2$ in  \eqref{eq:FhcT}.
Let  us consider
\begin{equation}
\frac{d\Gamma}{d \qz^2} = \frac{\al}{\pi} \left[ \frac{d\Gamma}{d \qz^2} \right]^{\!\!\LO} \big( A_0 \ln \des + C_0 \big) \ln m_{\ell} + \textrm{non-collinear} \;.
\end{equation}
The coefficients $A_0$ and $C_0$ are obtained  by integrating 
Eq.~(9) --using the boundary conditions implied by Eq.~(10) of~\cite{BIP16}--
w.r.t. to $x$  (which is our $z$ and moreover $1-\de^2 = \des$), 
and the expression  in \eqref{eq:FhcR} with $z(\Des) \to 1$ but finite $\des$. 
Not surprisingly, we find 
\begin{equation}
A_0 =A_0^{\textrm{INZ}}  =  A_0^{\textrm{BIP}}=  -4 \;.
\end{equation}
This is the universal coefficient of the soft--collinear singularity (double log), 
which is independent of the boundary conditions. 
Incidentally, we note that this  coefficient  is also the same for the 
$\frac{d\Gamma}{d q^2} $-distribution.  Low's theorem guarantees that 
the single $\ln \des$-term   is identical. 
 For the $C_0$ term, however, there are differences
\begin{alignat}{2}
& C_0^{\textrm{INZ}} &\;=\;& -\frac{25}{3}  + 2\ln\hat{q}_0^2 +2\frac{(1- \hat{m}_K^2+  \hat{q}_0^2)^2}{\hat{\la}} -  
R \ln \left[ \frac{1- \hat{m}_K^2+  \hat{q}_0^2  - \hat \la^{1/2}}{1- \hat{m}_K^2+  \hat{q}_0^2  + \hat \la^{1/2}}   \right]  + \ORD(\des) \nonumber \\[0.1cm]   
& C_0^{\textrm{BIP}} &\;=\;& \left( -3 - 4 \ln \left[ 1+ \frac{\hat{m}_K^2}{1- \hat{q}_0^2} \right] \right)  + \ORD(\des) \;,
\end{alignat}
where  $\hat{\la} \equiv \la(1, \hat{m}_K^2, \hat{q}_0^2)$ 
and
\begin{equation}
R =  2( 1+ \hat{q}_0^2 - \hat{m}_K^2) \left( \frac{1 }{\hat{\la}^{1/2}} - \frac{ 2 \hat{q}_0^2}{ \hat{\la}^{3/2}} \right) 
 \;.
\end{equation}
Note that in \cite{BIP16}, only the leading term in $m_K^2$ was kept in $(q_0^2)_{\max}$ 
and thus, for meaningful comparison, one has to assume the  $m_K \to 0$ limit
\begin{alignat}{2}
\label{eq:C0INZBIP}
& C_0^{\textrm{INZ}} &\;=\;&   -\frac{19}{3} + 8 \frac{\hat{q}_0^2}{ ( 1- \hat{q}_0^2)^2 }   +  
4   \frac{ (3- \hat{q}_0^2)}{(1-\hat{q}_0^2)^3} \hat{q}_0^4  \ln \hat{q}_0^2  + \ORD(\des) 
 ~  \stackrel{ \hat{q}_0^2  \to 1 }{\longrightarrow}  
  -3 + \ORD(\des) + \ORD( \hat{q}_0^2-1)   
\nonumber \\[0.1cm]   
& C_0^{\textrm{BIP}} &\;=\;&  -3   + \ORD(\des) \;.
\end{alignat}
Agreement is found at the kinematic endpoint  $\hat{q}_0^2  \to 1$  (including 
 $\ORD(\des)$-terms).
 This is to be expected since the  cut on $(q_0^2)_{\rm max}$ is independent 
 of the photon-emission angle, whereas differences are maximal at low $q_0^2$ values, consistent 
 with the numerical findings in \FIG\ref{fig:BIP-INZ} (bottom-right).
  
To better understand the agreement at large $\hat{q}_0^2$ illustrated in \eqref{eq:C0INZBIP}, consider \eqref{eq:z}, with $\de=\des$, which corresponds to the case where the photon becomes collinear with $\lone$. With a non-zero Kaon mass, $ (\hat{q}_0^2)_{\textrm{max}} =(1-\hat{m}_K)^2$, and thus the lower limit for the $z-$integration becomes
\begin{equation}
\label{eq:zINZ}
z_{\textrm{INZ}}(\des, (\hat{q}_0^2)_{\textrm{max}}, \clz) = 1-\frac{\des}{1-\hat{m}_K }  \;,
\end{equation}
where  the $\clz$-dependence drops (and thus the same $z$ limit applies for $\ltwo$).
Now, consider $q_0^2= q^2 + m_B  \des (E_{\qz}^{\FRone} + |\vec{q}_0^{\;\FRone}| \cos\theta_\gamma^{\;\FRone} )$, which is the defining principle behind Eq.~(10) of \cite{BIP16}, where $E_{\qz}^{\FRone} $ and $|\vec{q}_0^{\;\FRone}|$ are given in \eqref{eq:(1)}. Substituting  $q^2=z \qz^2$, one gets
\begin{equation}
\label{eq:zBIP}
z_{\textrm{BIP}}=1-\frac{m_B\, \des}{\qz^2} (E_{\qz}^{\FRone} + |\vec{q}_0^{\;\FRone}| \cos\theta_\gamma^{\;\FRone} )\stackrel{ \hat{q}_0^2  \to (\hat{q}_0^2)_{\textrm{max}} }{\longrightarrow} 1-\frac{\des}{1-\hat{m}_K },
\end{equation}
which matches \eqref{eq:zINZ}. 
This explains the agreement at large $\hat{q}^2_0$ in \eqref{eq:C0INZBIP} and in \FIG\ref{fig:BIP-INZ}.
Note that the $\theta_\gamma^{\;\FRone}$ dependence drops in the limit of $\hat{q}_0^2  \to  (\hat{q}_0^2)_{\textrm{max}}$, analogous to the $\clz$ dependence in \eqref{eq:zINZ}.  The $\clz$-independence  (or equivalently $\theta_\gamma^{\;\FRone}$) at $ (\hat{q}_0^2)_{\textrm{max}}$ happens since the 
Kaon's three-momentum vanishes and the  $\FRone$- and $\FRfour$-RF become equivalent 
and thus, there cannot be any non-trivial angular dependance.

On the other hand, the same argument does \textit{not} apply to the differential rate in $q^2$. As $q^2\to q^2_{\textrm{max}}$, the range of allowed photon energies becomes more and more restricted. The cut $\pBbar^2>m_B^2(1-\des)$ on its own is independent of $q^2$, and it is for this reason that one needs the maximum condition imposed on the lower limit of the $z$-integration in \eqref{eq:FhcT}. For larger $q^2$, the kinematic restriction on $z$, denoted by $z_\textrm{inc}$, becomes more important than the restriction on $z$ due to the photon energy cut $\des$. This is why the two INZ-curves in the bottom left plot in \FIG\ref{fig:BIP-INZ} approach  each other for large $q^2$.

 In summary, from the comparison of our work
with~\cite{BIP16} we may deduce the following two lessons or insights.
\begin{itemize}
\item[a)]  The indirect determination of  virtual logs in the photon-inclusive $d\Gamma/dq_0^2$-spectrum,
which is the key assumption behind both the approach of Ref.~\cite{BIP16} and PHOTOS~\cite{PHOTOS}, is correct.
\item[b)]  A meaningful comparison between theory and experiment (in a collider environment)
cannot be done by only considering  the two non-radiative variables ($\{ \qSaa,  \claa \}$) and 
the cut on the reconstructed $B$-meson mass, but it requires a detailed information on the (inevitable)
 photon-emission angle cut as their impact is sizeable.
\end{itemize}

Whereas point a) is reassuring in view of the current treatment of $R_K$, 
point b) indicates the necessity to build a Monte Carlo program
with a complete differential 
treatment of radiative corrections and an accurate parameterisation of the hadronic form factors 
(with the effective inclusion of long-distance effects, which we recall are {\em not} included in PHOTOS), 
in order to check the impact of the 
QED corrections on the kinematical distributions at the $\%$-level, with the explicit cuts 
applied in  experiments. This task, for which  this paper lays the groundwork,
is devoted to a future publication.

\section{Explicit Results of the Computation}
\label{app:AMP}

\subsection{Leading order differential rate 
}
\label{rate:LO}

The leading order amplitude rate is easily computed from \eqref{eq:d2Ga0} and the amplitude 
 $\Ampzero_{\Bin \to \Kout \lone \bltwo}$ \eqref{eq:A0} and is  rather simple
 \begin{alignat}{1}
\label{eq:ratell}
& \frac{d^2}{dq^2 d\cl}  \Gamma^{\LO}(q^2,\cl) \;= \frac{\rho_{\TT}|_{\pBbar^2 \to m_B^2}}{m_B} |\Ampzero|^2 = \;2|\geff|^2 
\frac{\rho_{\TT}|_{\pBbar^2 \to m_B^2}}{m_B} \times \nonumber \\[5pt]
&  \Bigg[ |\CV|^2\Big(\KallenB f_+(q^2)^2 (1-(\Demlh)^2-\frac{\Kallenl}{q^4} \cl^2) \; 
+
(\DemBKsq)^2(\Demlh)^2 f_0(q^2)^2 (1-\monetwoh^2) -\nonumber\\[5pt]
& \quad 2 \DemBKsqh\,\monetwoh\,\Demlh\, f_0(q^2)f_+(q^2) \KallenB^{1/2}\Kallenl^{1/2}\cl \Big)+
   |\CA|^2\Big(\monetwoh\leftrightarrow \Demlh\Big)\Bigg] \;,
\end{alignat}
where $\Kallenl = \la(q^2,m_{\lone}^2,m_{\ltwo}^2)$, 
$\Demlh=\bar{m}_{\lone}-\bar{m}_{\ltwo}$, $\monetwoh=\bar{m}_{\lone}+\bar{m}_{\ltwo}$, $\DemBKsq=m_B^2-m_K^2$, with $\rho_{\TT}$ as in \eqref{eq:barred}, 
and all barred quantities are dimensionless by division with $q$. In the limit of equal lepton masses ($\mlone=\mltwo\equiv m_{\ell}$), the above equation reduces to 
\begin{alignat}{2}
\label{eq:dG0}
& \frac{d^2}{dq^2 d\cl}  \Gamma^{\LO}(q^2,\cl) = 2 |\geff|^2 \frac{\rho_{\TT}|_{\pBbar^2 \to m_B^2}}{m_B}
  \Big(& &   |\CV|^2    (   \KallenB f_+(q^2)^2 (1-  \be_\ell^2 c_\ell^2))    +\\[0.1cm]
 & & &  |\CA|^2 (  \KallenB  f_+(q^2)^2  
(1- c_\ell^2) \be_\ell^2 + 4 f_0(q^2)^2 \bar{m}_\ell^2 (\DemBKsq)^2)       \Big)  \;,  \nonumber    
\end{alignat}
with $\be_\ell = \sqrt{1- 4 m_\ell^2/q^2}$ and $\KallenB = \la(m_B^2,q^2,m_K^2)$. 

\subsection{Virtual amplitude $\Amptwo_{\Bin \to \Kout \lone \bltwo }$}
\label{app:vir}

As the computation of the QED corrections including the tower 
of operators \eqref{eq:Lint_eff} is anew, to the best of our knowledge, we present
the explicit amplitudes prior to integration. 
The $B,K_{L_{1,2}}$ and $P_{L_{12}}$ and $P_{B,K}$ graphs are non-trivially amended. 
This is true in particular for the $P$-graphs.

The  $B L_{12}$-graphs read 
 \begin{eqnarray*}
 \label{eq:AB1}
\Amptwo_{B L_{1,2}}
 &=&
i \geff  \hat{Q}_{\Bin} e^2 \!\!
\int_k ( 2 p_B \pl k)^\nu \Delta_{\nu\rho}(k)  \Delta_B(l)  \tilde{H}_0^{\mu (B)}(\qz^2)   
 \bar u( \hat{Q}_{\lone}  \ga^\rho S_1(r)  \Gamma_\mu   \mi \hat{Q}_{\bltwo} \Gamma_\mu S_2(r) \ga^\rho )    v\;,
\end{eqnarray*} 
with shorthands $\bar u  \equiv \bar{u}(\lone) $, $v \equiv v(\ltwo)$,  $(2\pi)^4 \int_k = \int d^{4}k$, momentum assignments  $(r,l)\;=\;(  \pm [ \lone(\ltwo)+k] , p_B +k)$, 
 with  notation borrowed from the real emission case (cf. below \eqref{eq:realAmp}),
\begin{equation}
f_\pm(\qz^2) = \sum_{n \geq 0} \frac{ f_\pm^{(n)} (0)}{n!} ( \qz^2)^n \;, 
\quad P_n  \equiv   P_n(q^2,\qz^2) = \sum_{m=0}^n (q^2)^{(n-m)} (q_0^{2})^m\;,
\end{equation}
with $\qz = q + k$ and $k$ being the loop integration momentum $k$. 
Moreover $ \tilde{H}_0^{(B)}(\qz^2)  =  H_0(\qz^2)|_{p_B \to p_B+ k }$, and the 
$\qz^2$ in the argument indicates that the form factor is to be expanded as above
and   propagators are given by 
 \begin{eqnarray}
 \label{eq:prop}
 & & \Delta_M(k) = \frac{1}{k^2-m_M^2} \;, \quad S_i(k) = \frac{\slashed{k} + m_{\ell_i}}{k^2 - m_{\ell_i}^2} 
 \;, \quad \Delta_{\mu \nu}(k ) = - 
 \frac{g_{\mu\nu}}{k^2} + (1 - \gauge) \frac{k_\mu k_\nu}{k^4} \;.
  \end{eqnarray}
The $KL_{1,2}$-graphs are analogous     
 \begin{eqnarray*}
  \label{eq:AB2}
\Amptwo_{KL_{1,2}} &=& 
  - i \geff  \hat{Q}_{\Kout} e^2 \!\!
\int_k (2 \pK \mi k)^\nu 
 \Delta_{\nu\rho}(k) \Delta_K(l)   \tilde{H}_0^{\mu(K)}(\qz^2)   
   \bar{u}  (  \hat{Q}_{\lone}  \ga^\rho S_1(r)  \Gamma^\mu    \mi \hat{Q}_{\bltwo}  \Gamma^\mu S_2(r) \ga^\rho )    v \;,
\end{eqnarray*} 
where $(r,l) = ( \pm[ \lone(\ltwo)+k] , \pK-k)$ and
$ \tilde{H}_0^{(K)}(\qz^2)  =  H_0(\qz^2)|_{\pK \to \pK- k }$.
   The  $PL_{1,2}$ and $P_{B,K}$
graphs read
  \begin{alignat}{2}
  \label{eq:Pgraphs}
& \Amptwo_{PL_{1,2}} &\;=\;&  - i  \geff  e^2  \int_k F_{\mu \rho}^{L_{1,2}}
    \bar{u} ( \hat{Q}_{\lone}   \ga^\rho S_1(r)  \Gamma_\mu   - \hat{Q}_{\bltwo}  \Gamma_\mu S_2(r) \ga^\rho )    v\;,   \\[0.1cm]
&   \Amptwo_{P_{\Bga,\Kga}} &\;=\;&  i \geff  e^2    L_0^\mu 
\int_k     \hat{Q}_\Bin  F^{B}_{ \mu \rho } \,  \Delta_B(p_B+k) (2p_B +k)^\rho -    \hat{Q}_\Kout  \,  F^K_{ \mu \rho } \,   \Delta_K( \pK-k) (2 \pK- k)^\rho  \;, \nonumber
\end{alignat} 
 with  $ r = \pm ( \lone(\ltwo) + k) $, $L_0$ defined in \eqref{eq:L0}  and 
 the loop momentum $k$ enters the expressions
 \begin{equation*}
F_{\mu \rho}^{L_{1,2}} =  F_{\mu \rho}(p_B,\pK,\qz^2) \;, \quad 
F^{B}_{ \mu \rho } = F_{ \mu \rho }(p_B+ k,\pK,q^2) \;, \quad 
F^{K}_{ \mu \rho } = F_{ \mu \rho }(p_B,\pK - k,q^2) \;,
 \end{equation*}
 where the common functional form $F_{\mu\rho}$ is given by
 \begin{alignat}{2}
 \label{eq:FK-PL12}
 & F_{\mu \rho}(p_B,\pK,q^2) &\;=\;&   (\hat{Q}_{\Bin}\! \mp \!\hat{Q}_\Kout)  \Delta_{\mu\rho}(k)  f_\pm(q^2)  \nonumber \\[0.1cm]
&  &\;+\;&     (\hat{Q}_\Bin \!+\! \hat{Q}_\Kout) (p_B\! \pm \! p_K)_\mu
 (q + \qz )^\nu   \Delta_{\nu\rho}(k) \sum_{n \geq 1} \frac{f_\pm^{(n)}(0)}{n!} P_{n-1}      \;.  
  \end{alignat}
The  BK-vertex correction is given by
  \begin{eqnarray*}
 \label{eq:C1}
\Amptwo_{\Bga \Kga} =   i \geff \hat{Q}_\Bin \hat{Q}_\Kout e^2       \int_k  (2p_B - k)^\be \Delta_{\be\kappa}(k)  (2\pK - k)^\kappa 
L_0 \Cdot \tilde{H}^{(BK)}_0 \, \Delta_B(l)  \Delta_K(r) \;,
  \end{eqnarray*}
 where 
  $ l = ( p_B-k)$ and $r = ( \pK -k) $ and $\tilde{H}^{(BK)}_0 = H_0(q^2)|_{(p_B,\pK)\to   (p_B\mi k,\pK \mi k)}$. 
The lepton  vertex correction, which can be found in many textbooks,  reads 
 \begin{eqnarray}
 \label{eq:D1}
 \Amptwo_{L_1L_2} =  i  \geff \hat{Q}_{\lone} \hat{Q}_{\bltwo} e^2 H_0^\mu(q^2) 
    \int_k \Delta_{\al \be }(k) \, \bar{u}   \ga^\al S_1(l)\Gamma_\mu  S_2(r)   \ga^\be v
   \;,
 \end{eqnarray}
 with $l = k+ \lone-$ and $r = k-\ltwo $.

\subsection{Gauge invariance of the real amplitude $\Ampone_{\Bin \to \Kout \lone \bltwo \ga} $}
\label{app:GI}

The real amplitude is given in Eq.~\eqref{eq:realAmp}.
Explicit verification of gauge invariance of this amplitude  is instructive. 
In essence, we will flesh out the steps described at the end of \SEC\ref{sec:real}. 
Gauge invariance follows from the charge conservation \eqref{eq:charge} and inspecting the four 
terms in \eqref{eq:realAmp}, it is far from obvious how this will work out since the individual terms 
depend on the hadronic form factor in a non-uniform way e.g. $\hat Q_{\bltwo,\lone} H_0(q_0^2)$, 
$\hat{Q}_{B,K} \bar{H}^{(B,K)}(q^2), \dots$. 
 A special r\^ole is played by the contact terms arising from  
diagram $P$ in \FIG\ref{fig:real}. From the viewpoint of the effective Lagrangian, these terms 
arise from replacing ordinary derivatives with covariant ones and from the viewpoint of the Ward identity, 
they are induced by the derivatives acting on the $U(1)$ gauge transformation.

At first, we consider lines two and three of the amplitude 
 \begin{alignat}{2}
&\Ampone_{23} &\; \propto \;  &      \hat{Q}_{\Bin} \,   L_0 \Cdot \bar{H}^{(B)}_0(q^2)     \frac{\eps^{*} \Cdot p_B}{k \Cdot p_B}  \; + 
    \hat{Q}_{\Kout} \,  L_0 \Cdot \bar{H}^{(K)}_0(q^2)    \frac{\eps^{*} \Cdot p_K}{k \Cdot p_K }     
 \;
    \nonumber \\[0.1cm]
& & \; + \;   &   ( \hat{Q}_{\Bin} \!  -\!  \hat{Q}_{\Kout})  \,L_0 \Cdot \eps^* \, f_+(q^2) +  ( \hat{Q}_{\Bin}\!   +\! \hat{Q}_{\Kout})  \,L_0 \Cdot \eps^* \, f_-(q^2) \;   \nonumber \\[0.1cm]
& & \; \stackrel{\eps \to k}{\to}    \;   &   ( \hat{Q}_{\Bin}  + \hat{Q}_{\Kout}) L_0 \Cdot {H}_0(q^2)  \;, 
\end{alignat} 
and notice that a gauge transformation combines these two lines into an expression which 
will combine with the first line 
\begin{alignat}{2}
& \Ampone_{1}  &\; \propto \;&  
\bar u (\ltwo) \left[ \hat{Q}_{\lone}\,   \frac{2 \eps^* \Cdot \lone\pl  \slashed{\eps}^* \slashed{k}}{2 k \Cdot \lone }   \Gamma \Cdot H_0(\qz^2)    + 
   \hat{Q}_{\bltwo}  \,   \Gamma \Cdot H_0(\qz^2)  \frac{2 \eps^* \Cdot \ltwo\pl  \slashed{k} \slashed{\eps}^*}{2 k \Cdot \ltwo }  \right] v (\lone)    \;   \nonumber \\[0.1cm]
   & & \; \stackrel{\eps \to k}{\to}    \;   &   ( \hat{Q}_{\bltwo}  + \hat{Q}_{\lone}) L_0 \Cdot {H}_0(q_0^2)  \;,
\end{alignat}
 except that the argument of the form factors is $q_0^2$ in one case and $q^2$ in the other case. 
This is remedied, of course, by the fourth line
\begin{alignat}{2}
& \Ampone_{4}  &\; \propto \;&   ( \hat{Q}_{\Bin}\!   +\! \hat{Q}_{\Kout})  L_0 \Cdot (p_B \pm \pK)  ( 2 \eps^* \Cdot q) \sum_{n \geq 1} \frac{f^{(n)}_{\pm}(0)}{n!}  P_{n-1}  \;   \nonumber \\[0.1cm]
   & & \; \stackrel{\eps \to k}{\to}\;  &    ( \hat{Q}_{\Bin}\!   +\! \hat{Q}_{\Kout})  L_0 \Cdot (p_B \pm \pK) \sum_{n \geq 1} \frac{f^{(n)}_{\pm}(0)}{n!}    \QQ{2n}    \;   \nonumber \\[0.1cm]
 & &\;   =\;&   ( \hat{Q}_{\Bin}\!   +\! \hat{Q}_{\Kout})  L_0 \Cdot ( H_0(q_0^2)  - H_0(q^2) )  \;,
\end{alignat}
which follows from $\QQ2 = 2 q \Cdot k $ and $\QQ2 P_{n-1} = \QQ{2n}$ and 
$ \QQ{2n} \equiv  (q_0^2)^{n} - (q^2)^{n} $ as before. Adding them all together, 
one obtains
\begin{equation}
 \Ampone|_{\eps \to k} \propto  L_0 \Cdot H_0(q_0^2)   \sum_{i }  \hat Q_i = 0 \;,
\end{equation}
the explicit gauge invariance of the real amplitude.

\subsection{Cancellation of hard-collinear logs charge by charge}
\label{app:charge}

Whereas for the cancellation of soft divergences  charge conservation was not assumed, this is not true
for the hard-collinear logs $\ln m_\ell$ cf. \SEC\ref{sec:Fhc0}. 
The aim of this appendix is to show that this assumption is unnecessary, i.e.  that 
hard-collinear logs  cancel charge by charge combination. Charge conservation is though necessary for gauge 
invariance or conversely imposing gauge invariance implies charge conservation.  
Using charge conservation can still be  convenient such as for 
the photon-inclusive hard-collinear log formula 
\eqref{eq:magic}.

First, we focus on  the soft contribution $ {\cal F}^{\soft}(\Des )|_{\ln m_{\lone}}\equiv\sum_{i, j }   \hat{Q}_i \hat{Q}_j    {\cal F}^{\soft}_{ij}(\Des )|_{\ln m_{\lone}} $ to the hard-collinear log.
  In the limit of $\mlone\to 0$, using \EQs\eqref{eq:Fii}, \eqref{eq:smallbi}, \eqref{eq:FiB}, \eqref{eq:Fijhard} and \eqref{eq:Fsoftcoll}, one gets
\begin{alignat}{2}
	& {\cal F}^{\soft}(\Des )|_{\ln m_{\lone}}&\;=\;&\ln \mlone \left[ -\hat{Q}_{\lone}^2 +2\hat{Q}_{\lone}\left ( \hat{Q}_{\bltwo} \pl \hat{Q}_{\Bin}\pl \hat{Q}_{\Kout} \right )\ln\bar{z}(\Des) \right].
\end{alignat}
where we have used $\bar{z}(\Des)=\frac{\Des m_B}{2 E_{\lone}}$, as explained below \EQ\eqref{eq:Fsoftcoll}.
Next,  the virtual contribution, $ \tH|_{\ln m_{\lone}}\equiv \sum_{i, j }   \hat{Q}_i \hat{Q}_j   \left(  \tH^\soft_{ij} + \tH^\hc_{ij} \right)\Big{|}_{\ln m_{\lone}}$, using \EQs\eqref{eq:Z}, \eqref{eq:Hlead2} and \eqref{eq:hcvirt}, is given by
\begin{alignat}{2}
&\tH|_{\ln m_{\lone}}&\;=\;&\ln m_{\lone}\left[\frac{3}{2} \hat{Q}_{\lone}^2+2  \hat{Q}_{\lone}\left ( \hat{Q}_{\bltwo} \pl \hat{Q}_{\Bin}\pl \hat{Q}_{\Kout} \right )\right].
\end{alignat}
Moreover,  $ {\cal F}^{\hc}(\underline{\de} )|_{\ln m_{\lone}}\equiv\sum_{i, j }   \hat{Q}_i \hat{Q}_j    {\cal F}^{\hc}_{ij}(\underline{\de} )|_{\ln m_{\lone}}$ is given by
\begin{alignat}{2}
\label{eq:FhcCOLL}
& {\cal F}^{\hc}(\underline{\de} )|_{\ln m_{\lone}}&\;=\;& \ln m_{\lone}\left[  -\frac{1}{2}\hat{Q}_{\lone}^2-2\hat{Q}_{\lone}\left ( \hat{Q}_{\bltwo} \pl \hat{Q}_{\Bin}\pl \hat{Q}_{\Kout} \right )\left(1+\ln\bar{z}(\Des) \right)\right] \;.
\end{alignat} 
In obtaining \eqref{eq:FhcCOLL}, we followed  the procedure in \SEC\ref{sec:Fhc0} without using charge conservation in \EQ\eqref{eq:A1sqhardcoll}.

Finally,  adding  the three contributions, one finds (with ordering as above)
\begin{alignat}{2}
& \left[{\cal F}^{\soft}(\Des )+\tH+ {\cal F}^{\hc}(\underline{\de} )\right]\Big{|}_{\ln m_{\lone}}\!\!\!\!&\;=\;&  
 \left[  2 \ln \bar z(\Des) +2 - 2( 1+ \ln \bar z(\Des))   \right] \cdot \hat{Q}_{\lone}( \hat{Q}_{\bltwo} \pl \hat{Q}_{\Bin}\pl \hat{Q}_{\Kout} )
 \nonumber   \\[0.1cm] 
& &\;+\;& \left[-1+  \frac{3}{2}  - \frac{1}{2}\right] \cdot \hat{Q}_{\lone}^2 = 0 \;,
\,
\end{alignat}
that the  hard-collinear  cancel charge by charge (without the need for charge conservation).

 \section{Kinematics and other Conventions}
 \label{app:kinematics}


In this section, we collect a few conventions used throughout the paper to improve readability.
We make use of the abbreviation  $\claa = \cos \Tlaa$ and $s_a=\sin\Tlaa$ where the label $\AAA$ stands 
either for $\TT$ or $\RR$ and its meaning on the main kinematic variables is depicted in 
\eqref{eq:tr}.
The matrix elements 
$\matel{0}{B^\dagger(x) }{\bar{B}(p_B)}  =  e^{- i  p_B \cdot x}$,  
$ \matel{K(\pK)}{K^\dagger(x)}{0}  = e^{ i  \pK \cdot x}$ provide the link to 
the mesonic states $\Bin $ and $ \Kout$ of  valence quarks $b$ and $s$.  
Whenever there is no ambiguity, we use $p = \sqrt{p^2}$ and hatted 
quantities are understood to be divided by $m_B$ in order to render them dimensionless 
e.g. $\hat{q}^2 \equiv q^2/m_B^2$. We use dimensional regularisation with $d = 4- 2 \eps$.

\subsection{Kinematics in terms of the $\{q^2 ,\Tl\}$-variables} 
\label{app:q2}

The main frame is the $\pBbar$-RF, which will serve to define the  photon energy cut-off.
In this frame, the momenta are parametrised as 
follows\footnote{~All four-momenta are understood with lower Lorentz indices e.g. $(k^{\FRtwo})_{\mu}$.
It is understood that  $\Tg \equiv \Tg^{\FRtwo}, \Fg \equiv  \Fg^{\FRtwo}$
for brevity. If the angles do not refer to the $\FRtwo$-frame, then they will be indicated.}
\begin{alignat}{2}
\label{eq:kin2}
& k^{\FRtwo} &\;=\;&  (E_\ga^{\FRtwo}  , - \cos \Tg  |\vec{k}_\ga^{\FRtwo}| , - \sin \Tg \cos \Fg  |\vec{k}_\ga^{\FRtwo}| ,
- \sin \Tg \sin \Fg  |\vec{k}_\ga^{\FRtwo}| ) \;, \nonumber \\[0.1cm]
& \pBbar^{\FRtwo} &\;=\;& ( \pBbar,0,0,0)  \;,\quad q^{\FRtwo} =  (\pBbar- p_K)^{\FRtwo} =    (\pBbar- E_K^{\FRtwo}, | \vec{p}_K^{\,\FRtwo}|,0,0  ) = 
( E_q^{\FRtwo} , | \vec{p}_K^{\,\FRtwo}|,0,0  ) \;,
 \nonumber \\[0.1cm] 
& p_K^{\FRtwo} &\;=\;& (E_K^{\FRtwo},- | \vec{p}_K^{\,\FRtwo}| ,0,0)  \;, 
\end{alignat}
where
\begin{alignat}{5}
\label{eq:(2)}
& E_K^{\FRtwo} &\;=\;&  \sqrt{|\vec{p}^{\;\FRtwo}_K|^2 + m_K^2}&\;=\;&   \frac{1}{2 \pBbar} ( \pBbar^2 - q^2 + m_K^2) \;, 
\quad & &   |\vec{p}_K^{\;\FRtwo}| &\;=\;&  \frac{\la^{1/2}(\pBbar^2,q^2,m_K^2)}{2 \pBbar}  \;,
\nonumber \\[0.1cm] 
&  E_\ga^{\FRtwo} &\;=\;&  
\sqrt{  |\vec{k}_\ga^{\FRtwo}|^2 + \mga^2}  &\;=\;&  \frac{1}{2 \pBbar} \left( m_B^2 - \pBbar^2 - \mga^2 \right)   \;, \quad & & |\vec{k}_\ga^{\FRtwo}| &\;=\;&  \frac{\la^{1/2}(\pBbar^2,m_B^2,\mga^2)}{2 \pBbar}  \;,
 \nonumber   \\[0.1cm] 
& E_q^{\FRtwo} &\;=\;&   \sqrt{|\vec{p}^{\;\FRtwo}_K|^2 + q^2}  
&\;=\;&  \frac{1}{2 \pBbar} ( \pBbar^2 + q^2 - m_K^2)   \;,  
\end{alignat} 
consistent with $\pBbar- E_K^{\FRtwo} = E_{q}^{\FRtwo}$.  
The K\"all\'en function, 
\begin{equation}
\label{eq:Kallen}
\la(s,m_1^2,m_2^2) =  (s- (m_1 -m_2)^2)(s- (m_1 +m_2)^2)\;,
\end{equation}
 is related to the spatial momentum in $1 \to 2$ decay
\cite{PDG}.
The momenta $\lonetwo$ depend on the angle of the lepton $\lone$ w.r.t to the decay axis 
in the $q$-RF 
\begin{alignat}{2}
& \lone^{\FRtwo} &\;=\;& 
 (\ga ( E^{\FRthree}_{\lone} + \be \cos \Tl  \modveclone{\;\FRthree})   , \ga (   \be E^{\FRthree}_{\lone}+ \cos \Tl  \modveclone{\;\FRthree}), -  \modveclone{\;\FRthree} \sin \Tl ,0 )   \;, \nonumber \\[0.1cm]
& \ltwo^{\FRtwo} &\;=\;& 
 (\ga ( E^{\FRthree}_{\ltwo} - \be \cos \Tl  \modveclone{\;\FRthree})  , \ga (   \be E^{\FRthree}_{\ltwo} - \cos \Tl  
 \modveclone{\;\FRthree}) , + \modveclone{\;\FRthree} \sin \Tl ,0 )     \;, \end{alignat}
where the energy and momenta are defined in the $q$-RF and are given by
\begin{alignat}{5}
 & E_{\lonetwo}^{\FRthree} &\;=\;&  \sqrt{ \modveclone{\;\FRthree}^2 + m_{\lonetwo}^2} &\;=\;& \frac{1}{2 q}(q^2 + m_{\lonetwo}^2 - m_{\ltwoone}^2) \;,  \quad & & 
   \modveclone{\;\FRthree}  &\;=\;& \frac{\la^{1/2}(q^2,m_{\lone}^2,m_{\ltwo}^2)}{2 q}  \;,
\end{alignat}
and $q \equiv \sqrt{q^2}$, whenever it is clear that $q$ is not a vector, such as in 
 $E_q^{\FRthree} =E^{\FRthree}_{\lone}+E^{\FRthree}_{\ltwo} = q$.  
 The boost velocity $\be$  and $\ga$-factor are given by
\begin{equation}
\be =  \frac{ | \vec{p}^{\;\FRtwo}_K|}{E_q^{\FRtwo}} \;,
 \quad \ga = \frac{ E_q^{\FRtwo}}{q} \;, 
\end{equation}
where  $|\vec{q}| = |\vec{p_K}|$ was used.

\subsection{Kinematics in terms of the $\{\qz^2 ,\Tlz\}$-variables} 
\label{app:q02}

We start by defining kinematic variables in the $p_B-$RF, denoted by $\FRone$. Defining the $x$-axis along the direction of $ \vec{q}_0$, one has
\begin{alignat}{2}
\label{eq:(1)a}
 p_B^{\FRone} &\;=\;& ( m_B,0,0,0)  \;,  \quad  \qz^{\FRone} =  ( E_{\qz}^{\FRone} , | \vec{q}_0^{\;\FRone}|,0,0  )  \;,\quad p_K^{\FRone} \;=\;& (E_K^{\FRone },- | \vec{q}_0^{\;\FRone}| ,0,0)\;.
\end{alignat}
The momenta $\lone$, $\ltwo$, and $k$, will be defined in frame $\FRfour$, and
\begin{alignat}{5}
\label{eq:(1)}
&E_{K}^{\FRone} &\;=\;& m_B-E_{\qz}^{\FRone} &\;=\;&   \frac{1}{2 m_B} ( m_B^2-\qz^2+m_K^2) \;,\nonumber\\[5pt]
& E_{\qz}^{\FRone} &\;=\;&  \sqrt{|\vec{q}_0^{\;\FRone}|^2 + \qz^2}&\;=\;&   \frac{1}{2 m_B} ( m_B^2+\qz^2-m_K^2) \;, 
\quad & &   |\vec{q}_0^{\;\FRone}| &\;=\;&  \frac{\la^{1/2}(m_B^2,\qz^2,m_K^2)}{2 m_B}  \;.
\end{alignat} 
Frame $\FRone$ is useful for imposing the cut-off on the photon energy, c.f.~\EQ\eqref{eq:Emax}. For the phase space integration, we pick the independent variables $ |\vec{k}_\ga^{\FRfour}|$, $\Tg^{\FRfour}$, $\Fg^{\FRfour}$, all defined in the $\qz$-RF, which we denote as the $\FRfour$-frame. There, the four-momenta are given by
\begin{alignat}{2}
\label{eq:(4)}
& k^{\FRfour} &\;=\;&  (E_\ga^{\FRfour}  , - \cos \Tg^{\FRfour}  |\vec{k}_\ga^{\FRfour}| , - \sin \Tg^{\FRfour} \cos \Fg^{\FRfour}  |\vec{k}_\ga^{\FRfour}| ,
- \sin \Tg^{\FRfour} \sin \Fg^{\FRfour}  |\vec{k}_\ga^{\FRfour}| ) \;, \nonumber \\[5pt]
&p_B^{\FRfour }&\;=\;&  \gamma _{\qz}m_B \left (1,-\beta _{\qz},0,0 \right ) \;, 
\quad \qz^{\FRfour } = \left ( \qz,0,0,0 \right ),
 \nonumber\\[5pt]
&p_K^{\FRfour }&\;=\;& \gamma _{\qz} \left ( \left ( E_K^{\FRone }+\beta _{\qz}|\vec{q}_0^{\;\FRone}| \right ), -\left ( |\vec{q}_0^{\;\FRone}|+\beta _{\qz} E_K^{\FRone } \right ),0,0 \right ) \;,
\end{alignat}
where $E_\ga^{\FRfour}  =\sqrt{ |\vec{k}_\ga^{\FRfour}|^2 + m_{\ga}^2}$ and
the boost factors from the $p_B-$RF to the $\qz-$RF are given by
\begin{alignat}{2}
\beta _{\qz}\;=\;\frac{|\vec{q}_0^{\;\FRone}|}{E^{\FRone }_{\qz}} \;, \quad 
\gamma _{\qz}\;=\;\frac{E_{\qz}^{\FRone }}{\qz} \;.
\end{alignat}
We choose the axes such that $\veclone{\;\FRfour}$ lies in the $xy-$plane. Then
\begin{align}
\label{eq:boost4}
\lone^{\FRfour}=\left ( E_{\lone}^{\FRfour},|\veclone{\;\FRfour}| \clz ,- |\veclone{\;\FRfour}| \slz,0\right ) ,
\end{align}
where $\theta_0  $ (recall $c_0 \equiv \cos \theta_0$)  is the angle between $\veclone{\;\FRfour}$ and the x-axis in the $\qz$-RF (c.f.~\FIG\ref{fig:angles}), and $E_{\lone}^{\FRfour}= (|\veclone{\;\FRfour}|^2+\mlone^2)^{1/2}$. $\ltwo^{\FRfour}$ is found by momentum conservation via
$\ltwo^{\FRfour}=( \qz-\lone-k)^{\FRfour}$.
Solving for $|\veclone{\;\FRfour}|$ is quite complicated, and the explicit result is given by
\begin{align}
|\veclone{\;\FRfour}|=\frac{AB+\sqrt{D}}{C^2-B^2}\;,
\end{align}
where
\begin{alignat}{2}
\label{eq:ABCD}
& A &\;\equiv\;& \qz^2-2\qz E_{\gamma }^{\FRfour}+\mlone^2-\mltwo^2 + \mga^2 \nonumber\;,\\[5pt]
& B &\;\equiv\;&2E_{\gamma }^{\FRfour} \be_\ga \left (   \cos \Tg^{\FRfour}\;\clz\;-\;\sin \Tg^{\FRfour}\; \cos \Fg^{\FRfour}\;\slz  \right ) \nonumber\;,\\[5pt]
&C &\;\equiv\;& 2\qz-2E_{\gamma }^{\FRfour} \;, \nonumber \\[5pt]
& D  &\;\equiv\;& A^2 B^2+(C^2-B^2)(A^2-C^2 \mlone^2) \;,
\end{alignat}
where $\be_\ga = ((E_{\gamma }^{\FRfour})^2 - m_\ga^2)^{\frac{1}{2}}/E_{\gamma }^{\FRfour}$.
Using the above, one can also calculate 
$ \ff \equiv 2(|\vec{\lone}^{\FRfour}| E_q^{\FRfour} +  \partial_{ |\vec{\lone}^{\FRfour}|} [ \vec{k}\cdot \vec{\lone}]^{\FRfour}   E_{\lone}^{\FRfour}   ) $, needed in \eqref{eq:barred}. It reads
\begin{align}
\label{eq:ff}
\ff= 2 \left(|\vec{\lone}^{\FRfour}| (q_0- E_{\gamma }^{\FRfour})+E_{\lone}^{\FRfour}E_{\gamma }^{\FRfour} \be_\ga (  \sin \Tg^{\FRfour}\; \cos \Fg^{\FRfour}\;\slz \;-\; \cos \Tg^{\FRfour}\;\clz  )\right) \;.
\end{align}

\section{Soft Integral $\mathcal{F}^{\soft}_{ij}$}
\label{app:integral}

\subsection{IR sensitive part with photon mass and dimensional regularisation}
\label{app:soft1}

The $\mathcal{F}^{\soft}_{ij}$ integral is IR-divergent and has to be regulated. We discuss dimensional regularisation and photon mass regularisation in this appendix. The regularised integral, denoted by an ${\reg}$-subscript, 
is
\begin{alignat}{2}
\label{eq:Fzeroreg}
&\left [ \mathcal{F}^{\soft}_{ij} (\Des ) \right ]_{\reg} &=& 
 \int [d\Phi_\ga]_{\reg}  \left [ \frac{- (E_{\gamma}^{(n)})^2\,  p_i\cdot p_j}{(k\cdot p_i)(k\cdot p_j)} \right ]  = 
\frac{1}{2\pi } \int _0^{\EGAmaxn{n}}\frac{dE_{\gamma }^{(n)}}{E_{\gamma }^{(n)}}\rho^E _{\reg} \,
  I^{(\reg,n)}_{ij}(E^{(n)}_\ga) \;,
\end{alignat}
where 
\begin{equation}
 I^{\reg}_{ij}(E^{(n)}_\ga) \equiv  \int d\Omega_\ga^{(n)} \; \rho^{\Omega^{(n)}} _{\reg}
 \left [ \frac{- (E_{\gamma}^{(n)})^2\,  p_i\cdot p_j}{(k\cdot p_i)(k\cdot p_j)} \right ] \;,
\end{equation}
and
$\EGAmaxn{1,2}=\frac{\Des  m_B}{2}$ corresponds to the expression in \eqref{eq:Emax} 
with $\des \to \Des \ll 1$ for which the two frames become equivalent and 
\begin{align}
\rho^E _{\reg}&=\left\{  \begin{array}{ll} \frac{\Gamma (1-\epsilon )  }{\Gamma (1-2\epsilon )}\left ( \frac{E_{\gamma }^{(n)}  }{\sqrt{\pi} \mu } \right )^{-2 \epsilon }  & \textrm{dim-reg}  \\
\theta (E_{\gamma }^{(n)}-m_{\gamma }) & \mga    \end{array} \right. \;, \quad 
\rho^{\Omega^{(n)}}_{\reg}&=\left\{  \begin{array}{ll} \left ( \sin \Tg \,\sin \Fg \right )^{-2\epsilon } & \textrm{dim-reg}  \\
\frac{|\vec{k}_\ga^{(n)}|}{E_{\ga}^{(n)}} & \mga    \end{array} \right.\;,
\end{align}
and in addition one needs to set $\mga \to 0$ in dim-reg.
We will argue that the angular integral is Lorentz-invariant when the regulator is removed.
We may restore Lorentz invariance 
of \eqref{eq:Fzeroreg} by removing the photon energy cut-off.
In a second step, we remove 
the regulator, $\rho^E_\reg,\rho^{\Omega^{(n)}}_\reg   \to 1$. Then, the integral, which is frame- and scheme-independent, factorises into an energy integral $K$ and an angular integral $\Iijzero$, where the superscript $(0)$ indicates that the regulator has been removed. Since the energy integral is Lorentz invariant on its own, this implies the Lorentz-invariance of the finite $\Iijzero$-integral.

Focussing on the IR sensitive part, we keep $\rho^E_\reg$ to regulate the divergent energy integral 
and remove the angular regularisation $\rho^{\Omega^{(n)}}_\reg \to 1$ which is a useful limit as the integral still factorises into a doable energy integral and the Lorentz invariant $\Iijzero$-part,
\begin{align}
&\left [ \mathcal{F}^{\soft}_{ij} \right ]_{\reg}=-K_{\reg}(\Des )\; \Iijzero+ {\cal O}(f,\reg) \;,
\end{align}
where
\begin{align}
\label{eq:Iij0}
\Iijzero &=  I_{ij}^{(0,n)}  \equiv \int d \Omega_\ga^{(n)}\left [ \frac{- (E_{\gamma}^{(n)})^2\,  p_i\cdot p_j}{(k\cdot p_i)(k\cdot p_j)} \right ] = \eqref{eq:Iijzero} \;,
\end{align}
and we have used the Lorentz invariance of $ I_{ij}^{(0,n)}$. We note that while $\rho^{\Omega^{(n)}}_\reg \to 1$ captures all IR sensitive terms, it misses constant terms, indicated by ${\cal O}(f,\reg) $. These terms are determined in DR in the next section.

In DR,  the $K_{\reg}(\Des )$ integral  evaluates to
\begin{align}
K_{\epsilon }(\Des )&=\int _0^{\EGAmaxn{n}}\frac{dE_{\gamma }^{(n)}}{E_{\gamma }^{(n)}}\frac{\Gamma (1-\epsilon )  }{\Gamma (1-2\epsilon )}\left ( \frac{E_{\gamma }^{(n)}  }{\sqrt{\pi} \mu } \right )^{-2 \epsilon }=-\frac{1}{2}\rsoft+\ln \left ( \frac{\Des  m_B}{\mu } \right ) +\mathcal{O}(\epsilon  ),
\end{align}
whereas in photon mass regularisation the result is
\begin{align}
K_{\mga }(\Des  )&=\int _{\mga}^{\EGAmaxn{n}}\frac{dE_{\gamma }^{(n)}}{E_{\gamma }^{(n)}}=-\frac{1}{2}\rsoft+\ln \left ( \frac{\Des  m_B}{2\mu } \right ) +\mathcal{O}(\mga  ) \;,
\end{align}
and we note the additional factor of $2$ in the logarithm as compared to the DR result.

\subsection{Soft integrals in dimensional regularisation}
\label{app:SI}

In this section, we calculate the soft integrals fully analytically up $\mathcal{O}(\epsilon ^0)$ to using dimensional regularisation. We perform the integrals by introducing a soft cut-off $\Des$, and the result is obtained up to ${\cal O}(\Des)$ corrections, which can be safely neglected since $\Des\ll 1$.

The integrals have the general form
\begin{align*}
\mathcal{F}^{\soft}_{ij} (\Des ) =\frac{( \pi \mu^2 )^{ \epsilon } }{2\pi }\frac{\Gamma (1 \mi \epsilon )  }{\Gamma (1 \mi 2\epsilon )} \int _0^{\EGAmaxn{n}}\!\!\!\!\!
\frac{dE_{\gamma }^{(n)}}{\left ( E_{\gamma }^{(n)} \right )^{1+2\epsilon } }
\int_0^\pi  \frac{d\Tg}{ \sin^{2\epsilon-1 }\Tg}
 \int_0^\pi \frac{d\Fg}{ \sin^{2\epsilon }\Fg} \left [ \frac{- (E_{\gamma}^{(n)})^2\,  p_i\Cdot p_j}{(k\Cdot p_i)(k\Cdot p_j)} \right ] \;.
\end{align*}

We have a total of 10 soft integrals to evaluate, corresponding to the different cases of $i$ and $j$. Most of them can be evaluated using the results in the appendix of \cite{Beenakker:1988bq} and \cite{Harris:2001sx} (see also \cite{Somogyi:2011ir}). For $i=j$, we can write them as
\begin{align}
\label{eq:Fii}
\mathcal{F}^{\soft}_{ii} (\Des ) &= 
\left[ \frac{1}{2} \rsoft  -\ln \left( \frac{\Des m_B }{\mu} \right) \right] 
+\frac{1}{2\beta _{i}}\ln \left ( \frac{1+\beta _{i}}{1-\beta _{i}} \right ) \;,
\end{align}
where $\rsoft$ refers to the DR version in \eqref{eq:rsoft}, and all $\beta _{i}$ are measured in the $p_B$-RF, with $k=0$, since we are in the soft limit.\footnote{~The reason for measuring all $\beta _i$ in the $p_B$-RF is that it is the same frame in which we impose the cut-off on the photon energy, c.f.~\EQ\eqref{eq:Emax}.} We note that in the soft limit, the $\FRone$- and $\FRtwo$-frames are the same, and thus, we will use the two interchangeably in this section.  Further, we can isolate the  collinear logs in the case of small lepton masses by considering
\begin{align}
\label{eq:smallbi}
\frac{1}{2\beta _{i}}\ln \left ( \frac{1+\beta _{i}}{1-\beta _{i}} \right )= \frac{1}{2\beta _i}\ln \left ( \frac{\left ( 1+\beta _{i} \right )^2 }{1-\beta^2 _{i}} \right )\stackrel{m_i\to 0}{\longrightarrow} \frac{1}{2}\ln \frac{4E_i^2}{m_i^2}=-\ln m_i + \textrm{non-div} \;.
\end{align}
We now list the integrals corresponding to $i\neq j$.
The simplest one is 
\begin{align}
\label{eq:FiB}
\mathcal{F}^{\soft}_{i\, B} (\Des ) &=\left[ \frac{1}{2} \rsoft  -\ln \left( \frac{\Des m_B }{\mu} \right) \right]  I_{iB}^{(0)}  +\frac{1}{2\beta _{ i}}\left [ \Li \left ( \frac{2\beta _{ i}}{1+\beta _{ i}} \right ) +\frac{1}{4}\ln ^2 \left ( \frac{1+\beta _{ i}}{1-\beta _{ i}} \right )  \right ] \;,
\end{align}
where $I_{iB}^{(0)}$ can be obtained by using $j=B$ in \EQ\eqref{eq:Iijzero}.
 The 3 other non-diagonal integrals  require more work since they are not attributed to the frame in which the integral is evaluated. 
One can recast the remaining integrals as
\begin{align}
\label{eq:Fsoft}
\mathcal{F}^{\soft}_{ij} (\Des ) &=\left[ \frac{1}{2} \rsoft  -\ln \left( \frac{\Des m_B }{\mu} \right) \right]\Omega_{ij}  \;,
\end{align}
where $\Omega_{ij} = \Omega(\be_i,\be_j,\tau_{ij})$,
\begin{align}
\Omega (\be_i,\be_j,\tau_{ij})= &   P_{ij}                
 \int_0^\pi  \frac{d\Tg}{ \sin^{2\epsilon-1 }\Tg}
\int_0^\pi \frac{d\Fg}{ \sin^{2\epsilon }\Fg} \times \nonumber \\[0.1cm]
&  \left [ \frac{1}{(1-\be_i\cos \Tg )(1-\be_j \cos \Tg \cos \chi_{ij}-\be_j \sin \Tg \cos \Fg\sin \chi_{ij} )} \right ] \;,
\end{align}
where $\cos \chi_{ij}=2\tau_{ij}-1$, $\sin \chi_{ij}=\sqrt{1-\cos^2\chi_{ij}}$ 
and $P_{ij} = (1-\be_i \be_j (2 \tau_{ij}-1))/2\pi$.

Before matching $\be_i$, $\be_j$ and $\tau_{ij}$ to the cases we have, consider $\Omega (\be_i,\be_j,\tau_{ij})$. For $\be_i \neq1$ and $\be_j\neq 1$, the result to $\mathcal{O}(\epsilon )$ is not known in the literature. This is needed for isolating the collinear logs, since they arise from the $\mathcal{O}(\epsilon )$ part of the angular integrals multiplied by the $1/\eps$ from the $\rsoft$.

However, through \cite{gabor}, we were able to get an expression for $\Omega (\be_i,\be_j,\tau_{ij})$. The result is
\begin{alignat}{2}
&\Omega_{ij}&\;=\;& 
\frac{\pi\, P_{ij}}{2 \,\conetwo}\Bigg\{ \ln \left ( \frac{v_{ij}+\conetwo}{v_{ij}-\conetwo} \right ) +\epsilon \Bigg[ -\ln \left ( \frac{1-\coneone}{1+\coneone} \right )\ln\left ( \frac{\Rlog+\Slog}{\Rlog-\Slog} \right )+ \nonumber\\[5pt]
&&& \left ( \sum_{a, b=1 }^{4} \left [ -1+2\left ( \delta _{a2}+\delta_{a3}  \right )  \right ]\left [ 1-2\left ( \delta _{b3}+\delta _{b4} \right )  \right ] G(r^{(a)}_{ij},r^{(b)}_{ij},1) \right )   \Bigg ]  \Bigg \}\;.
\end{alignat}
The functions $G(a,b,1)$ are generalised polylogarithms of weight 2,
and for our parameters $a$ and $b$ the following representation holds
\begin{alignat}{2}
& G(a,b,1) &\;=\;&\text{Li}_2\left(\frac{b-1}{b-a}\right)-\text{Li}_2\left(\frac{b}{b-a}\right)+\ln
   \left(1-\frac{1}{b}\right) \ln \left(\frac{1-a}{b-a}\right) \;, \nonumber \\[0.1cm]
 &   G(a,a,1) &\;=\;& \frac{1}{2} \ln^2\left(1 - \frac{1}{a}\right) \;,
\end{alignat}
and
\begin{alignat}{6}
&r^{(1)}_{ij}&\;=\;&\frac{\glogf-\sqrt{\glogg}}{\glogh}\;,\quad& & r^{(2)}_{ij}&\;=\;&\frac{\glogf+\sqrt{\glogg}}{\glogh}\;,\nonumber\\[5pt]
&r^{(3)}_{ij}&\;=\;&r^{(1)}_{ij}|_{\beta _{i,j} \to -\beta _{i,j} }\;,\quad& & r^{(4)}_{ij}&\;=\;&r^{(2)}_{ij}|_{\beta _{i,j}\to -\beta_{i,j} }\;,\nonumber\\[5pt]
&\glogf&\;=\;&\beta _i\left ( \beta _j \left ( 1-2\tau _{ij} \right )+1  \right ) \;,\quad & & \glogh&\;=\;&\beta _i \left ( \beta _j +2-4\tau _{ij} \right )+\beta _j \;,\nonumber\\[5pt]
& \glogg&\;=\;&\beta _i^2\left ( 4\beta _j^2 \tau _{ij}\left ( \tau _{ij}-1 \right )+1  \right )+\beta _i \beta _j (2&-&4\tau _{ij})\!\!&\;+\;&\beta _j^2 \;,\nonumber\\[5pt]
&\Rlog&\;=\;&\coneone v_{ij}\ctwotwo-8v_{ii}v_{jj}+v_{ij}\;,\quad
& &\Slog&\;=\;&\left ( \coneone+\ctwotwo \right )\conetwo\;, \quad  \nonumber\\[5pt]
&\conetwo&\;=\;&\sqrt{v_{ij}^{2}-4v_{ii}v_{jj}}\;,\quad & & \coneone&\;=\;&\sqrt{1-4v_{ii}}\;,\quad & &\ctwotwo&\;=\;&\sqrt{1-4v_{jj}}\;, \nonumber\\[5pt]
&v_{ij}&\;=\;&\frac{1}{2}\left ( 1-\beta _i \beta_j \left ( 2\tau _{ij}-1 \right ) \right )\;,\quad& &v_{ii}&\;=\;&\frac{1}{4}\left ( 1-\beta _i^{2} \right )\;,\quad
& &v_{jj}&\;=\;&\frac{1}{4}\left ( 1-\beta _j^{2} \right )\nonumber
  \;,
\end{alignat}
with no summation over indices implied.
For the matching, we consider the momenta $\pK$, $\lone$ and $\ltwo$ in the $\FRtwo$-frame. Thus, for $\mathcal{F}^{\soft}_{K\lonetwo} (\Des )$, one has
\begin{align}
\beta _{K}=\frac{|\vec{p}_K^{\;\FRtwo}|}{E_{K}^{\FRtwo}}\;,\quad \beta _{\lonetwo}  =\frac{|\vec{\ell}_{1,2}^{\;\FRtwo}|}{E_{1,2}^{\FRtwo}}\;,\quad \tau_{K\lonetwo} =\frac{1}{2}\left ( 1-\frac{\ell_{1,2,x}^{\FRtwo}}{|\vec{\ell}_{1,2}^{\;\FRtwo}|} \right ) \; ,
\end{align}
where $\ell_{1,2,x}^{\FRtwo}$ corresponds to the $x-$component of $\lonetwo^{\FRtwo}$. 
Recall that the
$\be_i$'s can be evaluated either in the  $\FRone$-RF or $\FRtwo$-RF as these are equivalent 
in the $k \to 0$ limit assumed here. 

Finally, for $\mathcal{F}^{\soft}_{\lone \ltwo} (\Des )$, before the matching can be performed, one needs to perform a 3D rotation to eliminate the $y$-component of one of the momenta, for which we choose $\lone$. Thus, one has 
($\beta _{\lonetwo}$ is given above) 
\begin{align}
 \tau_{\lone\ltwo} =\frac{1}{2}\left ( 1+\frac{\ell_{2,x}^{\FRtwo}\cos\alpha-\ell_{2,y}^{\FRtwo}\sin \alpha  }{|\vec{\ell}_{2}^{\;\FRtwo}|} \right ) \;,
\end{align}
where, as before, the subscript on $\ltwo$ denotes the corresponding component of $\ltwo$. The angle of rotation $\alpha $ is defined via $\cos\alpha =\frac{\ell_{1,x}^{\FRtwo}}{\modveclone{\;\FRtwo}}$ and $\sin \alpha =\sqrt{1-\cos^2\alpha }$. Taking the limit of small lepton masses, one can isolate the collinear logs and obtain
\begin{align}
\label{eq:Fijhard}
\mathcal{F}^{\soft}_{\lone\ltwo} (\Des ) &=\left [ \frac{1}{2}\Dep -\ln \left (\Des m_B \right )  \right ]I_{\lone\ltwo}^{(0)}+
\left( \frac{1}{2}\ln^2 m_{\lone} -\ln m_{\lone} \ln \left ( 2E_{\lone}^{\FRone} \right ) +  \{1 \leftrightarrow 2\} \right)+\textrm{finite} \;, \nonumber\\[5pt]
\mathcal{F}^{\soft}_{K\lone} (\Des ) &=\left [ \frac{1}{2}\Dep -\ln \left (\Des m_B \right )  \right ]I_{K\lonetwo}^{(0)}+\frac{1}{2}\ln^2 m_{\lone}-\ln m_{\lone} \ln \left ( 2E_{\lone}^{\FRone} \right )+\textrm{finite} \;.
\end{align}
We now collect all single logs in $ {\cal F}^{\soft}(\Des )\equiv\sum_{i, j }   \hat{Q}_i \hat{Q}_j    {\cal F}^{\soft}_{ij}(\Des ) $. To this end, consider the divergent parts of the different limits of $I_{ij}^{(0)}$.
\begin{align}
\label{eq:Iijsmallml}
I_{ij}^{(0)}\to \Bigg\{
\begin{array}{lll}-\ln m_{i}&\quad m_{i} & \ll  \mK,m_B\\[5pt]
  -  \ln m_i - \ln m_j  &\quad m_{i}\simeq m_{j}& \ll \mK,m_B
\end{array} \;.
\end{align}
Assembling all bits and pieces, and using charge conservation,  we obtain 
\begin{alignat}{2}
& {\cal F}^{\soft}(\Des )|_{\ln m_{\lonetwo}}&\;=\;& \hat{Q}_{\lone}^2 \ln m_{\lone} ( 2\ln 2E_{\ltwo}^{\FRone}   -  (1  + 2 \ln \left ( \Des m_B \right )) +  \;
 \{ 1 \leftrightarrow 2 \}   \nonumber \\
 &  &\;=\;& \hat{Q}_{\lone}^2 \ln m_{\lone} \left [ -1-2\ln \left ( \bar{z}(\Des) \right )  \right ] +  \;
 \{ 1 \leftrightarrow 2 \}  \;,
 \label{eq:Fsoftcoll}
\end{alignat}

where we have used $2\hat{E}_{\lone}^{\FRone} \equiv 1-\hat{s}_{K\ltwo}$ to arrive at the final result, and $\bar{z}(\Des) \equiv 1-z(\Des)$ with $z(\Des)$ given in \EQ\eqref{eq:z}.

\section{Passarino-Veltman Functions}
\label{app:PV}

The aim of this appendix is to give a minimal self-contained discussion of the Passarino-Veltman functions 
appearing in our results. The integrals are defined in \cite{Denner:1991kt},
\begin{equation}
\label{eq:PaVeConvention}
I_n \equiv \frac{(2 \pi \mu)^{4\mi d}}{i \pi^{2}} \int \mathrm{d}^dl \frac{1}{(l^2 \mi m_0^2 \pl
  i0)((l\pl\lone)^2 \mi m_1^2  \pl i0)((l\pl\lone \pl \ltwo)^2 \mi m_2^2 \pl i0) \dots } \;,
\end{equation}
where $n=1,2,3,4$ form a complete $1$-loop basis and 
are usually referred to as $A_0,  B_0,C_0,D_0$ respectively.
 For our case, $n=1,2,3$ are sufficient.  
The $A_0$ and $B_0$ functions are given to ${\cal O}(\veps^0)$, with $d = 4 - 2 \veps$,
\begin{alignat}{2}
\label{eq:PaVe}
& A_0(m^2) &&=  m^2( \DepsUV + 1  - \ln \left(\frac{m^2}{\mu^2}\right) )+ {\cal O}(\veps) \;,   \\
&  B_0(s,m_0^2,m_1^2) && =    \left( \DepsUV  +    2 -  \ln \frac{m_0 m_1}{\mu^2}   +   \frac{ m_0^2-m_1^2}{s} \ln \frac{m_1}{m_0} - 
\frac{ m_0 m_1}{s} ( \frac{1}{r}-r) \ln r \right) + {\cal O}(\veps) \;,    \nonumber
\end{alignat}
where $ r = -\frac{1}{2}( -b + \sqrt{b^2-4})$ with $b = -\frac{s-m_0^2 - m_1^2 + i0}{ m_0m_1}$, and $\DepsUV$ is given in \EQ\eqref{eq:epshat}.

The $C_0$ function used is $C_0 (s,t,u,m_0^2,m_1^2,m_2^2)$, 
where the cuts of the momenta $\{s,t,u\}$  start at $\{ (m_0+m_1)^2,(m_0+m_2)^2,(m_1+m_2)^2\}$ 
respectively. This is the same convention used in FeynCalc \cite{FeynCalc1,FeynCalc2} and \cite{Denner:1991kt}.

The $C_0$ function can be found in the review paper \cite{Ditt} (\EQ B.5), valid for small photon mass (up to ${\cal O}(\mga^2)$ corrections) in mass regularisation and to $\mathcal{O}(\epsilon^0 )$ in DR,
\begin{eqnarray}
\label{eq:C0soft}
C_0 \! &=& \!
\frac{x_{ij}}{m_{i} m_{j} (1-x_{ij}^2)} \biggl\{
\left ( 	 \ln\left(\frac{m_{i} m_{j}}{\mu ^2}\right) -\rsoft \right ) \ln(x_{ij}) - \frac{1}{2}\ln^2(x_{ij})+ 2\ln(x_{ij})\ln(1-x_{ij}^2)
\nonumber \\[.5em]
&+& \!\! \frac{1}{2}\ln^2\left(\frac{m_{i}}{m_{j}}\right)
-\frac{\pi^2}{6}  + \Li(x_{ij}^2) + \Li \left(1-x_{ij}\frac{m_{i}}{m_{j}}\right)+\Li \left(1-x_{ij}\frac{m_{j}}{m_{i}}\right) \biggr\} \;,
\end{eqnarray}
where  $C_0 \equiv C_0( m_{i}^2, m_{j}^2, (\hat{p}_i \pl \hat{p}_j)^2, m_{i}^2, \mga^2 ,m_{j}^2 ) $,
$\rsoft$ is defined in \eqref{eq:rsoft}, and
\begin{equation}
x_{ij} \equiv  \frac{\sqrt{ y_{ij} }-1}{\sqrt{ y_{ij} }+1}  \;, 
\quad y_{ij} \equiv \frac{  (\hat{p}_i \pl \hat{p}_j)^2 \mi (m_i \pl m_{j})^2\pl i0} {(\hat{p}_i \pl \hat{p}_j)^2   \mi (m_i\mi m_{j})^2 \pl i0}  \;.
\end{equation}

\bibliographystyle{utphys}
\bibliography{References_QED}

\providecommand{\href}[2]{#2}\begingroup\raggedright\begin{thebibliography}{10}

\bibitem{Aaij:2014ora}
{\bfseries LHCb} Collaboration, R.~Aaij {\em et~al.}, ``{Test of lepton
  universality using $B^{+}\rightarrow K^{+}\ell^{+}\ell^{-}$ decays},''
  \href{http://dx.doi.org/10.1103/PhysRevLett.113.151601}{{\em Phys. Rev.
  Lett.} {\bfseries 113} (2014) 151601},
\href{http://arxiv.org/abs/1406.6482}{{\ttfamily arXiv:1406.6482 [hep-ex]}}.

\bibitem{Aaij:2017vbb}
{\bfseries LHCb} Collaboration, R.~Aaij {\em et~al.}, ``{Test of lepton
  universality with $B^{0} \rightarrow K^{*0}\ell^{+}\ell^{-}$ decays},''
  \href{http://dx.doi.org/10.1007/JHEP08(2017)055}{{\em JHEP} {\bfseries 08}
  (2017) 055},
\href{http://arxiv.org/abs/1705.05802}{{\ttfamily arXiv:1705.05802 [hep-ex]}}.

\bibitem{Aaij:2019wad}
{\bfseries LHCb} Collaboration, R.~Aaij {\em et~al.}, ``{Search for
  lepton-universality violation in $B^+\to K^+\ell^+\ell^-$ decays},''
  \href{http://dx.doi.org/10.1103/PhysRevLett.122.191801}{{\em Phys. Rev.
  Lett.} {\bfseries 122} no.~19, (2019) 191801},
\href{http://arxiv.org/abs/1903.09252}{{\ttfamily arXiv:1903.09252 [hep-ex]}}.

\bibitem{Bifani:2018zmi}
S.~Bifani, S.~Descotes-Genon, A.~Romero~Vidal, and M.-H. Schune, ``{Review of
  Lepton Universality tests in $B$ decays},''
  \href{http://dx.doi.org/10.1088/1361-6471/aaf5de}{{\em J. Phys. G} {\bfseries
  46} no.~2, (2019) 023001}, \href{http://arxiv.org/abs/1809.06229}{{\ttfamily
  arXiv:1809.06229 [hep-ex]}}.

\bibitem{PHOTOS}
N.~Davidson, T.~Przedzinski, and Z.~Was, ``{PHOTOS interface in C++: Technical
  and Physics Documentation},''
  \href{http://dx.doi.org/10.1016/j.cpc.2015.09.013}{{\em Comput. Phys.
  Commun.} {\bfseries 199} (2016) 86--101},
\href{http://arxiv.org/abs/1011.0937}{{\ttfamily arXiv:1011.0937 [hep-ph]}}.

\bibitem{BIP16}
M.~Bordone, G.~Isidori, and A.~Pattori, ``{On the Standard Model predictions
  for $R_K$ and $R_{K^*}$},''
  \href{http://dx.doi.org/10.1140/epjc/s10052-016-4274-7}{{\em Eur. Phys. J.}
  {\bfseries C76} no.~8, (2016) 440},
\href{http://arxiv.org/abs/1605.07633}{{\ttfamily arXiv:1605.07633 [hep-ph]}}.

\bibitem{PDG}
{\bfseries Particle Data Group} Collaboration, P.~Zyla {\em et~al.}, ``{Review
  of Particle Physics},'' \href{http://dx.doi.org/10.1093/ptep/ptaa104}{{\em
  PTEP} {\bfseries 2020} no.~8, (2020) 083C01}.

\bibitem{BFS01}
M.~Beneke, T.~Feldmann, and D.~Seidel, ``{Systematic approach to exclusive $B
  \to V l^+ l^-, V \gamma$ decays},''
  \href{http://dx.doi.org/10.1016/S0550-3213(01)00366-2}{{\em Nucl.Phys.}
  {\bfseries B612} (2001) 25--58},
\href{http://arxiv.org/abs/hep-ph/0106067}{{\ttfamily arXiv:hep-ph/0106067
  [hep-ph]}}.

\bibitem{DLZ2012}
M.~Dimou, J.~Lyon, and R.~Zwicky, ``{Exclusive Chromomagnetism in
  heavy-to-light FCNCs},''
  \href{http://dx.doi.org/10.1103/PhysRevD.87.074008}{{\em Phys.Rev.}
  {\bfseries D87} no.~7, (2013) 074008},
\href{http://arxiv.org/abs/1212.2242}{{\ttfamily arXiv:1212.2242 [hep-ph]}}.

\bibitem{LZ2013}
J.~Lyon and R.~Zwicky, ``{Isospin asymmetries in $B\to(K^*,\rho) \to l^+l^-$
  and $B\to Kl^+l^-$ in and beyond the standard model},''
  \href{http://dx.doi.org/10.1103/PhysRevD.88.094004}{{\em Phys.Rev.}
  {\bfseries D88} no.~9, (2013) 094004},
\href{http://arxiv.org/abs/1305.4797}{{\ttfamily arXiv:1305.4797 [hep-ph]}}.

\bibitem{Khodjamirian:2012rm}
A.~Khodjamirian, T.~Mannel, and Y.~Wang, ``{$B \to K \ell^{+}\ell^{-}$ decay at
  large hadronic recoil},''
  \href{http://dx.doi.org/10.1007/JHEP02(2013)010}{{\em JHEP} {\bfseries 1302}
  (2013) 010},
\href{http://arxiv.org/abs/1211.0234}{{\ttfamily arXiv:1211.0234 [hep-ph]}}.

\bibitem{Kubis:2010mp}
B.~Kubis and R.~Schmidt, ``{Radiative corrections in $K -> \pi \l^+ \l^-$
  decays},'' \href{http://dx.doi.org/10.1140/epjc/s10052-010-1442-z}{{\em Eur.
  Phys. J. C} {\bfseries 70} (2010) 219--231},
  \href{http://arxiv.org/abs/1007.1887}{{\ttfamily arXiv:1007.1887 [hep-ph]}}.

\bibitem{Ellis:1980wv}
R.~Ellis, D.~Ross, and A.~Terrano, ``{The Perturbative Calculation of Jet
  Structure in e+ e- Annihilation},''
  \href{http://dx.doi.org/10.1016/0550-3213(81)90165-6}{{\em Nucl. Phys. B}
  {\bfseries 178} (1981) 421--456}.

\bibitem{Weinberg:1995mt}
S.~Weinberg, {\em {The Quantum theory of fields. Vol. 1: Foundations}}.
\newblock Cambridge University Press,
2005.
\newblock

\bibitem{Muta:1998vi}
T.~Muta, {\em {Foundations of quantum chromodynamics. Second edition}},
  vol.~57.
\newblock 1998.

\bibitem{Sterman:1994ce}
G.~F. Sterman, {\em {An Introduction to quantum field theory}}.
\newblock Cambridge University Press, 8, 1993.

\bibitem{Ginsberg:1969jh}
E.~S. Ginsberg, ``{Radiative corrections to the k-l-3 +- dalitz plot},''
  \href{http://dx.doi.org/10.1103/physrev.187.2280.2,
  10.1103/PhysRev.162.1570}{{\em Phys. Rev.} {\bfseries 162} (1967) 1570}.
[Erratum: Phys. Rev.187,2280(1969)].

\bibitem{Catani:1996vz}
S.~Catani and M.~Seymour, ``{A General algorithm for calculating jet
  cross-sections in NLO QCD},''
  \href{http://dx.doi.org/10.1016/S0550-3213(96)00589-5}{{\em Nucl. Phys. B}
  {\bfseries 485} (1997) 291--419},
  \href{http://arxiv.org/abs/hep-ph/9605323}{{\ttfamily arXiv:hep-ph/9605323}}.
  [Erratum: Nucl.Phys.B 510, 503--504 (1998)].

\bibitem{Dittmaier:1999mb}
S.~Dittmaier, ``{A General approach to photon radiation off fermions},''
  \href{http://dx.doi.org/10.1016/S0550-3213(99)00563-5}{{\em Nucl. Phys. B}
  {\bfseries 565} (2000) 69--122},
  \href{http://arxiv.org/abs/hep-ph/9904440}{{\ttfamily arXiv:hep-ph/9904440}}.

\bibitem{Dittmaier_2008}
S.~Dittmaier, A.~Kabelschacht, and T.~Kasprzik, ``Polarized qed splittings of
  massive fermions and dipole subtraction for non-collinear-safe observables,''
  \href{http://dx.doi.org/10.1016/j.nuclphysb.2008.03.010}{{\em Nuclear Physics
  B} {\bfseries 800} no.~1-2, (Sep, 2008) 146–189}.
  \url{http://dx.doi.org/10.1016/j.nuclphysb.2008.03.010}.

\bibitem{Sch_nherr_2018}
M.~Schönherr, ``An automated subtraction of nlo ew infrared divergences,''
  \href{http://dx.doi.org/10.1140/epjc/s10052-018-5600-z}{{\em The European
  Physical Journal C} {\bfseries 78} no.~2, (Feb, 2018) }.
  \url{http://dx.doi.org/10.1140/epjc/s10052-018-5600-z}.

\bibitem{Harris:2001sx}
B.~W. Harris and J.~F. Owens, ``{The Two cutoff phase space slicing method},''
  \href{http://dx.doi.org/10.1103/PhysRevD.65.094032}{{\em Phys. Rev.}
  {\bfseries D65} (2002) 094032},
\href{http://arxiv.org/abs/hep-ph/0102128}{{\ttfamily arXiv:hep-ph/0102128
  [hep-ph]}}.

\bibitem{Bloch:1937pw}
F.~Bloch and A.~Nordsieck, ``{Note on the Radiation Field of the electron},''
  \href{http://dx.doi.org/10.1103/PhysRev.52.54}{{\em Phys. Rev.} {\bfseries
  52} (1937) 54--59}.

\bibitem{Ball:2004ye}
P.~Ball and R.~Zwicky, ``{New results on $B \to \pi, K, \eta$ decay formfactors
  from light-cone sum rules},''
  \href{http://dx.doi.org/10.1103/PhysRevD.71.014015}{{\em Phys. Rev. D}
  {\bfseries 71} (2005) 014015},
  \href{http://arxiv.org/abs/hep-ph/0406232}{{\ttfamily arXiv:hep-ph/0406232}}.

\bibitem{Bali_2019}
G.~S. Bali, V.~M. Braun, S.~Bürger, M.~Göckeler, M.~Gruber, F.~Hutzler,
  P.~Korcyl, A.~Schäfer, A.~Sternbeck, and et~al., ``Light-cone distribution
  amplitudes of pseudoscalar mesons from lattice qcd,''
  \href{http://dx.doi.org/10.1007/jhep08(2019)065}{{\em Journal of High Energy
  Physics} {\bfseries 2019} no.~8, (Aug, 2019) }.
  \url{http://dx.doi.org/10.1007/JHEP08(2019)065}.

\bibitem{Braun:2004vf}
V.~Braun and A.~Lenz, ``{On the SU(3) symmetry-breaking corrections to meson
  distribution amplitudes},''
  \href{http://dx.doi.org/10.1103/PhysRevD.70.074020}{{\em Phys. Rev. D}
  {\bfseries 70} (2004) 074020},
  \href{http://arxiv.org/abs/hep-ph/0407282}{{\ttfamily arXiv:hep-ph/0407282}}.

\bibitem{Ball:2005vx}
P.~Ball and R.~Zwicky, ``{SU(3) breaking of leading-twist K and K* distribution
  amplitudes: A Reprise},''
  \href{http://dx.doi.org/10.1016/j.physletb.2005.11.068}{{\em Phys. Lett. B}
  {\bfseries 633} (2006) 289--297},
  \href{http://arxiv.org/abs/hep-ph/0510338}{{\ttfamily arXiv:hep-ph/0510338}}.

\bibitem{Ball:2006fz}
P.~Ball and R.~Zwicky, ``{Operator relations for SU(3) breaking contributions
  to K and K* distribution amplitudes},''
  \href{http://dx.doi.org/10.1088/1126-6708/2006/02/034}{{\em JHEP} {\bfseries
  02} (2006) 034}, \href{http://arxiv.org/abs/hep-ph/0601086}{{\ttfamily
  arXiv:hep-ph/0601086}}.

\bibitem{Chetyrkin:2007vm}
K.~Chetyrkin, A.~Khodjamirian, and A.~Pivovarov, ``{Towards NNLO Accuracy in
  the QCD Sum Rule for the Kaon Distribution Amplitude},''
  \href{http://dx.doi.org/10.1016/j.physletb.2008.02.031}{{\em Phys. Lett. B}
  {\bfseries 661} (2008) 250--258},
  \href{http://arxiv.org/abs/0712.2999}{{\ttfamily arXiv:0712.2999 [hep-ph]}}.

\bibitem{Greco:1975rm}
M.~Greco, G.~Pancheri-Srivastava, and Y.~Srivastava, ``{Radiative Corrections
  for Colliding Beam Resonances},''
  \href{http://dx.doi.org/10.1016/0550-3213(75)90304-1}{{\em Nucl. Phys. B}
  {\bfseries 101} (1975) 234--262}.

\bibitem{DAmbrosio:1994bks}
G.~D'Ambrosio and G.~Isidori, ``{K $\to$ pi pi gamma decays: A Search for novel
  couplings in kaon decays},'' \href{http://dx.doi.org/10.1007/BF01578672}{{\em
  Z. Phys. C} {\bfseries 65} (1995) 649--656},
  \href{http://arxiv.org/abs/hep-ph/9408219}{{\ttfamily arXiv:hep-ph/9408219}}.

\bibitem{DAmbrosio:1996jmq}
G.~D'Ambrosio, G.~Ecker, G.~Isidori, and H.~Neufeld, ``{K $\to$ pi pi pi gamma
  in chiral perturbation theory},''
  \href{http://dx.doi.org/10.1007/s002880050554}{{\em Z. Phys. C} {\bfseries
  76} (1997) 301--310}, \href{http://arxiv.org/abs/hep-ph/9612412}{{\ttfamily
  arXiv:hep-ph/9612412}}.

\bibitem{BBS17}
M.~Beneke, C.~Bobeth, and R.~Szafron, ``{Enhanced electromagnetic correction to
  the rare $B$-meson decay $B_{s,d} \to \mu^+ \mu^-$},''
\href{http://arxiv.org/abs/1708.09152}{{\ttfamily arXiv:1708.09152 [hep-ph]}}.

\bibitem{Beneke:2020vnb}
M.~Beneke, P.~B\"oer, J.-N. Toelstede, and K.~K. Vos, ``{QED factorization of
  non-leptonic $B$ decays},'' \href{http://arxiv.org/abs/2008.10615}{{\ttfamily
  arXiv:2008.10615 [hep-ph]}}.

\bibitem{Beneke:2019slt}
M.~Beneke, C.~Bobeth, and R.~Szafron, ``{Power-enhanced leading-logarithmic QED
  corrections to $B_q \to \mu^+\mu^-$},''
  \href{http://dx.doi.org/10.1007/JHEP10(2019)232}{{\em JHEP} {\bfseries 10}
  (2019) 232}, \href{http://arxiv.org/abs/1908.07011}{{\ttfamily
  arXiv:1908.07011 [hep-ph]}}.

\bibitem{Carrasco:2015xwa}
N.~Carrasco, V.~Lubicz, G.~Martinelli, C.~T. Sachrajda, N.~Tantalo,
  C.~Tarantino, and M.~Testa, ``{QED Corrections to Hadronic Processes in
  Lattice QCD},'' \href{http://dx.doi.org/10.1103/PhysRevD.91.074506}{{\em
  Phys. Rev.} {\bfseries D91} no.~7, (2015) 074506},
\href{http://arxiv.org/abs/1502.00257}{{\ttfamily arXiv:1502.00257 [hep-lat]}}.

\bibitem{Sachrajda:2019uhh}
C.~Sachrajda, M.~Di~Carlo, G.~Martinelli, D.~Giusti, V.~Lubicz, F.~Sanfilippo,
  S.~Simula, and N.~Tantalo, ``{Radiative corrections to semileptonic decay
  rates},'' in {\em {37th International Symposium on Lattice Field Theory}}.
\newblock 10, 2019.
\newblock \href{http://arxiv.org/abs/1910.07342}{{\ttfamily arXiv:1910.07342
  [hep-lat]}}.

\bibitem{Giusti:2017dwk}
D.~Giusti, V.~Lubicz, G.~Martinelli, C.~Sachrajda, F.~Sanfilippo, S.~Simula,
  N.~Tantalo, and C.~Tarantino, ``{First lattice calculation of the QED
  corrections to leptonic decay rates},''
  \href{http://dx.doi.org/10.1103/PhysRevLett.120.072001}{{\em Phys. Rev.
  Lett.} {\bfseries 120} no.~7, (2018) 072001},
  \href{http://arxiv.org/abs/1711.06537}{{\ttfamily arXiv:1711.06537
  [hep-lat]}}.

\bibitem{DiCarlo:2019thl}
M.~Di~Carlo, D.~Giusti, V.~Lubicz, G.~Martinelli, C.~Sachrajda, F.~Sanfilippo,
  S.~Simula, and N.~Tantalo, ``{Light-meson leptonic decay rates in lattice
  QCD+QED},'' \href{http://dx.doi.org/10.1103/PhysRevD.100.034514}{{\em Phys.
  Rev. D} {\bfseries 100} no.~3, (2019) 034514},
  \href{http://arxiv.org/abs/1904.08731}{{\ttfamily arXiv:1904.08731
  [hep-lat]}}.

\bibitem{Portelli:2019}
A.~Portelli, ``{Electromagnetic corrections to leptonic decays}.''. {talk given
  at 37th International Symposium on Lattice Field Theory, Wuhan, China, June
  2019}.

\bibitem{Gratrex:2015hna}
J.~Gratrex, M.~Hopfer, and R.~Zwicky, ``{Generalised helicity formalism, higher
  moments and the $B \to K_{J_K}(\to K \pi) \bar{\ell}_1 \ell_2$ angular
  distributions},'' \href{http://dx.doi.org/10.1103/PhysRevD.93.054008}{{\em
  Phys. Rev.} {\bfseries D93} no.~5, (2016) 054008},
\href{http://arxiv.org/abs/1506.03970}{{\ttfamily arXiv:1506.03970 [hep-ph]}}.

\bibitem{Cetal01}
V.~Cirigliano, M.~Knecht, H.~Neufeld, H.~Rupertsberger, and P.~Talavera,
  ``{Radiative corrections to K(l3) decays},''
  \href{http://dx.doi.org/10.1007/s100520100825}{{\em Eur. Phys. J.} {\bfseries
  C23} (2002) 121--133},
\href{http://arxiv.org/abs/hep-ph/0110153}{{\ttfamily arXiv:hep-ph/0110153
  [hep-ph]}}.

\bibitem{CGH08}
V.~Cirigliano, M.~Giannotti, and H.~Neufeld, ``{Electromagnetic effects in
  K(l3) decays},'' \href{http://dx.doi.org/10.1088/1126-6708/2008/11/006}{{\em
  JHEP} {\bfseries 11} (2008) 006},
\href{http://arxiv.org/abs/0807.4507}{{\ttfamily arXiv:0807.4507 [hep-ph]}}.

\bibitem{deBoer:2018ipi}
S.~de~Boer, T.~Kitahara, and I.~Nisandzic, ``{Soft-Photon Corrections to
  $\bar{B} \to D \tau^{-} \bar{\nu}_{\tau}$ Relative to $\bar{B} \to D \mu^{-}
  \bar{\nu}_{\mu}$},''
  \href{http://dx.doi.org/10.1103/PhysRevLett.120.261804}{{\em Phys. Rev.
  Lett.} {\bfseries 120} no.~26, (2018) 261804},
  \href{http://arxiv.org/abs/1803.05881}{{\ttfamily arXiv:1803.05881
  [hep-ph]}}.

\bibitem{Bobeth2004}
C.~Bobeth, P.~Gambino, M.~Gorbahn, and U.~Haisch, ``Complete nnlo qcd analysis
  ofb xsl l and higher order electroweak effects,''
  \href{http://dx.doi.org/10.1088/1126-6708/2004/04/071}{{\em Journal of High
  Energy Physics} {\bfseries 2004} no.~04, (Apr, 2004) 071–071}.
  \url{http://dx.doi.org/10.1088/1126-6708/2004/04/071}.

\bibitem{HLMW05}
T.~Huber, E.~Lunghi, M.~Misiak, and D.~Wyler, ``{Electromagnetic logarithms in
  $\bar B \to X_s l^+ l^-$},''
  \href{http://dx.doi.org/10.1016/j.nuclphysb.2006.01.037}{{\em Nucl. Phys.}
  {\bfseries B740} (2006) 105--137},
\href{http://arxiv.org/abs/hep-ph/0512066}{{\ttfamily arXiv:hep-ph/0512066
  [hep-ph]}}.

\bibitem{Peskin:1995ev}
M.~E. Peskin and D.~V. Schroeder, {\em {An Introduction to quantum field
  theory}}.
\newblock Addison-Wesley, Reading, USA, 1995.

\bibitem{Hiller:2013cza}
G.~Hiller and R.~Zwicky, ``{(A)symmetries of weak decays at and near the
  kinematic endpoint},'' \href{http://dx.doi.org/10.1007/JHEP03(2014)042}{{\em
  JHEP} {\bfseries 03} (2014) 042},
  \href{http://arxiv.org/abs/1312.1923}{{\ttfamily arXiv:1312.1923 [hep-ph]}}.

\bibitem{Bordone:2015nqa}
M.~Bordone, A.~Greljo, G.~Isidori, D.~Marzocca, and A.~Pattori, ``{Higgs Pseudo
  Observables and Radiative Corrections},''
  \href{http://dx.doi.org/10.1140/epjc/s10052-015-3611-6}{{\em Eur. Phys. J. C}
  {\bfseries 75} no.~8, (2015) 385},
  \href{http://arxiv.org/abs/1507.02555}{{\ttfamily arXiv:1507.02555
  [hep-ph]}}.

\bibitem{Beenakker:1988bq}
W.~Beenakker, H.~Kuijf, W.~L. van Neerven, and J.~Smith, ``{QCD Corrections to
  Heavy Quark Production in p anti-p Collisions},''
\href{http://dx.doi.org/10.1103/PhysRevD.40.54}{{\em Phys. Rev.} {\bfseries
  D40} (1989) 54--82}.

\bibitem{Somogyi:2011ir}
G.~Somogyi, ``{Angular integrals in d dimensions},''
  \href{http://dx.doi.org/10.1063/1.3615515}{{\em J. Math. Phys.} {\bfseries
  52} (2011) 083501},
\href{http://arxiv.org/abs/1101.3557}{{\ttfamily arXiv:1101.3557 [hep-ph]}}.

\bibitem{gabor}
G.~Somogyi, ``{private communication}.''.

\bibitem{Denner:1991kt}
A.~Denner, ``{Techniques for calculation of electroweak radiative corrections
  at the one loop level and results for W physics at LEP-200},''
  \href{http://dx.doi.org/10.1002/prop.2190410402}{{\em Fortsch. Phys.}
  {\bfseries 41} (1993) 307--420},
\href{http://arxiv.org/abs/0709.1075}{{\ttfamily arXiv:0709.1075 [hep-ph]}}.

\bibitem{FeynCalc1}
R.~Mertig, M.~Bohm, and A.~Denner, ``{FEYN CALC: Computer algebraic calculation
  of Feynman amplitudes},''
\href{http://dx.doi.org/10.1016/0010-4655(91)90130-D}{{\em Comput. Phys.
  Commun.} {\bfseries 64} (1991) 345--359}.

\bibitem{FeynCalc2}
V.~Shtabovenko, R.~Mertig, and F.~Orellana, ``{New Developments in FeynCalc
  9.0},'' \href{http://dx.doi.org/10.1016/j.cpc.2016.06.008}{{\em Comput. Phys.
  Commun.} {\bfseries 207} (2016) 432--444},
\href{http://arxiv.org/abs/1601.01167}{{\ttfamily arXiv:1601.01167 [hep-ph]}}.

\bibitem{Ditt}
S.~Dittmaier, ``{Separation of soft and collinear singularities from one loop N
  point integrals},''
  \href{http://dx.doi.org/10.1016/j.nuclphysb.2003.10.003}{{\em Nucl. Phys.}
  {\bfseries B675} (2003) 447--466},
\href{http://arxiv.org/abs/hep-ph/0308246}{{\ttfamily arXiv:hep-ph/0308246
  [hep-ph]}}.

\end{thebibliography}\endgroup

\end{document}